\documentclass[prb,twocolumn,floatfix]{revtex4}

\usepackage{amsfonts}
\usepackage{amsmath}
\usepackage{verbatim}
\usepackage[dvips]{graphicx}
\usepackage[dvips]{color}
\usepackage{subfigure}

\begin{document}

\title{Two-dimensional heavy fermions on the strongly correlated boundaries of Kondo topological insulators}
\author{Predrag Nikoli\'c$^{1,2}$}
\affiliation{$^1$School of Physics, Astronomy and Computational Sciences,\\George Mason University, Fairfax, VA 22030, USA}
\affiliation{$^2$Institute for Quantum Matter at Johns Hopkins University, Baltimore, MD 21218, USA}
\date{\today}


\begin{abstract}

Samarium hexaboride (SmB$_6$), a representative Kondo insulator, has been characterized recently as a likely topological insulator. It is also a material with strong electron correlations, evident by the temperature dependence of its bandgap and the existence of a nearly flat collective mode whose energy lies within the bandgap. Similar strong correlations can affect or even destabilize the two-dimensional metallic state of topological origin at the crystal boundary. Here we construct the minimal lattice model of the correlated boundary of topological Kondo insulators, and make phenomenological predictions for its possible ground states. Depending on the microscopic properties of the interface between the topological Kondo material and a conventional insulator, the boundary metal can exhibit a varied degree of hybridization between the $d$ and $f$ orbitals of the rare earth element, yielding a rich two-dimensional heavy fermion phenomenology. A pronounced participation of the $f$ orbitals is expected to create a heavy fermion Dirac metal, possibly unstable to a spin density wave, electron localization or even superconductivity. The opposite limit of ``localized magnetic moments'' helped by the partial Kondo screening on the crystal boundary can bring about a non-Fermi liquid of $d$ electrons that exhibits two-dimensional quantum electrodynamics, or other unconventional states. In addition, ultra-thin films made from topological Kondo insulators could open the possibility of creating exotic incompressible quantum liquids with non-Abelian fractional excitations, whose dynamics shaped by the strong Rashba spin-orbit coupling resembles that of fractional quantum Hall systems.

\end{abstract}

\maketitle

\section{Introduction}

Topological insulators (TI) are materials whose crystal boundaries host a metallic state protected by the time-reversal (TR) symmetry, while the bulk is (ideally) insulating \cite{Hasan2010, Moore2010, Qi2010a}. The ideal crystals of all well-characterized ``strong'' TI materials, such as Bi$_2$Se$_3$ and Bi$_2$Te$_3$, or half-Heusler compounds \cite{Hsin2010, Lin2010b}, are \emph{uncorrelated} band-insulators shaped by the spin-orbit coupling. The bulk is a band-insulator identified by multiple Z$_2$ topological invariants \cite{Fu2007, Moore2007, Chen2009b}, while the surface hosts a two-dimensional metallic state born out of the spin-momentum-locked massless Dirac spectrum. As long as the TR symmetry is respected, the surface metal is protected \cite{Fu2007} against disorder, interactions and any other source of electron back-scattering that causes Anderson localization in two-dimensional metals \cite{abrahams79}.

Very promising materials that combine interactions and a strong spin-orbit coupling are Kondo insulators (SmB$_6$, YbB$_{12}$, Ce$_3$Bi$_4$Pt$_3$, CeNiSn, CeRhSb, Ce$_3$Pt$_3$Sb$_3$, UNiSn, etc.) \cite{Menth1969, Nickerson1971, Hundley1990, Riseborough1992, Alekseev1993, Nyhus1995, Sera1996, Okamura1998, Bouvet1998, Gorshunov1999, Riseborough2000} and iridium oxides (Pr$_2$Ir$_2$O$_7$, Sr$_2$IrO$_4$, Na$_2$IrO$_3$, etc.) \cite{Kim2009, Shitade2009, Pesin2010, Wang2011b, Kim2012c, Gretarsson2013, Mazin2013}. The former are insulating heavy fermion systems that were recently theoretically characterized as likely three-dimensional strong TIs \cite{Dzero2010, Takimoto2011, Dzero2012, Alex2013, Dzero2013} or topological crystal insulators \cite{Ye2013}. All heavy fermion materials feature some degree of hybridization between the $f$ and $d$ orbitals of their rare earth element. The $f$ orbitals have a very small intrinsic bandwidth in the crystal environment, while the $d$ orbitals are broadly dispersing. Coulomb interactions tend to localize electrons in the $f$ orbitals when they become dense, and produce Kondo singlet correlations between the $f$ and $d$ electrons. The resulting ground state of Kondo TIs is a strongly correlated insulator.

The most studied Kondo insulator material is samarium hexaboride (SmB$_6$). The Fermi energy of a Kondo insulator lies within a small temperature-dependent bandgap \cite{Hundley1994}. Neutron scattering experiments have revealed a gapped coherent collective mode with a fairly narrow dispersion that also resides inside the bandgap \cite{Alekseev1993, Kasuya1994, Bouvet1998, Fuhrman2014}. The slave boson theory of Kondo materials views this mode as an exciton, which is created by the Coulomb interactions among the narrow-band $f$ electrons and protected against decay by the material's bandgap \cite{Riseborough1992, Riseborough2000}. In addition to this evidence of correlations, some Kondo materials exhibit a characteristic insulating temperature dependence of the DC conductivity until the lowest temperatures where the conductivity saturates at a finite value \cite{Kasuya1979, Kebede1996, Flachbart2001, Barla2005, Derr2008}. While this may or may not be related to the existence of a metallic surface, the experimental support for SmB$_6$ being a TI is growing \cite{Zhang2013, Wolgast2013, Kim2013, Kim2013a, Xiang2013, Rossler2013, Denlinger2013, Thomas2013, Neupane2013, Jiang2013, Phelan2014, Fuhrman2014}. This motivates new theoretical studies of the band-structure and correlation phenomena in Kondo insulators \cite{Werner2013, Kang2013, Lu2013, Si2013, Werner2014, Baruselli2014, Alex2014, Legner2014}.

The purpose of this paper is to construct the minimal interacting lattice model of Kondo TI's crystal boundaries, and then discuss some possible correlation phenomena that arise from it. The Anderson model treated with the slave boson method \cite{Dzero2010, Dzero2012} is adapted for the description of two opposite Kondo TI's surfaces with an odd number of Dirac points each. The obtained Hamiltonian is an effective theory with a useful low-energy sector constrained by symmetries, low-energy degrees of freedom and any known properties of the spectrum. Its parameters can be fitted to an experimentally or numerically \cite{Nickerson1971, Farberovich1983, Takimoto2011, Kang2013, Lu2013} determined surface band-structure. The lattice formulation of an effective theory at the Kondo TI boundary is more appropriate for analyzing strongly correlated states than its continuum limit formulation \cite{Roy2014}.

The metallic two-dimensional surface state of a Kondo TI is only more susceptible to interactions and quantum fluctuations \cite{Efimkin2014} than its insulating three-dimensional bulk. Our main prediction is that the boundaries of Kondo TIs can in certain regimes exhibit correlation phenomena analogous to those found in heavy fermion metals \cite{Steglich1979, Taillefer1988, Maple1994, Lohneysen1994, Mathur1998, Schroder1998, Stockert1998,  Stewart2001, McCollam2005, Kadowaki2006}, but with features specific to the two-dimensional geometry and the existence of protected Dirac quasiparticles (whenever the TR symmetry is not broken). A two-dimensional Dirac metal of hybridized $d$ and $f$ electrons is susceptible to spin or charge density wave instabilities, and perhaps even superconductivity. The argument is largely based on the slave boson theory of Kondo materials, which seemingly describes in great detail \cite{Fuhrman2014} the collective mode seen by inelastic neutron scattering in SmB$_6$, and then predicts similar quantum fluctuations on the Kondo TI's boundary. Additional non-trivial phases with localized $f$ electrons can be stabilized by strong Coulomb interactions acting in two dimensions. Among them are spin liquids of localized $f$ moments. They can arise from the Kondo singlet fluctuations, which frustrate the correlations of $f$ moments along the surface. We discuss a particular surface regime in which the underlying spin liquid dynamics of localized $f$ electrons produces a ``marginal'' non-Fermi liquid metal of $d$ electrons described by the two-dimensional quantum electrodynamics. Which particular phase is realized at the crystal surface depends on the microscopic surface properties, and may be controlled to some extent by interface engineering. Even the surface orientation with respect to the crystal can affect the surface phase through its specific symmetry and structure of Dirac points.

Topological insulator quantum wells (TIQW) or ultra-thin films made from Kondo TIs are another system of interest in this paper. Their added virtues are tunability in gated heterostructures and the potential to introduce instabilities in the spin-triplet channel. We discuss unconventional triplet condensates within a simplified model of TIQWs, and find that they break the translation symmetry either by a pair density wave or a vortex lattice. The latter is analogous to the Abrikosov vortex lattice of superconductors in magnetic fields: the Rashba spin-orbit coupling on the TI's surface, which creates its Dirac quasiparticle spectrum, is equivalent to an external SU(2) Yang-Mills flux \cite{Frohlich1992}, a non-Abelian analogue of the ordinary U(1) magnetic field. The fully gapped quasiparticle spectrum of TIQWs opens a possibility of stabilizing incompressible quantum liquids with fractional excitations, by quantum melting of the mentioned vortex lattice \cite{Nikolic2011a, Nikolic2012a}. Quantum wells made from certain Kondo TI materials could have all the necessary ingredients for such exotic physics: strong gauge flux, gapped spectrum, and flat heavy-fermion surface bands sensitive to Coulomb interactions. Note that fractional quantum Hall systems have the same fundamental properties, only with the SU(2) spin-orbit flux being replaced by a much weaker U(1) magnetic flux (the realistic Rashba spin-orbit coupling can be equivalent to about $1000 \textrm{ T}$ magnetic fields, at least in bismuth-based TIs).

The existence of fractional incompressible quantum liquids in the phase diagram of strongly correlated TIQWs is supported by fundamental physical principles \cite{Nikolic2011a, Nikolic2012b}. The Rashba spin-orbit coupling is expected to naturally shape fractional states with non-Abelian statistics in the TIQWs \cite{Nikolic2012}, which have the kind of many-body quantum entanglement needed for quantum computing. The experimental exploration of TIs as a platform for topological quantum computing has begun very recently \cite{Kasumov1996, Koren2011, Qu2011, Sacepe2011, Yang2011, Zhang2011, Veldhorst2012, Wang2012, Wang2012a, Williams2012, Yang2012}, starting off with the Fu-Kane idea to create zero-energy Majorana quasiparticles using the proximity-induced superconducting state on the TI surface \cite{Fu2008}. In contrast, the possible fractional incompressible quantum liquids in TIQWs could become the platform for quantum computing analogous to that envisioned in non-Abelian quantum Hall states \cite{Kitaev2003, Nayak2008}. Its advantage over the Majorana system are fully gapped non-Abelian quasiparticles that maintain their exchange statistics at short distances and likely allow universal quantum computing.

This paper is organized as follows. Section \ref{secKondo} reviews the essential physics of Kondo materials and the slave boson method. The model and properties of protected Kondo TI's boundaries are then discussed in section \ref{secBoundary}. The phenomenology of Kondo TI quantum wells is analyzed in section \ref{secTIQW}. Finally, all conclusions are summarized in section \ref{secConcl}.

\section{The basic physics of heavy fermion materials}\label{secKondo}

By way of introduction, we review here some essential properties of Kondo insulators and other heavy fermion materials through the lenses of the Anderson model. Section \ref{secModel} justifies this model in the context of Kondo material band-structures, and section \ref{secCorrel} surveys the competing correlation effects that emerge from Coulomb interactions. The introduction concludes by a brief review of the slave boson method in section \ref{secSB}, which provides the framework for several predictions of this paper.

\subsection{The minimal model}\label{secModel}

The fundamental degrees of freedom in Kondo insulators are electrons that originate from the atomic $d$ and $f$ orbitals of a rare earth element. These orbitals spread into bands in the crystal environment, and hybridize to avoid crossing each other at the same momentum. The resulting electronic states have a broad energy dispersion as a function of crystal momentum when their character is predominantly $d$-like, and a fairly flat dispersion giving rise to a large effective mass when their character is predominantly $f$-like. The intrinsic spin-orbit coupling of $d$ electrons is usually neglected, so their spin $\sigma=\pm1$ is considered a good quantum number. In contrast, the internal quantum number $\alpha=1,\dots,N_{\alpha}$ of $f$ electrons labels states within a degenerate multiplet that arises from the crystal electric fields and spin-orbit coupling. In general, multiple $d$ and $f$ orbitals may significantly contribute to dynamics. However, the hybridization between $d$ and $f$ orbitals, due to both crystal fields and interactions, opens a narrow bandgap that contains the Fermi energy in Kondo insulators. Only one Kramers-degenerate pair of effective $d$ and $f$ orbitals each may be sufficient to capture the lowest energy dynamics of particle and hole excitations. Coulomb interaction is appropriately defined for local charged degrees of freedom such as electrons in the atomic orbitals, but we may approximately emphasize its influence only among the $f$ electrons because their intrinsic bandwidth (i.e. kinetic energy) is very small. The dominant effect of the Coulomb repulsion is to suppress the double occupancy of any lattice site by $f$ electrons, so we may model it as a simple on-site potential $U$.

\begin{figure}
\subfigure[{}]{\includegraphics[height=1.2in]{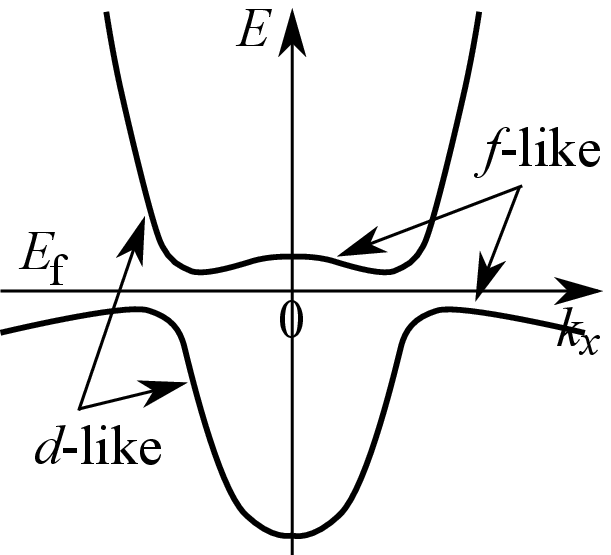}}\hspace{0.2in}
\subfigure[{}]{\includegraphics[height=1.2in]{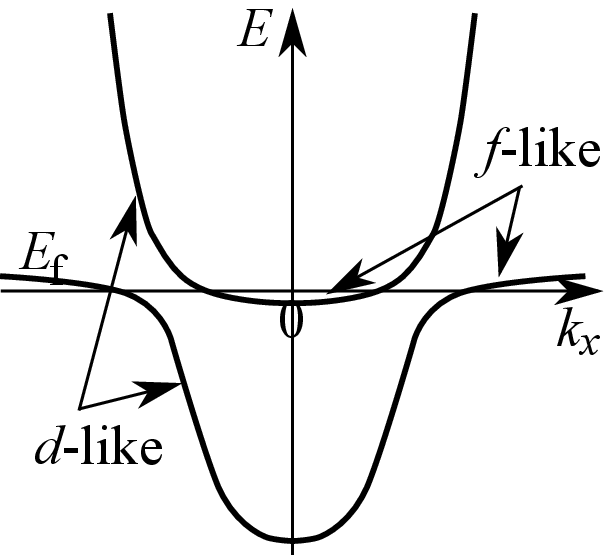}}
\caption{\label{HybGap}(a) A schematic band-structure of a Kondo insulator. Electron states are dominated by either $d$ or $f$ orbitals away from avoided crossings. Band inversion must occur at all places in the first Brillouin zone where the $f$ and $d$ orbitals hybridize. (b) Hybridization does not produce a bandgap without band inversion.}
\end{figure}

The minimal model of a Kondo insulator that captures the above features is given by the following second-quantized tight-binding Hamiltonian involving $d$ and $f$ electron field operators $d_{\sigma{\bf R}}$ and $f_{\alpha{\bf R}}$ respectively, or their Fourier transforms \cite{Dzero2010, Dzero2012}:
\begin{eqnarray}\label{MinMod}
H &=& \sum_{\sigma}\int\limits _{\textrm{1BZ}}\frac{d^{3}k}{(2\pi)^{3}}\,\xi_{{\bf k}}^{\phantom{\dagger}}
  d_{\sigma{\bf k}}^{\dagger}d_{\sigma{\bf k}}^{\phantom{\dagger}}
+\sum_{\alpha}\int\limits _{\textrm{1BZ}}\frac{d^{3}k}{(2\pi)^{3}}\,\epsilon_{{\bf k}}^{\phantom{\dagger}}
  f_{\alpha{\bf k}}^{\dagger}f_{\alpha{\bf k}}^{\phantom{\dagger}} \nonumber \\
&& +\sum_{\sigma\alpha}\sum_{{\bf R}{\bf R}'}\Bigl(V_{\sigma\alpha;{\bf R}-{\bf R}'}^{\phantom{\dagger}}
  d_{\sigma{\bf R}}^{\dagger}f_{\alpha{\bf R}'}^{\phantom{\dagger}}+h.c.\Bigr) \nonumber \\
&& +U\sum_{{\bf R}}\sum_{\alpha\beta}f_{\alpha{\bf R}}^{\dagger}f_{\alpha{\bf R}}^{\phantom{\dagger}}f_{\beta{\bf R}}^{\dagger}
  f_{\beta{\bf R}}^{\phantom{\dagger}} \ .
\end{eqnarray}
This is a version of the Anderson model. The sites $\bf R$ may form a simple cubic lattice found in Kondo insulators. The hybridization couplings $V_{\sigma\alpha;{\bf R}-{\bf R}'}$ arise from crystal fields and can be calculated microscopically. All important features of the band-structure near the hybridization gap can be captured by the nearest $t_1$, next-nearest $t_2$ and third-neighbor $t_3$ hopping:
\begin{eqnarray}
\xi_{{\bf k}}&=&-2t_{d1}\sum_{i}^{x,y,z}\cos(k_{i}a)
   -2t_{d2}\sum_{i\neq j}^{x,y,z}\cos(k_{i}a)\cos(k_{j}a) \nonumber \\
&& -2t_{d3}\cos(k_{x}a)\cos(k_{y}a)\cos(k_{z}a)-\mu \ ,
\end{eqnarray}
and similarly for the $f$ electrons. The dispersion $\epsilon_{{\bf k}}$ of the bare $f$ electrons is sometimes modeled by a flat band, $\epsilon_{{\bf k}} \approx \epsilon_{f} - \mu$, where $\epsilon_f$ is the relative energy shift of the $f$ orbitals with respect to the $d$ orbitals, and $\mu$ is the chemical potential (Fermi energy). However, a hybridization bandgap featured in all Kondo insulators exists only because the $f$ orbitals have an ``inverted'' dispersion with a finite effective mass ($m_{f}\sim100m_{0}$ in SmB$_{6}$, where $m_{0}$ is the bare electron mass), as shown in Fig.\ref{HybGap}. Such band inversions are necessary (but not sufficient) for the emergence of a topological insulator.

\subsection{Correlations due to Coulomb interactions}\label{secCorrel}

If the Coulomb interaction $U$ were weak, it would renormalize the band-structure through electron self-energy but yield no qualitative departures from the physics of a Fermi liquid or a band-insulator. However, distinct correlation phenomena emerge in the realistic limit $U \gg V \gg t_{f}$. They can be understood easily from the Hubbard model perspective, by regarding the hybridization coupling $V$ as an inter-orbital ``hopping'' analogous to the lattice hopping $t_{f}$ for $f$ electrons. If the occupation of $f$ orbitals is one electron per lattice site ($n_f=1$), then a sufficiently large Coulomb interaction localizes the $f$ electrons and prohibits their hopping both on the lattice and between orbitals. This situation is most easily envisioned in the light fermion metallic state, where the intrinsic $f$ orbitals are completely buried below the chemical potential by the amount of energy $\mu-\epsilon_f$. Charge-carrying excitations involving $f$ electrons are pushed to high energies due to the large energy cost $U-(\mu-\epsilon_f)$ of double occupancy, or the cost $\mu-\epsilon_f$ of removing an $f$ electron. The spin of $f$ electrons remains for now a low energy degree of freedom.

In the absence of hybridization, virtual electron hopping correlates the $f$ electron spins into an antiferromagnetic Neel state, governed by the spin-exchange coupling $J \sim t_f^2/U$ at the second order of perturbation theory. By the same mechanism, the inter-orbital hopping due to hybridization correlates the spins of $f$ and $d$ electrons. This effect is even stronger than the correlation between different lattice sites ($V\gg t_f$), and results in the formation of inter-orbital (Kondo) singlets. However, the competition between the Kondo and lattice spin exchange for the spins of $f$ electrons can create very complicated ground states. The low-energy dynamics of conduction $d$ electrons and localized $f$ moments is captured by the Kondo lattice model in this regime \cite{Doniach1977}. The $f$ electrons whose spin is screened by the formation of strong Kondo singlets are hardly available for the formation of antiferromagnetic orders on the lattice, but their number $n_k$ per site is limited by the number of itinerant $d$ electrons \cite{Nozieres1985a}. The situation $n_k<n_f=1$ is largely analogous to doping a Neel antiferromagnet of $f$ electrons by $n_k$ (chargeless) holes per site, where the terminal points of Kondo singlets reside. These holes are mobile and frustrate the residual spin correlations on the lattice.

The $f$ orbital is not exactly half-filled ($n_f<1$) in Kondo insulators and heavy-mass metallic states. This requires the chemical potential $\mu$ to cross the energy range of $f$ orbitals, either through the hybridization bandgap as in Kondo insulators or through a flat portion of the hybridized band as in heavy fermion metals. The ensuing mobility of $f$ electrons invalidates the Kondo lattice model of dynamics in the strict sense because the $f$ electrons can now contribute their charge to the Fermi surface. However, the quantum fluctuations resulting from the evolution of Kondo singlets continue to play an essential role in the strongly correlated dynamics \cite{Varma1976a, Varma1985, Varma1986, Auerbach1986, Millis1987a}. Magnetic ordering without localization can emerge out of this metallic state through the spin density wave instability \cite{Hertz1976, Moriya1985, Millis1993, Kadowaki2006}. Note that a different type of quantum criticality, where the $f$ electrons localize at $n_f=1$, is seen more often in heavy fermion compounds \cite{Maple1994, Maple1995, Aronson1995, Steglich1997, Aoki1997, Stockert1998, Schroder1998, Schroder2000, Grosche2000, Trovarelli2000, Fisher2002x, Custers2003}. Understanding this unconventional quantum criticality is a major theoretical challenge \cite{Si2001, Si2006, Coleman2003y, Pepin2005, Rech2006, Senthil2004b, Senthil2005c}.

Instead of controlling the $f$ electron localization by changing the chemical potential or ``doping'' as in the previous scenario, we can use the hybridization $V$ as another theoretical tuning parameter. If $V$ becomes comparable or larger than $U-(\mu-\epsilon_f)$ or $\mu-\epsilon_f$, then the $f$ electrons cannot be localized in the inter-orbital sense. It now costs virtually no energy to promote an $f$ electron to a $d$ orbital and move it across a large distance before recombining it back to the $f$ orbital. Consequently, $f$ electrons become delocalized, even if their intra-orbital hopping $t_f$ is still too small (the $d$ orbitals provide a shunt). Just before such delocalization takes place, the spin dynamics of weakly localized $f$ electrons can be highly frustrated by ever-increasing range of their effective spin-exchange couplings. Such conditions are friendly to exotic states of matter. Kondo insulators are at least close to being in this regime simply because their chemical potential must reside within the hybridization bandgap that splits open the $f$ ``band'' ($\mu\approx\epsilon_f$). 

\begin{figure}
\subfigure[{}]{\includegraphics[width=3.3in]{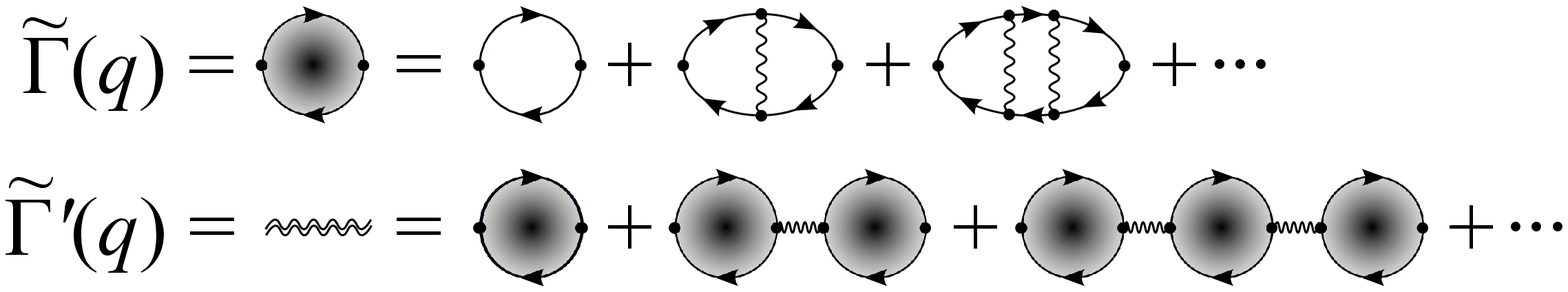}}
\subfigure[{}]{\includegraphics[width=1.4in]{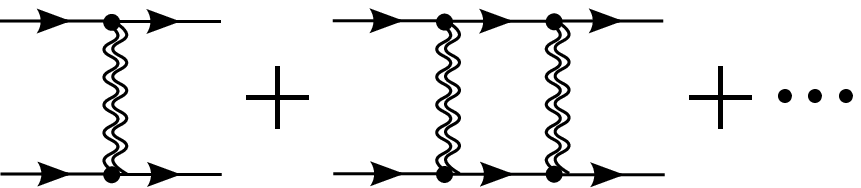}}
\caption{\label{CollectiveMode}(a) The Feynman diagrams for the processes that produce collective exciton modes in Kondo materials. The wiggly line represents the Kondo singlet propagator, which mediates an attractive interaction between an electron and a hole represented by solid lines. The ensuing exciton bound state $\Gamma(q)$ acquires self-energy renormalization $\Gamma'(q)$. The outcome is a coherent low-energy excitation in Kondo insulators such as SmB$_6$, protected against decay by having energy that lies within the particle-hole bandgap. (b) The process mediated by collective modes that generates a pairing glue for Cooper pairs.}
\end{figure}

If not in the ground state, correlation effects are visible in the excitation spectrum. The electron band-structure itself is renormalized and effectively shows temperature dependence \cite{Hundley1994}. More importantly, coherent collective modes of the paramagnon type have been seen in Kondo insulators \cite{Alekseev1993, Kasuya1994, Bouvet1998, Fuhrman2014} like SmB$_6$. These modes can be understood as excitons of hybridized particle-hole pairs \cite{Riseborough1992, Riseborough2000} whose binding is mediated by the quantum fluctuations of Kondo singlets, as illustrated by the Feynman diagrams in Fig.\ref{CollectiveMode}. Even though the modes are gapped, they are protected from decay by energy conservation because their energy lies within the particle-hole bandgap. The condensation of such a mode would typically correspond to a spin density wave instability. The analogous fluctuations in Kondo metals are more likely damped, but they can still be involved in magnetic instabilities. On the other hand, before a low-energy paramagnon mode gets a chance to condense, it may produce an entirely different instability in the Cooper channel, through the process illustrated in Fig.\ref{CollectiveMode}, especially in two-dimensions \cite{Metlitski2012, Metlitski2014}.

The Hamiltonian parameters that we considered in the above survey are hardly tunable, so any particular material will realize only one concrete scenario that we discussed. However, our main interest here are \emph{topological} Kondo insulators, which have protected metallic surface states. Microscopically, the dynamics of these surface states is governed by a properly modified two-dimensional version of the model (\ref{MinMod}) that we will construct later. Much of the heavy fermion physics becomes possible on the surface of a Kondo TI, but with even more enhanced quantum fluctuation effects due to the reduced dimensionality. Furthermore, some aspects of the dynamics can be tuned through interface or heterostructure design.

\subsection{Slave boson approximation}\label{secSB}

The main difficulty of analyzing the correlation effects in Kondo materials is that the Coulomb interaction is too strong for a perturbative treatment. On the other hand, the most important effect of Coulomb interactions is to suppress the double occupation of any lattice site by $f$ electrons. Slave boson approximation is a convenient way to remove the high-energy states with double occupancy from the low-energy effective theory. In the first step, it represents an $f$ electron as a bound state of two auxiliary degrees of freedom, a slave boson and a slave fermion. The $f$ electron creation operator is written as a product
\begin{equation}\label{SlaveBoson}
f_{\alpha{\bf R}}^{\dagger}=\psi_{\alpha{\bf R}}^{\dagger}b_{{\bf R}}^{\phantom{\dagger}}
\end{equation}
where $\psi_{\alpha{\bf R}}^{\dagger}$ creates a slave fermion at the site ${\bf R}$ and $b_{{\bf R}}^{\phantom{\dagger}}$ annihilates a slave boson at the same site. Slave bosons are a surplus degree of freedom that enlarge the Hilbert space beyond that of the physical states. Therefore, we need a local constraint that projects out all unphysical states. The constraint is designed to mimic the physics at large Coulomb interactions $U$ and prohibit double occupation of any lattice site by $f$ electrons:
\begin{equation}\label{LocalConstraint}
\sum_{\alpha}\psi_{\alpha{\bf R}}^{\dagger}\psi_{\alpha{\bf R}}^{\phantom{\dagger}}+b_{{\bf R}}^{\dagger}b_{{\bf R}}^{\phantom{\dagger}}=1 \ .
\end{equation}
Now, the number of slave bosons and slave fermions must add up to one on every site, so an empty site contains one slave boson. An attempt to create two $f$ electrons on an initially empty site involves two slave boson annihilations where only one boson is present, and thus results in the zero probability amplitude.

Since an $f$ electron is mathematically represented as a composite particle, there are multiple ways in which its charge can be distributed among its constituents. Most generally, the slave boson and slave fermion can have charges $q$ and $q-1$ respectively in the units where the electron charge is $-1$. The value of $q$ is not fixed in the Hamiltonian, even by the coupling between the $f$ electrons and the electromagnetic field that we do not explicitly write in this paper. Instead, $q$ has to be regarded as a variational parameter selected by the dynamics in an approximate slave boson ground state. Note that fixing the value of $q$ in this manner is equivalent to removing the unphysical local symmetry of the slave boson Hamiltonian under the transformation $b_{\bf R} \to e^{i\lambda_{\bf R}} b_{\bf R}$, $\psi_{\alpha{\bf R}} \to e^{i\lambda_{\bf R}} \psi_{\alpha{\bf R}}$, which was introduced by (\ref{SlaveBoson}). We are unable to calculate $q$, so we can only discuss its phenomenology.

The slave boson can be electrically neutral ($q=0$) while the slave fermion takes the full electron charge. The phase angle of slave bosons is then fixed by the lack of the corresponding global U(1) symmetry in the Hamiltonian (i.e. there can be no Goldstone modes). Neutral slave bosons are fluctuations related to Kondo singlets; they have the same quantum numbers of a neutral spinless particle, and the presence of Kondo singlets is generally reflected in $\langle b_{\bf R}^2 \rangle \neq 0$. The constraint enables vibrant slave boson dynamics in quantum states with delocalized $f$ electrons. This is where the slave boson approach is most useful as a practical approximation, because it gives birth to a slave boson ``condensate''. A slave boson order parameter $\langle b_{\bf R} \rangle \neq 0$ is found to renormalize the quasiparticle spectrum even at the mean-field level, while its quantum fluctuations can effectively give rise to collective modes and instabilities. 

Given that spin liquids and superconducting states can also arise from frustrated spin dynamics, we should consider another use of the slave boson theory. Spin-charge separation that takes place in spin liquids can be captured by a charged slave boson ($q=1$) and a neutral slave fermion. Uncondensed slave bosons represent a spin liquid, whose fermionic spinon excitations arise from the dynamics of the neutral slave fermions. A charged slave boson condensate is an exotic fractionalized superconducting phase \cite{senthil00}. The slave boson theory allows a transparent description of these phases, but it is usually quantitatively unclear what microscopic conditions are needed to establish any one of them.

The slave boson method first becomes an approximation if we take the $U\to\infty$ limit. This can be remedied in the large but finite $U$ limit by first integrating out the high-energy states with double occupancy, using for example the degenerate perturbation theory, and then applying the slave boson method on the resulting effective Hamiltonian without $f$ electron double-occupancy. This would generate additional Kondo and Ruderman-Kittel-Kasuya-Yosida (RKKY) couplings with exchange energy scales $V^2/U$ and $t_f^2/U$ respectively. However, the effects of these terms may be small in comparison to the other aspects of dynamics captured by the slave boson method. The main approximation step comes in the implementation of the constraint, which cannot be done exactly. At least when the slave fermions and bosons are highly mobile, it is sensible to implement the constraint softly on the average densities:
\begin{equation}\label{SoftConstraint}
\left \langle \sum_{\alpha}\psi_{\alpha{\bf R}}^{\dagger}\psi_{\alpha{\bf R}}^{\phantom{\dagger}}
  +b_{{\bf R}}^{\dagger}b_{{\bf R}}^{\phantom{\dagger}} \right\rangle = 1 \ .
\end{equation}
The slave boson can condense, $\langle b_{\bf R} \rangle = B$, and we may even use the mean field theory to estimate band-structure renormalization from the Hamiltonian:
\begin{eqnarray}\label{SBMF}
H_{\textrm{mf}}\!\!&=&\!\!\sum_{\sigma}\int\limits_{\textrm{1BZ}}\frac{d^{3}k}{(2\pi)^{3}}\,
   \xi_{{\bf k}}^{\phantom{\dagger}}d_{\sigma{\bf k}}^{\dagger}d_{\sigma{\bf k}}^{\phantom{\dagger}} \\
&& +\sum_{\alpha}\int\limits_{\textrm{1BZ}}\frac{d^{3}k}{(2\pi)^{3}}\,
     \Bigl(\epsilon_{{\bf k}}^{\phantom{\dagger}}|B|^2+\epsilon_f^{\phantom{\dagger}}-\mu\Bigr)
     \psi_{\alpha{\bf k}}^{\dagger}\psi_{\alpha{\bf k}}^{\phantom{\dagger}} \nonumber \\
&& +\sum_{\sigma\alpha}\sum_{{\bf R}{\bf R}'}\Bigl(V_{\sigma\alpha;{\bf R}-{\bf R}'}^{\phantom{\dagger}}
  B^{*\phantom{|}}\!d_{\sigma{\bf R}}^{\dagger}\psi_{\alpha{\bf R}'}^{\phantom{\dagger}}+h.c.\Bigr) \ . \nonumber
\end{eqnarray}
Note that the boson commutation relation $b_{{\bf R}}^{\phantom{\dagger}}b_{{\bf R}}^{\dagger}=1+b_{{\bf R}}^{\dagger}b_{{\bf R}}^{\phantom{\dagger}}\to1+|B|^{2}$ yields the additional constant $\epsilon_{f}-\mu$ term in the renormalized spectrum of $f$ electrons. The Coulomb coupling is eliminated by the no-double-occupancy constraint. This theory of fermionic excitations is formally non-interacting, but the value of $|B|$ has to be determined self-consistently by minimizing the ground state energy under the soft constraint (\ref{SoftConstraint}). The renormalized spectrum of hybridized electrons takes the form
\begin{equation}\label{SBspect}
E_{s\lambda{\bf k}}=\frac{\xi_{{\bf k}}^{\phantom{i}}+\epsilon'_{{\bf k}}}{2}
  +\lambda\sqrt{\left(\frac{\xi_{{\bf k}}^{\phantom{i}}-\epsilon'_{{\bf k}}}{2}\right)^{2}+V_{{\bf k}}^{2}|B|^{2}} \ ,
\end{equation}
where
\begin{eqnarray}
\epsilon'_{{\bf k}}&=&\epsilon_{{\bf k}}^{\phantom{\dagger}}|B|^2+\epsilon_f^{\phantom{\dagger}}-\mu \\
V_{{\bf k}}&=&\sqrt{\sum_\sigma\sum_{\alpha}
    \left\vert \sum_{{\bf R}-{\bf R}'}
    V_{\sigma\alpha;{\bf R}-{\bf R}'}^{\phantom{\dagger}}\, e^{-i{\bf k}({\bf R}-{\bf R}')} \right\vert ^{2}} \ . \nonumber
\end{eqnarray}
The hybridized states are labeled by the conduction/valence band index $\lambda=\pm 1$ and the Kramers degeneracy index $s=\pm 1$.

\section{Topologically protected boundary states}\label{secBoundary}

Samarium hexaboride has been identified as a likely candidate for a strong topological insulator \cite{Dzero2010, Dzero2012}. Other Kondo insulator materials are potential candidates too. A Kondo TI crystal will normally have a two-dimensional metallic state at its entire boundary that surrounds the insulating bulk. This ``helical'' metal has only one low-energy spin-projection mode, where the spin and momentum vectors are orthogonal and related by the right-hand rule. The energy dispersion of quasiparticles in the helical metal of a TI has an odd number of massless Dirac points in the first Brillouin zone, although no symmetry binds the chemical potential to any one of them. These properties of the boundary spectrum can be formally attributed to the Rashba spin-orbit coupling, and their robustness against perturbations is the result of topology and TR symmetry.

The argument in favor of SmB$_{6}$ being a TI was based on the band-structure analysis. It neglected various consequences of quantum fluctuations. We will argue that Coulomb interactions might give rise to various manifestations of strong correlations on the Kondo TI boundary. Both the interactions and the spin-orbit coupling enhance dynamics at the cut-off length scales, so we will need a lattice effective model of the Kondo TI's surface to study correlated states. We develop such a model in sections \ref{secModel2D} and \ref{secKondoModel}, and then survey its various possible ground states in section \ref{secCorrel2D}.

\subsection{Lattice models of topological insulator boundaries}\label{secModel2D}

Our ideal goal here is to construct a two-dimensional lattice model of a single flat TI surface. This turns out to be impossible. The main problem is capturing an odd number of Dirac points in the two-dimensional band-structure. Constructing a continuum limit with a single Dirac point at zero momentum is rather simple:
\begin{equation}
H_s = v \hat{\bf z} ({\bf S} \times {\bf p}) \ ,
\end{equation}
where $\bf S$ and $\bf p$ are the electron's spin and momentum operators respectively, $\hat{\bf z}$ is the unit vector perpendicular to the surface, and $v$ is a coupling with units of velocity. The hint for constructing a lattice theory with this continuum limit is found by ``rewriting'' the Hamiltonian as a gauge theory:
\begin{equation}\label{CL}
H'_s = \frac{({\bf p} - \tau^z \boldsymbol{\mathcal{A}})^2}{2m} \quad,\quad \boldsymbol{\mathcal{A}} = -mv (\hat{\bf z} \times {\bf S}) \ .
\end{equation}
The SU(2) gauge field $\boldsymbol{\mathcal{A}}$ is a static background representing a non-zero Yang-Mills ``magnetic'' flux created by the Rashba spin-orbit coupling. Electrons carry SU(2) charge $\tau^z = \pm 1$ with respect to this gauge field, where the opposite surfaces of the TI carry opposite charges with a global interpretation of $\hat{\bf z}$. The surface Hamiltonians $H_s$ and $H'_s$ differ only by a $p^2/2m$ term and a constant, which makes $H'_s$ more realistic. It is now straight-forward to propose a simple lattice theory in the second-quantized form
\begin{equation}\label{LL}
H_l = - \sum_{{\bf r}{\bf r}'} t_{{\bf r}-{\bf r}'}^{\phantom{\dagger}} c_{\bf r}^\dagger\, e^{i\tau^z\mathcal{A}_{{\bf r},{\bf r}'}}
  c_{{\bf r}'}^{\phantom{\dagger}} \ ,
\end{equation}
with a lattice SU(2) gauge field defined on lattice bonds:
\begin{equation}\label{GF}
\mathcal{A}_{{\bf r},{\bf r}'}=-\mathcal{A}_{{\bf r}',{\bf r}} \quad;\quad
  \mathcal{A}_{{\bf r},{\bf r}+\hat{\bf x}} = a \sigma^y \quad,\quad \mathcal{A}_{{\bf r},{\bf r}+\hat{\bf y}} = -a \sigma^x \ .
\end{equation}
Indeed, the continuum limit of this tight-binding Hamiltonian describes the $\Gamma$ point of the two-dimensional first Brillouin zone the same way as (\ref{CL}). However, there is an even number of Dirac points in the first Brillouin zone of the square lattice, so we are not describing the surface of a TI.

The spin-orbit coupling can lift the two-fold spin-degeneracy of bands at almost any point in the first Brillouin zone. However, the TR symmetry protects the degeneracy of bands at high-symmetry points, $\Gamma$ ${\bf k}=(0,0)$, M ${\bf k}=(\pi,\pi)$, and two X points ${\bf k}=(\pi,0),(0,\pi)$. There are four high-symmetry points in the Brillouin zone of the square lattice, so four Dirac points are unavoidably pinned to them for any finite SU(2) gauge field that captures the Rashba spin-orbit coupling. Additional Dirac points in the interior of the first Brillouin zone can be generated by introducing non-zero gauge fields associated to hopping beyond the nearest-neighbor sites. Note that gauge invariance does not require the existence of such extended-range gauge fields in any circumstances.

The above argument in favor of the four Dirac points is very fundamental: four Dirac points are protected by the TR and square lattice symmetries. Of course, disorder easily violates the lattice symmetry, so all Dirac cones can be gapped out in pairs. There is no way to formulate a strictly local lattice theory of a single TI surface. This is similar to a quantum anomaly: the continuum limit of a single TI surface exists \cite{Roy2014}, but its direct lattice regularization is not possible.

We still need a lattice theory. Some regularized effective theory that captures the dynamics of only the low-energy surface states must exist. The hint to finding such a theory lurks in the fact that the surface states envelop the entire crystal boundary and cannot be terminated. Considering a slab crystal geometry, it may not be possible to adequately describe a single surface, but it must be possible to describe two opposite surfaces in a single theory. The lattice Hamiltonian (\ref{LL}) already labels the two opposite surfaces by $\tau^z$, so we can promote $\tau^z$ into an operator (Pauli matrix). The lattice symmetries along the slab still protect the four Dirac points, so the only way to gap out one or three of them, without introducing high-energy bulk degrees of freedom in our effective theory, is to include couplings between the two surfaces. For example,
\begin{equation}\label{LL2}
H'_l = \sum_{{\bf r}{\bf r}'} c_{\bf r}^\dagger \Bigl( - t_{{\bf r}-{\bf r}'}^{\phantom{\dagger}} \, e^{i\tau^z\mathcal{A}_{{\bf r},{\bf r}'}}
  + \Delta_{{\bf r}-{\bf r}'}^{\phantom{\dagger}} \tau^x \Bigr) c_{{\bf r}'}^{\phantom{\dagger}}
\end{equation}
with
\begin{equation}\label{DPgap}
\sum_{\delta{\bf r}} \Delta_{\delta{\bf r}} e^{i{\bf k}\delta{\bf r}} =
  \Delta_{\textrm{M}} \sin^2\left(\frac{k_x}{2}\right) \sin^2\left(\frac{k_y}{2}\right)
\end{equation}
gaps out just one Dirac cone at the M point of the Brillouin zone. Note again that the phenomenology of TIs does not leave us with any options to construct a qualitatively different lattice theory of just TI boundaries. We are then left with the problem of justifying the necessary couplings like $\Delta$, which appear to connect the two opposite surfaces of a TI across arbitrarily large distances through the insulating bulk.

\begin{figure}
\includegraphics[width=1.7in]{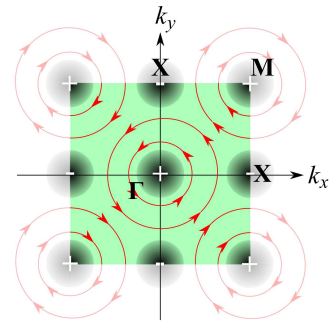}
\caption{\label{SpinTexture}The texture of spin-momentum locking of the Hamiltonian (\ref{LL}) in the first Brillouin zone. Every Dirac cone is a vortex singularity of the vector field formed by the local spin orientations. The periodic boundary condition of the Brillouin zone requires that the total vortex charge be zero, a total of four Dirac points.}
\end{figure}

In a nutshell, a two-dimensional spectrum $E({\bf k})$ with an odd number of Dirac points cannot connect to itself across the first Brillouin zone, as required by the zone's periodic boundary condition. This can be visualized by plotting the texture of spin-momentum locking at energies above the Dirac points in the first Brillouin zone, as in Fig.\ref{SpinTexture}. The vector field of spin orientations features a vector-vortex singularity at every Dirac point. The periodic boundary condition of the first Brillouin zone requires that the total vortex charge be zero. A vector vortex is symmetric under rotations but the antivortex is not. We may place a vortex at the $\Gamma$ point, but we cannot compensate it with a single antivortex without violating the rotational symmetry of the square lattice. The solution is to put two antivortices at the two symmetry-related X points, and then add another vector-vortex singularity at the M point to achieve vortex charge neutrality. We have four symmetry-protected Dirac points that we identified earlier. Even if we relax the lattice symmetry requirement, we cannot have an odd number of Dirac points in the Brillouin zone with periodic boundary conditions.

Since an odd number of Dirac points prohibits the 2D spectrum from independently connecting to itself across the Brillouin zone boundaries, it is necessary for this spectrum to cross into the high-energy regions somewhere in the Brillouin zone, where it overlaps with the bulk bands. An odd number of Dirac points must appear at such high energies. There is now a channel through the bulk for the coupling between such Dirac points on the opposite crystal's surfaces. This is how they become gapped, and our effective theory (\ref{LL2}) captures precisely this process.

Considered by itself, the inter-surface coupling $\Delta$ brings about some long-range entanglement between the two surfaces. The ensuing physical consequence is a half-quantized Hall conductivity of a TI's surface in the presence of a magnetic field \cite{Qi2010a}. No simple two-dimensional non-interacting lattice model can produce a non-integer Chern number. The lattice model of two surfaces (\ref{LL2}) must also produce an integer Chern number, but a half of it can be associated to one and the other half to the other surface.

It is satisfying to observe that the same problem, and essentially the same solution, are found in an attempt to formulate a lattice theory of just the edge states in integer quantum Hall systems. A tight-binding dispersion of an edge state must be $E(k)\sim\sin(ka)$ in order to reproduce $E(k)\propto k$ in the continuum limit. However, this generates two low energy modes near the chemical potential $\mu\approx 0$, one with the correct and one with a wrong chirality. The ``wrong'' mode cannot be formally removed from a well-defined lattice theory of a single edge. The solution is to write a theory of two opposite edges, and introduce formal inter-edge couplings at near-the-cutoff momenta $k\approx 2\pi/a$ that gap out the undesired edge mode. This can be simply justified by considering the continuum-limit Landau gauge, where a momentum change corresponds to a lateral spatial shift of the state's ``guiding center''. If the system size is $L_x \times L_y$ and the applied magnetic field is $B$, then the Landau level degeneracy is $N=BL_{x}L_{y}/2\pi$, and the change of momentum by the cut-off amount $\Delta k=2\pi/a=2\pi N/L_{x}$ implies the guiding center displacement $\Delta y=\Delta k/B=L_{y}$ across the entire sample between the two opposite edges. Therefore, the ``wrong'' modes of the two edges should belong to the bulk where they must be a part of the high-energy spectrum.

\subsection{Lattice models of topological Kondo insulator boundaries}\label{secKondoModel}

Here we apply the insight from the previous section to construct models of Kondo TI surfaces. Since we are looking for an effective theory, we could focus on the protected metallic state of \emph{hybridized} electrons on the crystal boundary and write a Hamiltonian like (\ref{LL2}) to describe their dynamics. This may be sufficient for dealing with fermionic quasiparticle excitations, but makes it very hard to study correlation effects due to interactions.

Instead, we will construct a lattice model of the TI's surface that contains both $d$ and $f$ electrons as microscopic degrees of freedom. Then we will be able to use approximations such as the slave boson method to explore correlation phenomena. The Hamiltonian we seek is the two-dimensional analogue of (\ref{MinMod}) enhanced by the SU(2) gauge fields that implement a Rashba-type spin-orbit coupling:
\begin{eqnarray}\label{MinMod2D}
H_{\textrm{2D}} &=& \sum_{{\bf r}{\bf r}'}\Bigl\lbrack d_{\bf r}^\dagger\Bigl(-t_{{\bf r}-{\bf r}'}^{d}\,
     e^{i\tau^z\mathcal{A}_{{\bf r},{\bf r}'}^d}+\Delta_{{\bf r}-{\bf r}'}^{d}\tau^x\Bigr)d_{{\bf r}'}^{\phantom{\dagger}} \nonumber \\
&& + f_{\bf r}^\dagger\Bigl(-t_{{\bf r}-{\bf r}'}^{f}\,
     e^{i\tau^z\mathcal{A}_{{\bf r},{\bf r}'}^f}+\Delta_{{\bf r}-{\bf r}'}^{f}\tau^x\Bigr)f_{{\bf r}'}^{\phantom{\dagger}} \nonumber \\
&& + \Bigl(d_{{\bf r}}^{\dagger}\,V_{{\bf r},{\bf r}'}^{\phantom{\dagger}}f_{{\bf r}'}^{\phantom{\dagger}}+h.c.\Bigr)\Bigr\rbrack \\
&& +U\sum_{{\bf r}}\Bigl(f_{{\bf r}}^{\dagger}f_{{\bf r}}^{\phantom{\dagger}}\Bigr)^2 \ . \nonumber
\end{eqnarray}
The square lattice sites are labeled by $\bf r$, and we organized the field operators into spinors:
\begin{equation}
d_{\bf r} = \left(\begin{array}{c} d_{\uparrow{\bf r}} \\ d_{\downarrow{\bf r}} \end{array} \right) \quad,\quad
f_{\bf r} = \left(\begin{array}{c} f_{1{\bf r}} \\ \vdots \\ f_{N_f{\bf r}} \end{array} \right) \ .
\end{equation}
It will be sufficient for our purposes to work with a doublet of $f$ orbitals, $N_f=2$. A background SU(2) lattice gauge field of the type (\ref{GF}) is associated to every electrons' direct hopping path between two sites. Note that the hybridization term $V$ now involves a matrix that can also be gauged. The $f$ electrons do not have spin as a good quantum number, but their multiplet index $\alpha=1,\dots,N_f$ transforms non-trivially under TR like a generalized angular momentum, and so can be involved in the spin-orbit coupling. This two-dimensional model describes two opposite surfaces of a TI; the Pauli matrix operators $\tau^z,\tau^x$ operate on the surface index. The unusual inter-surface couplings $\Delta$ are needed to ensure an odd number of Dirac cones in the first Brillouin zone, as we discussed in the previous section.

\begin{figure}
\subfigure[{}]{\includegraphics[height=1.6in]{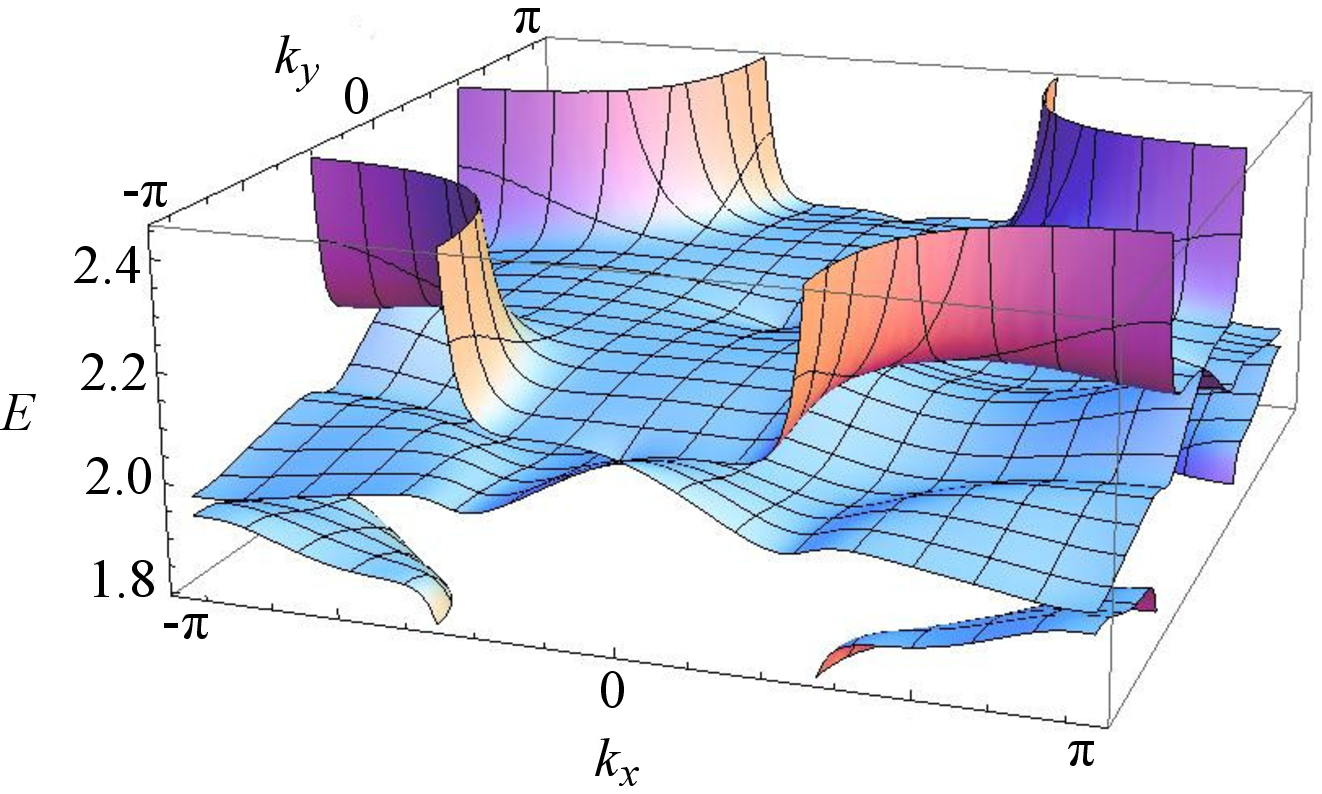}}
\subfigure[{}]{\includegraphics[height=1.2in]{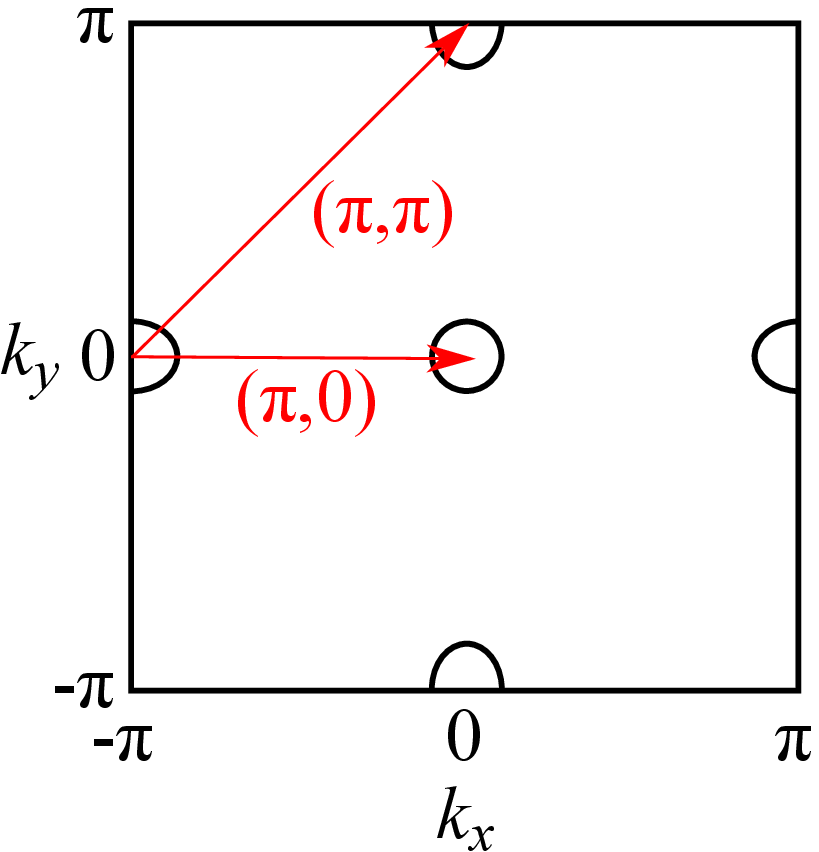}}
\caption{\label{SmB6surf}A modeled band-structure (a) and Fermi surface (b) of the metallic SmB$_6$ boundary states that quantitatively reproduce the important low-energy features of the LDA+Gutzwiller ab-initio spectrum in Ref.\cite{Lu2013}. Electron hopping takes place between the nearest-neighbor and next-nearest-neighbor sites for $d$ electrons $t_1^d = 1$, $t_2^d=-0.5$ and up to 3$^{\textrm{rd}}$ neighbor sites for a doublet of $f$ electrons $t_1^f = -0.05$, $t_2^f=0.0165$, $t_3^f=-0.015$ (the units are $\textrm{eV}$). The chemical potentials are $\epsilon_f=2.0$, $\mu=2.09$. The SU(2) gauge field (\ref{GF}) is assigned only to the $f$ electron hopping, with the parameter $a=0.9$ (the spin matrices in (\ref{GF}) are taken to act on the orbital space of the $f$ electron doublet). The Fourier transform of the hybridization term is given by $V_{\bf k} = V_0 \lbrack s^x \sin(k_x) + s^y \sin(k_y) \rbrack$, where $V_0=0.4$, and $s^x,s^y$ are Pauli matrices that convert the $d$ electron spin to the $f$ electron multiplet index. The inter-surface couplings (\ref{DPgap}) are implemented for both $d$ and $f$ electrons, with $\Delta_0^d=4$, $\Delta_0^f=0.1$. There is a large freedom to choose different values of various couplings without significantly affecting the depicted low-energy features of the spectrum.}
\end{figure}

At this point, the only remaining task is to choose the values for various coupling constants that reproduce a desired surface band-structure at low energies. One example is shown in Fig.\ref{SmB6surf}. There are several rules to consider in choosing the parameter values. First of all, the SU(2) gauge fields should vanish on all but the nearest-neighbor lattice bonds in order to have the minimal number of Dirac points. Otherwise, additional Dirac points arise in the interior of the first Brillouin zone. For the same reason, either $d$ or $f$ electrons, but not both, should be coupled to the SU(2) gauge field. The hopping parameters determine the relative energies of the Dirac points through the overall shape of the electron dispersion. The ab-initio calculation in Ref.\cite{Lu2013} suggests that the Dirac points at X lie about $5-8\textrm{ meV}$ above that at $\Gamma$, while the spurious Dirac point at M can be gapped out only if immersed in the bulk band that spreads at energies more than $15-20\textrm{meV}$ below the energy of the X Dirac points. These features were taken into account in the example from Fig.\ref{SmB6surf}, but require up to third-neighbor hopping in the $f$ orbitals. In this particular case, the $d$ orbitals are very broad, so most details of their dispersion controlled by the hopping parameters are irrelevant as they appear at high energies. The parameter $\epsilon_f$ strongly affects the momentum extent of hybridized states that are dominated by the $f$ orbitals and thus disperse weakly.

\subsection{Correlations on the Kondo TI boundaries}\label{secCorrel2D}

Here we make qualitative predictions for a few strongly correlated states that could arise on the Kondo TI boundaries. The best way to summarize them is to say that a Kondo TI can exhibit ``helical'' two-dimensional heavy fermion physics on its boundary even though its bulk is insulating. 

The correlated states on the boundary may depend sensitively on the surface quasiparticle spectrum. It should be appreciated now that the detailed properties of the surface band-structure are not universal. They depend on the precise conventional insulator that is interfaced with the Kondo TI, as well as the orientation of the surface with respect to the bulk crystal (e.g. the 100 cut of the cubic lattice is a square lattice, but the 111 cut is a triangular lattice). Various impurities that have affinity for the surface or the bulk can also alter the spectrum. Pressure and temperature are the usual controllable parameters. The only universal feature of the TI's surface spectrum is the presence of an odd number of Dirac points (if the TR symmetry is not broken).

\begin{figure}
\subfigure[{}]{\includegraphics[height=1.2in]{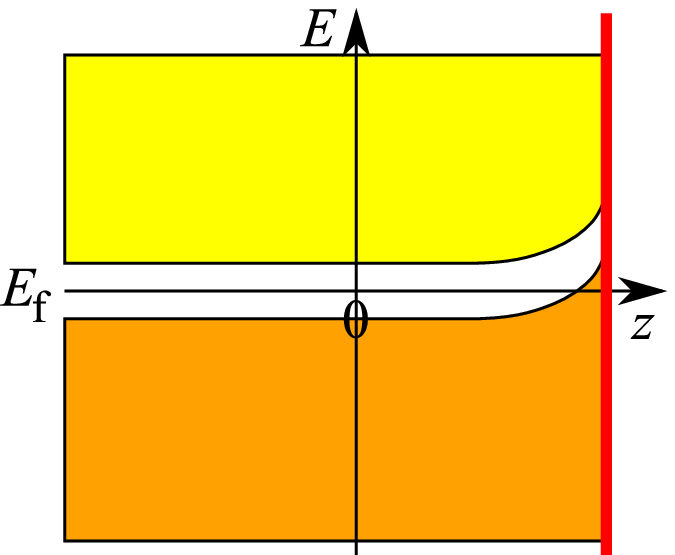}}
\subfigure[{}]{\includegraphics[height=1.2in]{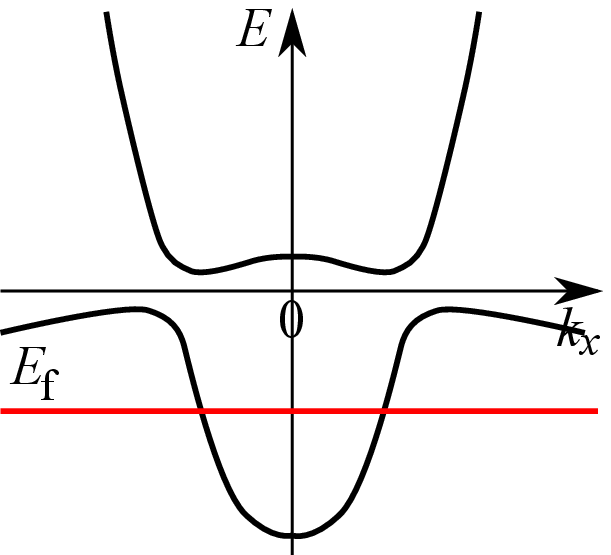}}
\subfigure[{}]{\includegraphics[height=1.2in]{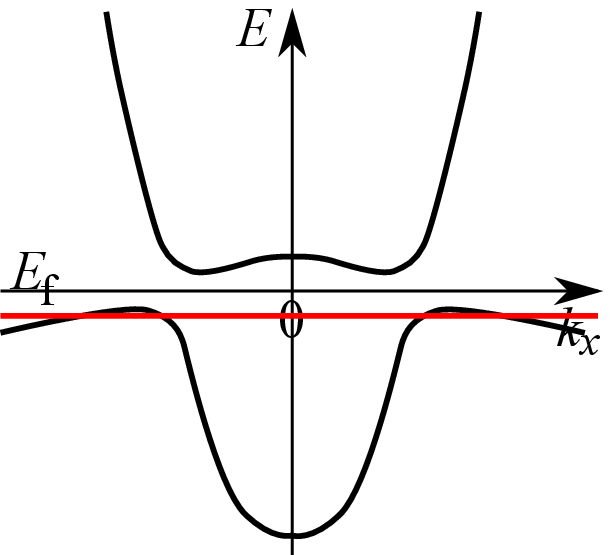}}
\subfigure[{}]{\includegraphics[height=1.2in]{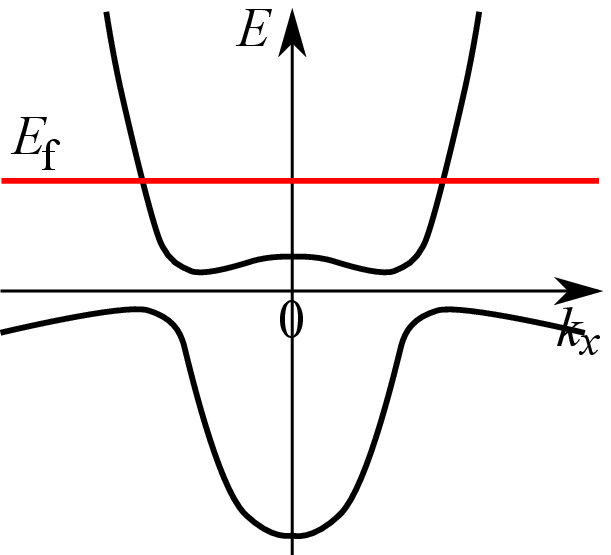}}
\caption{\label{KIbands}(a) Bending of the Kondo insulator's bands near the crystal boundary. The amount of deformation depends on the microscopic surface (interface) properties, and determines the relative placement of the chemical potential $E_f$ in the bulk band-structure near the surface. (b) Weakly correlated metallic boundary dominated by $d$ electrons, without appreciable Kondo singlet fluctuations. (c) Strongly correlated states of hybridized $d$ and $f$ electrons. (d) Strongly correlated metallic boundary dominated by light $d$ orbitals, influenced by the Kondo singlet dynamics amid the half-filled $f$ orbital.}
\end{figure}

One non-universal phenomenon at the crystal boundary that creates different regimes for correlations is band bending. Even though band bending in not responsible for the existence of metallic surface states in TIs, it affects their spectrum and occurs in general. Fig.\ref{KIbands} illustrates three characteristic circumstances that we will discuss. The least interesting regime, Fig.\ref{KIbands}(b), is obtained if the chemical potential crosses the bulk $d$ orbital at energies well below the $f$ orbital in the bent bulk spectrum near the surface. Then, the metallic state at the crystal boundary consists primarily of $d$ electrons, while the $f$ orbitals are depleted. There is hardly any substance to establish Kondo singlets or other correlations through Coulomb interactions. The surface metallic state is conventional. The second regime is shown in Fig.\ref{KIbands}(c). Now the chemical potential crosses the energy range where the $d$ and $f$ orbitals hybridize. Quantum fluctuations are abundant and Coulomb interactions are highly influential. It is extremely difficult to formulate a controlled approximation that describes this regime, but we may use the dynamical slave boson theory to approximately describe the dynamics at least when the $f$ electrons are not localized. Indeed, there is no generic reason for the $f$ electrons to localize in this regime given that their orbital is only partially populated. We will briefly discuss the ensuing possible ground states in section \ref{secHybridCorrel}. Finally, Fig.\ref{KIbands}(d) illustrates the third characteristic regime, where $d$ electrons again dominate the surface transport, but not in a weakly correlated fashion. Here the $f$ orbital is at half-filling, so the $f$ electrons are localized. Still, they can form Kondo singlets with $d$ electrons, which affects both the metallic properties of the conduction $d$ electrons and the spin dynamics of the localized $f$ electrons. We will discuss some interesting possible consequences of these correlations in section \ref{secKondoCorrel}.

It should be noted that the above naive picture of band bending has nothing to do with the size of the Fermi surface at the crystal boundary. The strength of the Rashba spin-orbit coupling, and the energy difference between the chemical potential and the Dirac points, determine the Fermi surface size. Therefore, the Fermi surface at the sample boundary can be either large or small regardless of whether it is dominated by $d$ or $f$ electrons. Fig.\ref{KIbands} schematically represents the bulk bands and should not be used to visualize the size of the Fermi surface at the crystal boundary.

\subsubsection{Hybridized correlated regime}\label{secHybridCorrel}

Here we discuss the boundary states of Kondo TIs in the regime depicted in Fig.\ref{KIbands}(c) where band bending is small. Surface states are formed by hybridized electrons and have an appreciable $f$ orbital content. The $f$ orbital is partially populated so Coulomb interactions create frustration by disallowing two $f$ electrons on the same lattice site. Since $f$ electrons are not localized, we are able to qualitatively capture the dynamics using the slave boson theory with a soft constraint. To that end, we introduce the slave bosons and fermions by (\ref{SlaveBoson}) on every site $\bf r$ of the two-dimensional lattice that models the TI's surface. The ensuing slave boson Hamiltonian (\ref{MinMod2D}) of the Kondo TI's surface:
\begin{eqnarray}\label{SB2D}
H_{\textrm{sb}} &=& \sum_{{\bf r}{\bf r}'}\Bigl\lbrack d_{\bf r}^\dagger\Bigl(-t_{{\bf r}-{\bf r}'}^{d}\,
     e^{i\tau^z\mathcal{A}_{{\bf r},{\bf r}'}^d}+\Delta_{{\bf r}-{\bf r}'}^{d}\tau^x\Bigr)d_{{\bf r}'}^{\phantom{\dagger}} \nonumber \\
&& + b_{\bf r}^{\phantom{\dagger}}b_{{\bf r}'}^\dagger\,\psi_{\bf r}^\dagger\Bigl(-t_{{\bf r}-{\bf r}'}^{f}\,
     e^{i\tau^z\mathcal{A}_{{\bf r},{\bf r}'}^f}+\Delta_{{\bf r}-{\bf r}'}^{f}\tau^x\Bigr)\psi_{{\bf r}'}^{\phantom{\dagger}} \nonumber \\
&& + \Bigl(b_{{\bf r}'}^\dagger d_{{\bf r}}^{\dagger}\,V_{{\bf r},{\bf r}'}^{\phantom{\dagger}}\psi_{{\bf r}'}^{\phantom{\dagger}}
     +h.c.\Bigr)\Bigr\rbrack
\end{eqnarray}
ought to be diagonalized with the restriction:
\begin{equation}
\frac{1}{N}\sum_{\bf r}\Bigl\langle \psi_{\bf r}^\dagger \psi_{\bf r}^{\phantom{\dagger}}
  +b_{\bf r}^\dagger b_{\bf r}^{\phantom{\dagger}}\Bigr\rangle = 1 \ ,
\end{equation}
where $N$ is the number of 2D lattice sites. The choice of the Hamiltonian parameter values is discussed in section \ref{secKondoModel}. It seems most appropriate to form Dirac points predominantly on $f$ orbitals, so that only $\mathcal{A}^f$ should be non-zero among the lattice SU(2) gauge fields in (\ref{SB2D}).

The mobility of slave bosons allows them to condense, $\langle b_{\bf r} \rangle = B \neq 0$. The condensate alone renormalizes the quasiparticle spectrum according to the expression like (\ref{SBspect}) but with different quantum numbers and lifted degeneracy. In practice, various properties of the realistic surface band-structure may be determined experimentally or numerically, and then they already include renormalizations due to Coulomb interactions. Hence, it is prudent to fit the renormalized band-structure (\ref{SBspect}) to the measured one, and obtain $B$ by ``reverse engineering'' instead of the more complicated self-consistent approach which assumes the knowledge of microscopic parameters.

The lowest energy excitations are hybridized quasiparticles whose Fermi surface and Fermi velocity are dominated by the Dirac cones rather then their significant $f$-character. The slave boson condensate does not contribute Goldstone modes to the low-energy spectrum in any scenario that we discussed in section \ref{secCorrel}. However, slave boson fluctuations beyond the condensate can give rise to collective modes that live at finite low energies, or even cause instabilities.

\begin{figure}
\subfigure[{}]{\includegraphics[height=0.5in]{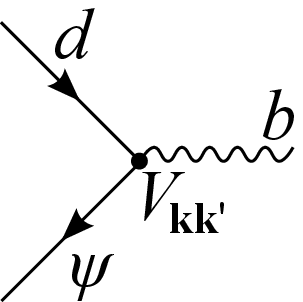}}\hspace{0.2in}
\subfigure[{}]{\includegraphics[height=0.5in]{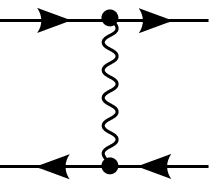}}\hspace{0.2in}
\subfigure[{}]{\includegraphics[height=0.5in]{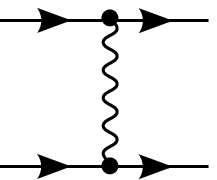}}
\caption{\label{Diagrams}The fundamental processes of the slave-boson perturbation theory. (a) The hybridization vertex in which a slave $f$ fermion is converted to a $d$ electron via the emission of a slave boson. (b) Exciton pairing process: an electron and a hole attract each other by exchanging slave bosons. (c) Cooper pairing process: slave boson exchange can take advantage of the nesting between different Fermi pockets to produce a superconducting phase.}
\end{figure}

In order to analyze fluctuations, we rewrite the slave boson operator as $b_{\bf r} = B + \delta b_{\bf r}$, and rewrite the Hamiltonian as a theory of hybridized electrons $c_{n{\bf r}}$ in the presence of the mean-field condensate, which can now interact by exchanging slave bosons $\delta b_{\bf r}$. As long as the typical fluctuations $\delta b_{\bf r}$ are small and random enough to not change the average slave boson density, and not create locally unphysical states, we may treat them perturbatively without any further concern about the constraint. This also allows us to switch to the momentum space representation of the interacting Hamiltonian:
\begin{eqnarray}\label{SB2Db}
&& \!\!\!\!\!\!\! H_{\textrm{2D}}=\int\limits_{\textrm{1BZ}}\frac{d^2 k}{(2\pi)^2}\left(\sum_n E_{n{\bf k}}^{\phantom{\dagger}}
   c_{n{\bf k}}^\dagger c_{n{\bf k}}^{\phantom{\dagger}}+u\,\delta b_{\bf k}^\dagger \delta b_{\bf k}^{\phantom{\dagger}}\right) \\
&& \!\!\! +\int\limits_{\textrm{1BZ}}\frac{d^{2}k}{(2\pi)^{2}}\frac{d^{2}k'}{(2\pi)^{2}}\sum_{nn'}\left(
   V_{n{\bf k},n'{\bf k}'}^{\phantom{\dagger}}c_{n{\bf k}}^{\dagger}c_{n'{\bf k}'}^{\phantom{\dagger}}
   \delta b_{{\bf k}'-{\bf k}}^{\dagger}+h.c.\right) \ . \nonumber
\end{eqnarray}
The index $n$ labels eigenstates in the mean-field version of the Hamiltonian (\ref{SB2D}), where every occurrence of $b_{\bf r}$ is replaced by $B$. The corresponding mean-field quasiparticle spectrum $E_{n{\bf k}}$ is given by a two-dimensional extension of (\ref{SBspect}) that takes into account the spin-momentum locking. Various technicalities, presented elsewhere, are hidden behind the simple appearance of the interaction term. For example, the fluctuations that are longitudinal and transversal to the condensate acquire different couplings to the fermions, but one can neglect or integrate out the higher-energy longitudinal fluctuations. As an approximation, we handle only one Feynman diagram vertex in the perturbation theory shown in Fig.\ref{Diagrams}(a), in which a slave $f$ fermion is converted to a $d$ electron via the emission of a slave boson, or the other way round. The microscopic theory produces another vertex, the mutual scattering of a slave fermion and a slave boson, but this process is negligible because its amplitude is inversely proportional to the $f$ electron effective mass.

The perturbation theory is made difficult by two features. First, slave bosons are artificial degrees of freedom without intrinsic (bare) dynamics. Polarization bubble diagrams are entirely responsible for producing an effective dynamics of slave bosons. A slave boson propagator acquired in this manner mediates interactions between fermionic quasiparticles. The most important processes are fermion scattering in the particle-hole and Cooper channels, shown in Fig.\ref{Diagrams}(b-c). The former contributes self-consistently to the slave boson propagator, which can be calculated at least at the RPA level by summing infinite series' of ladder and bubble diagrams (see Fig.\ref{CollectiveMode}). There lurks the second difficulty: the ladder diagrams, which contain repeated exchanges of slave bosons between two propagating fermions, are hard to calculate because both the slave boson propagator and the vertex function have non-trivial momentum and frequency dependence. The RPA summation of ladder diagrams deals with two-body correlation functions, making it a task equivalent to solving a non-local three-dimensional partial differential equation.

Various approximations put in place to solve the above problems enable calculating the energy dispersion of a collective paramagnon mode represented by the $\widetilde{\Gamma}'$ diagram in Fig.\ref{CollectiveMode}. In essence, certain parts of the full vertex function and slave boson propagator are approximated by a few phenomenological parameters that depend on the self-consistent renormalization. The simplified vertex and propagator are then amenable to further analytical and numerical treatment. There is little point in trying to calculate these phenomenological parameters starting from the microscopic formulation of the problem (\ref{SB2Db}), because they depend on the poorly known microscopic details of the surface band-structure, interactions, etc. Instead, it is more useful to treat them as fitting parameters in the theory.

The perturbative slave boson calculation of this kind has been done for the bulk SmB$_6$ crystal, where a weakly dispersing collective paramagnon mode is seen by neutron scattering. The parameters were fitted to match the calculated and measured mode energy everywhere in the first Brillouin zone (where data is available) \cite{Fuhrman2014}. The calculated spectral weight qualitatively follows the measured neutron scattering intensity as a function of momentum transfer. Therefore, despite the high level of uncontrolled approximations, the perturbative slave boson theory may be able to reveal the correct physical picture of strong correlations in bulk Kondo insulators.

By a direct formal analogy, we can predict the existence of similar collective modes in the surface spectrum of Kondo TIs. Only now, the surface is a two-dimensional metal instead of a three-dimensional insulator. Two-dimensional systems with gapless excitations are surely more sensitive to quantum fluctuations and susceptible to instabilities than a bulk insulator, where a coherent collective mode is still created by strong interactions. The mere existence of a gapped surface paramagnon is not expected to qualitatively alter the helical Fermi liquid behavior of the surface states, but should renormalize their dynamics. However, this collective mode can condense, or stimulate other kinds of instabilities such as superconductivity. Below we discuss some possibilities.

It is well known that repulsive interactions between electrons on a nested Fermi surface can lead to spin density wave (SDW) instabilities. The most direct effect of slave boson fluctuations is indeed to generate repulsive interactions between hybridized electrons. Looking at the LDA+Gutzwiller Fermi surface of the 100 cut of SmB$_6$, approximately depicted in Fig.\ref{SmB6surf}(b), nesting is possible at wavevectors $(\pi,\pi)$ and $(\pi,0)$, $(0,\pi)$. The Fermi pockets may be small, but the bulk collective mode is found \cite{Fuhrman2014} to have the lowest energy and exceedingly largest spectral weight at wavevectors $(\pi,\pi,\pi)$ and $(\pi,0,0)$, which project to the ``nesting'' wavevectors of the 100 crystal boundary. Given that our system is in the strong-coupling limit, it is fairly likely that an SDW instability takes place in this hybridized regime. The outcome is a ground state that breaks TR symmetry and thus \emph{gaps out} the Dirac points of the surface heavy fermion metal. Should this happen, it could be experimentally tested by raising temperature and observing a restoration of the protected surface metal when the SDW is thermally destroyed. A harder test would be to look for lattice translation symmetry breaking on the crystal boundary, or Goldstone modes associated with the broken spin-rotation symmetry. Some amount of ferromagnetic polarization could also occur due to the Rashba spin-orbit coupling, which introduces Dzyaloshinskii-Moriya type of couplings between lattice spins.

Another interesting possibility is a superconducting phase. This is the two-dimensional analogue of a superconducting state that hides the quantum critical point between magnetic and heavy-fermion metallic phases of heavy fermion materials \cite{Mathur1998}. The slave boson mediates repulsive interactions between fermions, but this can still lead to unconventional superconductivity by having sign changes of the order parameter between different Fermi pockets. The prominent candidate states in the Fermi surface from Fig.\ref{SmB6surf}(b) are a $d$-wave and a sign-changing $s$-wave superconductor. The former features opposite signs of the order parameter on the two symmetry-related X points $(\pi,0)$, $(0,\pi)$ of the Brillouin zone, and requires slave boson transfers at the ``nesting'' wavevector $(\pi,\pi)$. The latter features the same order parameter sign on the two X points, and the opposite sign on the $\Gamma$ point, being stimulated by the slave bosons with $(\pi,0)$, $(0,\pi)$ wavevectors. All of these ``singlet'' superconducting states effectively screen the charge of fermionic quasiparticles, but allow spin to be transported by the surviving helical Dirac metal. The promising conditions for superconductivity are indeed found near the SDW quantum critical point, where the slave boson disguised as a collective paramagnon mode is about to condense.

Making more than these phenomenological predictions is a rather daunting task, in many ways comparable to attempts to solve the problem of high-temperature superconductivity in cuprates. Specifically, we can hardly tell which particular instability is the most prominent and takes place. This is largely a microscopic question of competing orders. Precise shapes and sizes of the Fermi pockets, as well as the collective mode dispersion details, play a major role in selecting the wining instability. The only thing we can reliably say is that different instabilities may occur on different cuts or sides of the Kondo TI crystal. It may be even feasible to stimulate one instability over another by interface engineering.

As a last topic in this section, we briefly mention the possibility of obtaining exotic boundary states featuring electron fractionalization.
Strong interactions in a two-dimensional geometry can localize particles into a Mott insulator at a lattice-commensurate density of $p/q$ particles per site \cite{balents05}. The simplest kind of such an insulator is a charge density wave (CDW). However, no spinful particle can experience backscattering on the surface of a TI unless the TR symmetry is broken. A charge Mott insulator can exist either if some neutral spinful fermions remain delocalized to form Dirac points, or if the TR symmetry is spontaneously broken (exceptions to the latter have been recently identified \cite{Metlitski2013, Bonderson2013, Senthil2014, Senthil2014a}). The former is an exotic ``algebraic'' spin liquid state with spin-charge separation, in which the TI's surface is metallic for spin and not for charge.

Fractionalized states of matter are captured in our formalism by a slave boson field that takes the full charge of an $f$ electron and leaves behind a neutral spinon as the slave fermion excitation. The process of fractionalization cannot be described perturbatively, but could be encouraged by the vibrant spin dynamics of dense and strongly interacting $f$ electrons, especially through the quantum motion of Kondo singlets which frustrates the spatial spin correlations. Delocalized fractionalized $f$ electrons would allow the slave boson to condense and form an exotic fractionalized superconducting state \cite{senthil00}. Alternatively, localized $f$ electrons could suppress the condensation of slave bosons and instead produce a spin liquid ground state.

\subsubsection{Localized moment regime}\label{secKondoCorrel}

This section focuses on the Kondo TI boundary regime with large band bending shown in Fig.\ref{KIbands}(d). The $f$ electrons are localized by Coulomb interactions and their orbital is essentially half-filled. A three-dimensional heavy fermion metal in this regime could be expected to exhibit an antiferromagnetic Neel order of $f$ electrons and a conduction sea of $d$ electrons with a small Fermi surface. But, in two dimensions there are fewer lattice bonds per site for spatial spin correlations, and Kondo singlets are more competitive. Some $f$ electron moments can be consumed by Kondo singlets, and some involved in establishing correlations across separated lattice sites. Kondo singlets act as dopants that destabilize the prospects of $f$ electrons to form long range magnetic or valence-bond solid (VBS) orders. The resulting spin dynamics depends on how well the $f$ electron spins are screened via the Kondo mechanism.

One way of exploring the dynamics in this regime is provided by the well-known Kondo lattice model:
\begin{equation}\label{KondoLat}
H = -\sum_{{\bf r}{\bf r}'}t_{{\bf r}{\bf r}'}^{\phantom{\dagger}}d_{{\bf r}}^{\dagger}d_{{\bf r}'}^{\phantom{\dagger}}
    +\frac{J_{K}}{2}\sum_{{\bf r}}{\bf S}_{{\bf r}}^{\phantom{\dagger}}d_{{\bf r}}^{\dagger}\boldsymbol{\sigma}d_{{\bf r}}^{\phantom{\dagger}}
    +\sum_{{\bf r}{\bf r}'}J_{{\bf r}{\bf r}'}^{\phantom{\dagger}}
      {\bf S}_{{\bf r}}^{\phantom{\dagger}}{\bf S}_{{\bf r}'}^{\phantom{\dagger}}+\cdots \nonumber
\end{equation}
This is a low-energy effective theory of localized moments $\bf S$ coupled to the conduction $d$ electrons.
 
If the short-range RKKY exchange $J_{{\bf r}{\bf r}'}$ is sufficiently larger than the Kondo exchange $J_K$, then the local moments prefer to establish spatial correlations among themselves rather than participate in Kondo singlets. Antiferromagnetic or any long-range order of local moments that breaks the TR symmetry will gap out the Dirac points of conduction electrons on the TI surface, and potentially but not necessarily produce an insulating surface. If the TR symmetry is not broken, then the metallic state surely survives on the TI boundary with a protected odd number of Dirac points. The charge transport properties of the surface are qualitatively the same as in uncorrelated TI's, except that the Fermi surface (and the structure of Dirac points) may be reconstructed due to the order of local moments. The local moments can themselves add charge-neutral low-energy excitations to the spectrum.

At least in SmB$_6$, the $d$-$f$ hybridization energy scale $V$ seems to be considerably larger than the bandwidth $t_f$ of the $f$ orbitals. Then, we may naively expect the opposite $J_K \gg J_{{\bf r}{\bf r}'}$ limit. The local $f$ moments are screened via Kondo singlets whenever possible in such circumstances. Over-screening occurs if the number of local moments is smaller than the number of $d$ electrons. Since the over-screened moments are entirely consumed by Kondo singlets, their spatial correlations are short-ranged, featureless and accompanied by gapped excitations (broken singlets), while the surplus $d$ electrons can conduct currents on the Kondo TI's surface as an uncorrelated metal. The opposite and realistic under-screened limit opens new possibilities, as sufficiently many local $f$ moments may be left alone to strengthen correlations among themselves \cite{Nozieres1985a}. The possibilities for correlations range from magnetic orders that gap out the Dirac points, to metallic VBS and spin liquid states.

A spin liquid of localized $f$ electrons is a real possibility in the under-screened Kondo singlet regime. Every lattice site temporarily caught in the state of having exactly one $d$ electron will neutralize one local moment through a Kondo singlet. This becomes a mobile ``magnetic hole'' in the $f$ orbital which frustrates the two-dimensional spatial correlations of the local moments. If the outcome of frustration is a spin liquid, we can most easily describe it within the TI-surface slave boson formalism (\ref{SB2D}) based on the Anderson model (\ref{MinMod2D}). The spin liquid is captured by a non-condensed \emph{charged} slave boson field that separates the charge of physical $f$ electrons from the neutral slave fermions $\psi$. The slave boson energy gap is the charge excitation gap of localized $f$ electrons. The slave fermions are localized via the slave-boson constraint, but their gapped hole excitations are mobile and represent spinons. We will now work out the feedback of this spin liquid dynamics on the charge transport properties of conduction electrons.

The average number $f^\dagger_{\bf r} f^{\phantom{\dagger}}_{\bf r} \lesssim 1$ of $f$ electrons on a surface lattice site is close to but smaller than one in the localized moment regime. It cannot exceed one in our effective Anderson model (\ref{MinMod2D}) when $U \to \infty$, but can be reduced below one by virtual transfers of $f$ electrons to the $d$ orbital due to the hybridization term $V$. Once we switch to the slave boson Hamiltonian (\ref{SB2D}), we can use the exact local constraint
\begin{equation}\label{ConstrSB2D}
\psi_{\bf r}^\dagger \psi_{\bf r}^{\phantom{\dagger}} + b_{\bf r}^\dagger b_{\bf r}^{\phantom{\dagger}} = 1
\end{equation}
and the commutator $\lbrack b_{\bf r}^{\phantom\dagger}, b_{\bf r}^\dagger \rbrack = 1$ to express the number of $f$ electrons on a site:
\begin{equation}
f^\dagger_{\bf r} f^{\phantom{\dagger}}_{\bf r}
= b^{\phantom{\dagger}}_{\bf r} b^\dagger_{\bf r} \psi^\dagger_{\bf r} \psi^{\phantom{\dagger}}_{\bf r}
= \left( 1 + b^\dagger_{\bf r} b^{\phantom{\dagger}}_{\bf r} \right)
  \left( 1 - b^\dagger_{\bf r} b^{\phantom{\dagger}}_{\bf r} \right)
= 1 - \left( b^\dagger_{\bf r} b^{\phantom{\dagger}}_{\bf r} \right)^2 \ . \nonumber
\end{equation}
The number of slave bosons $b^\dagger_{\bf r} b^{\phantom{\dagger}}_{\bf r}$ on every site is close to zero, but still finite. Without the hybridization between the $d$ and $f$ orbitals, there would be no Kondo singlets and strictly no slave bosons on any site in the ground state. However, since the hybridization term does not conserve the slave boson number while conserving charge, we generally have $\langle b_{\bf r}^\dagger b_{\bf r}^{\phantom{\dagger}} \rangle > 0$ without necessarily having a superconducting condensate of slave bosons. The system remains an insulator much like the QED vacuum despite its virtual electron-positron fluctuations.

We begin by translating the Hamiltonian (\ref{SB2D}) to an imaginary-time path integral with the action:
\begin{eqnarray}\label{MinAct2D}
S_{\textrm{sb}} &=& \int d\tau \biggr\lbrack \sum_{\bf r} \left(
      d_{\tau\bf r}^* \frac{\partial}{\partial\tau} d_{\tau\bf r}^{\phantom{*}}
     +b_{\tau\bf r}^{\phantom{*}}\psi_{\tau\bf r}^* \frac{\partial}{\partial\tau} b_{\tau\bf r}^*\psi_{\tau\bf r}^{\phantom{*}} \right)
     \nonumber \\
&& +H_{\textrm{sb}} \Bigl( d_{\tau\bf r}^{\phantom{*}}, d_{\tau\bf r}^* ; \psi_{\tau\bf r}^{\phantom{*}}, \psi_{\tau\bf r}^* ;
      b_{\tau\bf r}^{\phantom{*}}, b_{\tau\bf r}^*  \Bigr) \biggr\rbrack \ .
\end{eqnarray}
Here, $d_{\tau\bf r}^{\phantom{*}}, d_{\tau\bf r}^*, \psi_{\tau\bf r}^{\phantom{*}}, \psi_{\tau\bf r}^*$ are Grassmann numbers, $b_{\tau\bf r}^{\phantom{*}}, b_{\tau\bf r}^*$ are complex numbers, and $H_{\textrm{sb}}(\cdots)$ is the Hamiltonian (\ref{SB2D}) with all operators replaced by their corresponding Grassmann or complex fields. We may implement the constraint (\ref{ConstrSB2D}) either directly in the path-integral measure or through a Lagrange multiplier, the details of which are not important in the following discussion.

Since the fields $b_{\tau\bf r}$ live in a Mott-like insulating state at the sample surface, their local amplitude fluctuations $|b_{\tau\bf r}|$ are suppressed as high-energy excitations. However, their phase $\theta_{\tau\bf r}$ fluctuations in
\begin{equation}
b_{\tau\bf r} = |b_{\tau\bf r}| e^{i\theta_{\tau\bf r}}
\end{equation}
are hardly restricted. Of special interest to us will be the vortex configurations of $\theta_{\tau\bf r}$, which proliferate in our two-dimensional state of uncondensed bosons. A certain energy cost is associated to a vortex core because the slave boson density must be depleted there. This disturbance of the optimal $\langle b_{\bf r}^\dagger b_{\bf r}^{\phantom{\dagger}} \rangle > 0$ represents a local expulsion of the Kondo singlets from the vortex core region.

The Hamiltonian part $H_{\textrm{sb}}$ part of the action (\ref{MinAct2D}) reads:
\begin{eqnarray}
H_{\textrm{sb}} \!\!&=&\!\! \sum_{{\bf r}{\bf r}'}\biggl\lbrack d_{\tau\bf r}^*\,K_{{\bf r}{\bf r}'}^d\,d_{\tau{\bf r}'}^{\phantom{*}}
   +b_{\tau{\bf r}'}^* d_{{\tau\bf r}}^{*}V_{{\bf r}{\bf r}'}^{\phantom{*}}\,
     \psi_{{\tau\bf r}'}^{\phantom{*}}+h.c. \nonumber \\
&& +b_{\tau\bf r}^{\phantom{*}} b_{\tau{\bf r}'}^*
     \psi_{\tau\bf r}^*\,K_{{\bf r}{\bf r}'}^f\,\psi_{{\tau\bf r}'}^{\phantom{*}} \biggr\rbrack \ .
\end{eqnarray}
We collected all details of the SU(2) gauged hopping and inter-surface tunneling of electrons into the $K^d$ and $K^f$ symbols. We are free to carry out a gauge transformation
\begin{equation}
d_{{\tau\bf r}} \to \widetilde{d}_{{\tau\bf r}} e^{-i\theta_{\tau\bf r}} \quad,\quad
b_{{\tau\bf r}} \to \widetilde{b}_{{\tau\bf r}} e^{i\theta_{\tau\bf r}}
\end{equation}
by a change of variables in the path-integral. Slave fermions are not affected because they are neutral. Note that electrons and slave bosons carry opposite charges with respect to the physical electromagnetic gauge field, which is also transformed but not shown in this discussion due to being less important. Our choice of the gauge transformation will depend on the field configuration in a way that makes the new slave boson field purely real and positive, $\widetilde{b}_{{\tau\bf r}} = |\widetilde{b}_{{\tau\bf r}}| \in \mathbb{R}$. Then, after dropping the tilde symbols, the Hamiltonian written in terms of the new $d$ and $b$ fields becomes:
\begin{eqnarray}
&& \!\!\!\!\!\!\!\! H_{\textrm{sb}} = \sum_{{\bf r}{\bf r}'}e^{i(\theta_{\tau\bf r}^{\phantom{l}}-\theta_{{\tau\bf r}'})}\biggl\lbrack 
    d_{\tau\bf r}^*\,K_{{\bf r}{\bf r}'}^d\,d_{{\tau\bf r}'}^{\phantom{*}} \\
&& +|b_{\tau{\bf r}'}^{\phantom{*}}|\,d_{{\tau\bf r}}^{*}V_{{\bf r}{\bf r}'}^{\phantom{*}}\,
     \psi_{{\tau\bf r}'}^{\phantom{*}}+h.c.
   +\,|b_{\tau\bf r}^{\phantom{*}}b_{\tau{\bf r}'}^{\phantom{*}}|\,
   \psi_{\tau\bf r}^*\,K_{{\bf r}{\bf r}'}^f\,\psi_{{\tau\bf r}'}^{\phantom{*}} \biggr\rbrack \nonumber \ .
\end{eqnarray}
Its form is reminiscent of a lattice gauge theory if we interpret
\begin{equation}\label{CompactGF}
A_{\tau{\bf r}, \tau'{\bf r}'} = \theta_{\tau\bf r}-\theta_{{\tau'\bf r}'}
\end{equation}
as a compact U(1) gauge field that lives on the lattice bonds. Even though $A_{\tau{\bf r}, \tau'{\bf r}'}$ looks like a pure gauge, it is impossible to trivially absorb it into matter fields when we keep the slave boson $|b_{\tau{\bf r}'}|$ strictly real, and the slave fermions are neutral. Consequently, $A_{\tau{\bf r}, \tau'{\bf r}'}$ must have some physical effect. It serves merely as a convenient way to separate the abundant fluctuations of the slave boson phase $\theta_{\tau{\bf r}}$ from the high-energy amplitude fluctuations $|b_{\tau{\bf r}'}|$ that we want to integrate out.

In order to make progress, we ought to temporarily discretize the imaginary time $\tau \to \Delta\tau \times \textrm{integer}$. This merely corresponds to identifying a high energy cut-off $\Lambda \sim (\Delta\tau)^{-1}$. The above Hamiltonian introduces the gauge field on the spatial links, while the discretized time derivative terms in the action (\ref{MinAct2D}) introduce the gauge field on the temporal links of the space-time lattice.

We are now ready to integrate out the gapped fluctuations of the slave fermions $\psi$ and the slave boson amplitude $|b|$, which are correlated by the local constraint (\ref{ConstrSB2D}). Instead of doing it in detail, we rely on the gauge symmetry to restrict the form of the resulting effective action on the space-time lattice sites $i\equiv({\tau,\bf r})$. Our first naive guess is:
\begin{eqnarray}\label{MinAct2Db}
S'_{\textrm{sb}} &=& \sum_i d_i^* \Bigl( \Delta_\tau - iA_{i,i+\Delta\tau} \Bigr) d_i^{\phantom{*}}
   + \sum_{\langle i,j \rangle} e^{iA_{i,j}} d_i^{*}K_{i,j}^{\phantom{f}}d_j^{\phantom{*}} \nonumber \\
&& - \mathcal{K}' \sum_\square \cos\Bigl(\textrm{curl}(A_{i,j}\Bigr) \ ,
\end{eqnarray}
which keeps treating the gauge field as a compact one. The discrete time derivative means $\Delta_\tau d_{\tau\bf r} = d_{\tau+\Delta\tau,\bf r} - d_{\tau\bf r}$, and the symbol $\langle i,j \rangle$ indicates the summation over the nearest-neighbor lattice sites at the same time $\tau$. The last Maxwell summation runs over all lattice plaquettes and takes the lattice curls of the gauge field on them: $\textrm{curl}(A_{i,j})$ is defined as the sum of $A_{i,j}$ on the four plaquette bonds oriented in a circular clockwise sense.

It turns out, however, that our guess (\ref{MinAct2Db}) is too naive because it microscopically treats (\ref{CompactGF}) as a pure gauge. It follows from (\ref{CompactGF}) that $A_{i,j}$ must have a quantized $2\pi n$ flux on every lattice plaquette, so that the compact Maxwell term is just a constant. The remaining appearances of $A_{i,j}$ can be completely removed by a gauge transformation, and we end up with a theory in which the slave boson phase fluctuations have absolutely no effect.

\begin{figure}
\includegraphics[height=1.5in]{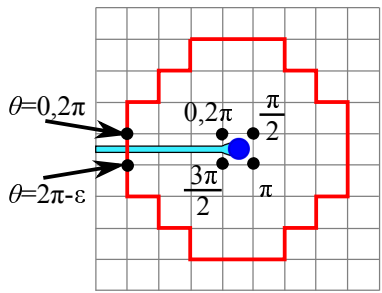}
\caption{\label{Interp}Illustration of the two interpretations of a superfluid phase in the vicinity of a vortex. The thick red loop is a path on the lattice on which the phase factor $e^{i\theta}$ gradually rotates by $2\pi$ in the complex plane. The phase is single-valued on all lattice sites, but the angle $\theta$ can be interpreted either as having a jump across a semi-infinite string that emanates from the singularity, or as being smooth but not single-valued far away from the singularity.}
\end{figure}

Still, the overall form of the effective action is restricted by the gauge symmetry. We will now argue that the theory we seek is a \emph{non-compact} gauge theory. We have no means to mathematically derive this result, so the argument will be phenomenological. Consider a quantized slave boson vortex on the lattice. We can interpret the corresponding $\theta_{i}$ configuration in two different ways. In the first interpretation, $\theta_{i}$ is considered single-valued, so that the gauge field $A_{i,j}$ is a pure gauge. Alternatively, the phase $\theta_{i}$ can be interpreted as a gradually varying quantity on the lattice, at the expense of being defined up to an additive value $2\pi n$, $n\in\mathbb{Z}$. Then, the gauge field $A_{i,j}$ is allowed to have a non-zero quantized flux
\begin{equation}
\Phi_{C} = \sum_{\langle i\to j \rangle}^C A_{i,j} = 2\pi n \quad,\quad n\in\mathbb{Z}
\end{equation}
through any oriented closed loop $C$ on the space-time lattice. The two interpretations are absolutely equivalent on the lattice because the lattice derivatives (including the discretized time derivative) are always applied on $e^{i\theta}$ which is insensitive to the changes of $\theta$ by $2\pi$. This is illustrated in Fig.\ref{Interp}. However, we will have to adopt the second interpretation when constructing the continuum limit. The lattice gradients of $\theta_{i}$ (i.e. $A_{i,j}$) must be small for taking the continuum limit, even far away from a vortex core where the first interpretation would make $\theta_{i}$ jump by almost $2\pi$ on a single lattice link somewhere on the loop $C$ that encloses the singularity. In other words, we must wind $\theta_{i}$ and give a finite flux to $A_{i,j}$ on the loops $C$ around a vortex in order to consistently describe it in the continuum limit.

The appropriate continuum limit effective action restricted by the gauge symmetry can be formally constructed from (\ref{MinAct2Db}) by keeping only the lowest order terms in the gradient expansion:
\begin{eqnarray}\label{MinAct2Dc}
S_{\textrm{sb}}^{\textrm{eff}} &=& \int d\tau\,d^2 r \Bigl\lbrace d^* (\partial_\tau - iA_\tau) d
   + \mathcal{K} (\epsilon^{\mu\nu\lambda} \partial_\nu A_\lambda)^2 \nonumber \\
&& + \Bigl\lbrack(\boldsymbol{\nabla} - i{\bf A})d\Bigr\rbrack^* \widetilde{K} \Bigl\lbrack(\boldsymbol{\nabla} - i{\bf A})d\Bigr\rbrack 
    \Bigr\rbrace \ .
\end{eqnarray}
We used the convenient Einstein's notation for the summation over repeated indices $\mu\in\lbrace \tau, x, y \rbrace$ and the Levi-Civita symbol $\epsilon^{\mu\nu\lambda}$ to represent the Maxwell term $\mathcal{K}$, now a non-compact 2+1D curl in the continuum space-time. Strictly speaking, the continuum gauge field $A_\mu = (A_\tau, {\bf A})$ inherits perfectly quantized flux lines (now infinitely thin line singularities) from its definition (\ref{CompactGF}). However, these singularities are mobile and proliferate in the spin liquid state, so a simple coarse-graining (integrating out some high-energy fluctuations) relaxes this constraint and captures flux diffusion. Even though we did not microscopically derive the non-compact Maxwell term, there is a clear physical justification for it. The rapid singular variations of $\theta$ near a vortex core cannot be accurately described by the continuum limit in which we keep only the lowest-order gradients. Instead of introducing higher-orders in the gradient expansion, we phenomenologically associate some extra energy cost $\mathcal{K}$ to the regions where we have a finite vortex density $(\epsilon^{\mu\nu\lambda} \partial_\nu A_\lambda)^2 \neq 0$. This takes into account the energy cost of vortex cores.

The effective theory (\ref{MinAct2Dc}) describes a 2D Fermi surface of ``helical'' $d$ electrons coupled to a 2D fluctuating U(1) gauge field. It has an additional ``photon'' degree of freedom to the $d$ electrons. This is a collective mode of mobile Kondo singlets, made possible by the ``magnetic hole'' doping of the $f$ orbitals. The $d$ electrons cannot screen the magnetic fields of $A_\mu$, so the ``photon'' mode is gapless. A yet more complete theory includes the coupling of the two-dimensional $d$ electrons to the physical three-dimensional electromagnetic field.

The Rashba spin-orbit coupling locks the spin of $d$ electrons to their momentum and leaves a single spin mode at low energies. A theory equivalent to (\ref{MinAct2Dc}) but with two degenerate spin modes of electrons coupled to a 2D U(1) gauge field was extensively studied in the context of $\nu=\frac{1}{2}$ quantum Hall states. A two-component 2D metal in the presence of gauge fields is known to be only marginally different from a standard Fermi liquid \cite{Halperin1993, Altshuler1994, Kim1994a, Kim1995, Kim1995a, Kim1996}. Gauge invariance protects the conventional Fermi-liquid-like behavior of the most usually observed equilibrium and transport properties of this system, but various thermodynamic and transport properties exhibit sub-leading deviations from the Fermi liquid behavior \cite{Kim1994a}.

An open question here is whether our one-component ``helical'' metal coupled to a gauge field has any different behavior from the two-component gauged metal. One obvious difference is that large momentum transfers across the Fermi surface by gauge bosons are suppressed in the helical metal in comparison to those of the ordinary metal. Namely, two-dimensional photons have only one state of polarization and thus cannot transfer spin. The helical spin-momentum locking implies that a large momentum transfer across the Fermi surface must be accompanied by a spin flip, which cannot be generated by a 2D photon. Such momentum transfers in ordinary metals are not a problem because they need not be accompanied by spin flips. Therefore, the two types of non-Fermi liquids may have different charge and spin responses to perturbations with spatial periodicity of the order of $k_f^{-1}$. This is in addition to every aspect of the helical response caused by the non-trivial topology.

\section{Kondo TI Quantum Wells}\label{secTIQW}

Quantum wells made from Kondo TIs are a potentially interesting and tunable platform for creating novel strongly correlated states of matter. The Rashba spin-orbit coupling is strong enough to make lattice details important for dynamics, i.e. the amount of flux per plaquette is not small (equivalent to about $1000 \textrm{ T}$ magnetic fields in bismuth-based TIs).

Here we will focus on the Kondo TIQWs within the hybridized regime, which has a rich phenomenology. Two opposite surfaces of the Kondo TI contribute their Dirac quasiparticle bands to the low-energy spectrum, but all Dirac points are gapped by the inter-surface tunneling. We will assume that the ensuing bandgap in the Dirac spectrum is sufficiently smaller than the bandgap of the bulk 3D crystal. This two-dimensional insulator of hybridized $d$ and $f$ electrons is still strongly affected by the Rashba spin-orbit coupling and Coulomb interactions. The slave boson model of this system is given by (\ref{SB2D}), where $\Delta_{{\bf r}{\bf r}'} = \Delta_0 + \cdots$ acquires a constant tunneling term $\Delta_0$ in each orbital. An improved model can also take into account the ``extended range'' Coulomb interactions between (at least $f$) electrons on the opposite TIQW surfaces.

On one hand, the existence of a gap in the ``surface'' spectrum can hinder some surface instabilities that were discussed in section \ref{secHybridCorrel}. On the other hand, a quantum well can be embedded in a gated heterostructure device that allows controlling the chemical potential placement within its band-structure. The chemical potential can be raised or lowered toward or into a band by tuning the gate voltage. This is a practical way to control the phases and drive phase transitions in a Kondo lattice material.

The possible instabilities of a Kondo TIQW are a superset of those anticipated on a single Kondo TI surface. The Coulomb repulsion
\begin{eqnarray}\label{ScatProc}
&& \!\!\!\!\!\!\!\! \sum_{\tau\tau'} U_{\tau\tau'}^{\phantom{1}} \sum_{\sigma\sigma'} \sum_{\bf r} 
      c_{\sigma\tau,{\bf r}}^\dagger c_{\sigma\tau,{\bf r}}^{\phantom{\dagger}}
      c_{\sigma'\tau',{\bf r}}^\dagger c_{\sigma'\tau',{\bf r}}^{\phantom{\dagger}} \\
&& \!\!\!\!\! = \sum_{ijmn} \int \frac{d^2 k}{(2\pi)^2}  \frac{d^2 k'}{(2\pi)^2} \frac{d^2 q}{(2\pi)^2} U_{{\bf k},{\bf k}',{\bf q}}^{ijmn}
      c_{i,{\bf k}}^\dagger c_{j,{\bf k}'}^\dagger c_{m,{\bf k}'+{\bf q}}^{\phantom{\dagger}} c_{n,{\bf k}-{\bf q}}^{\phantom{\dagger}} \nonumber
\end{eqnarray}
between local electrons of spin $\sigma$ and surface index $\tau$ has finite overlaps $U_{{\bf k},{\bf k}',{\bf q}}^{ijmn}$ with almost generic TR-respecting scattering processes expressed in terms of the eigenstate field operators $c_{i,{\bf k}}$ of the Kondo TIQW. The eigenstate label $i$ is a two-state index at any fixed wavevector and energy; it morphs into the surface index at high energies away from the gapped Dirac points while spin remains locked to momentum. The conduction and valence bands of gapped Dirac quasiparticles generally have multiple valleys of low-energy excitations (e.g. surrounding the $\Gamma$ and X points in SmB$_6$ quantum wells). The Coulomb scattering can create low-energy particle-hole pairs across the bandgap and optionally between different valleys. It takes an arbitrary weak amount of interaction to create an exciton bound state in two-dimensions, which lives as a coherent gapped excitation at an energy inside the bandgap \cite{Nikolic2010}. An instability at which such a collective mode condenses requires a finite interaction strength.

The intra-surface and inter-surface Coulomb interactions $U_{++}=U_{--}$, $U_{+-}=U_{-+}$ respectively do not depend on the spins of interacting electrons. Furthermore, electron pairs can anti-symmetrize their wavefunctions through the surface index instead of spin, and form small spin-triplets confined by the quantum well potential. Neither the spin nor the surface index are good quantum numbers in the conduction and valence band valleys. Therefore, the scattering processes in (\ref{ScatProc}) will explore exciton pairing in essentially all channels. The kinds of instabilities that were anticipated on a single surface of a TI are still possible, especially if the chemical potential is raised into the conduction or valence band to recreate Fermi pockets. Let us refer to these as ``singlet'' instabilities, even though spin is not conserved. New ``triplet'' instabilities specific to quantum wells may preempt the ``singlet'' ones in insulating ground states.

A triplet exciton or Cooper pair is a spinful bosonic particle which necessarily experiences the strong Rashba spin-orbit coupling of the TI's boundary. This can be formally understood from the gauge principle, since the Rashba spin-orbit coupling effectively introduces a background SU(2) gauge field to any particle with an internal degree of freedom that transforms non-trivially under TR \cite{Nikolic2011a}. Therefore, triplet excitons and Cooper pairs exhibit spin-momentum locking. They acquire a mode whose energy decreases when its momentum increases (the Rashba spin-orbit coupling can be viewed as a spin-dependent Zeeman effect). Even when this triplet mode is costly or unstable at small momenta, a large-momentum triplet (e.g. near the cut-off momentum scale) can be a low energy excitation and even condense.

If this scenario occurs in the Cooper channel, a superconducting order parameter that lowers the ground state energy must have different signs on the valleys between which electron pairs are scattered (given that interactions are repulsive). Pair scattering of the Cooper type occurs within a single band, and becomes stronger when the chemical potential approaches or enters that band. Since the band quantum numbers are mixtures of spin and surface indices, triplet pairing has a finite amplitude in the general scattering process and then becomes dynamically enhanced by the Rashba spin-orbit coupling. An example of the ensuing triplet superconductivity in a TIQW can be found in Ref.\cite{Nikolic2012a}.

\begin{figure}[!b]
\centering
\subfigure[{}]{\includegraphics[height=1.8in]{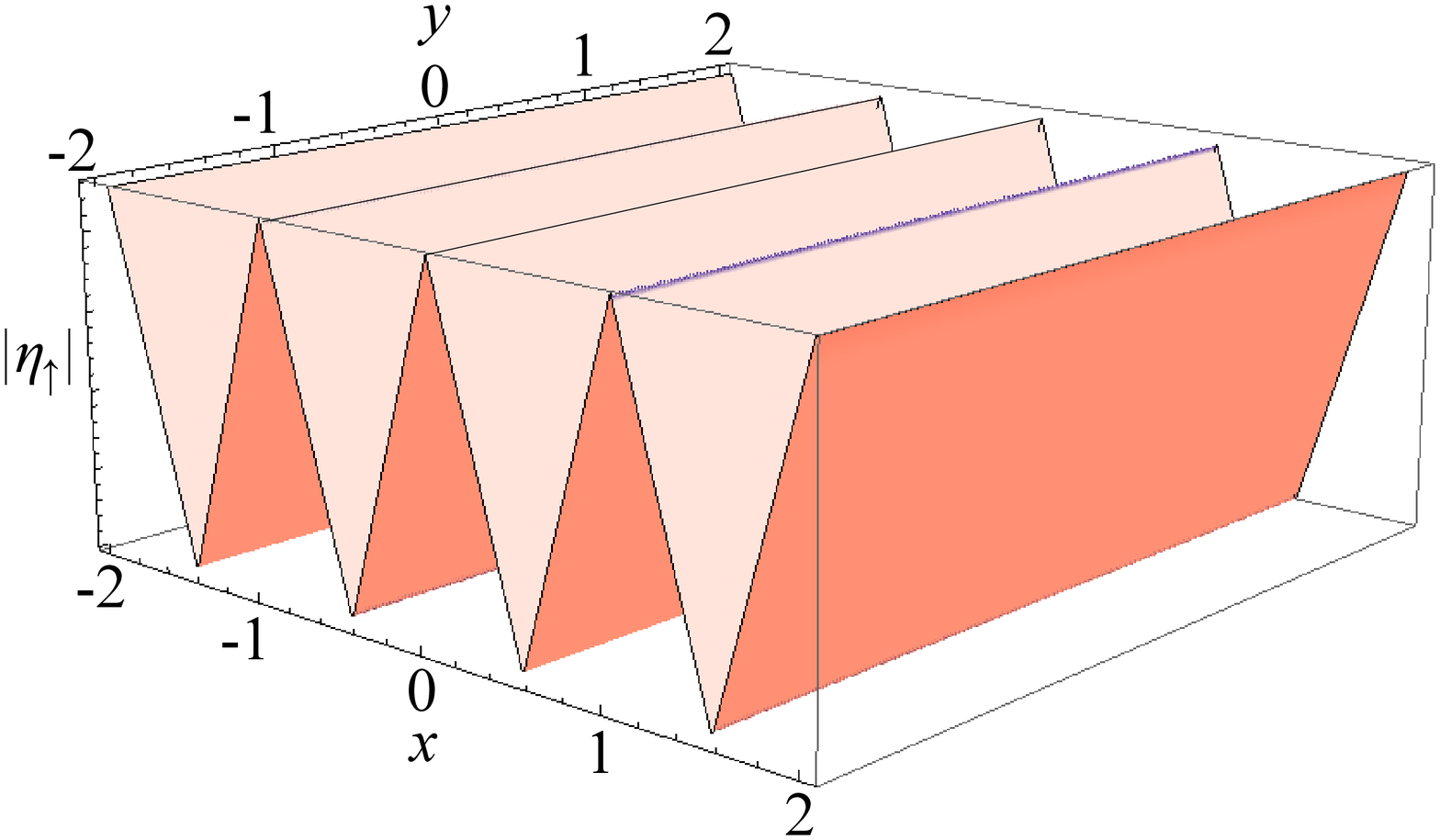}}
\hspace{0.01in}
\subfigure[{}]{\includegraphics[height=1.8in]{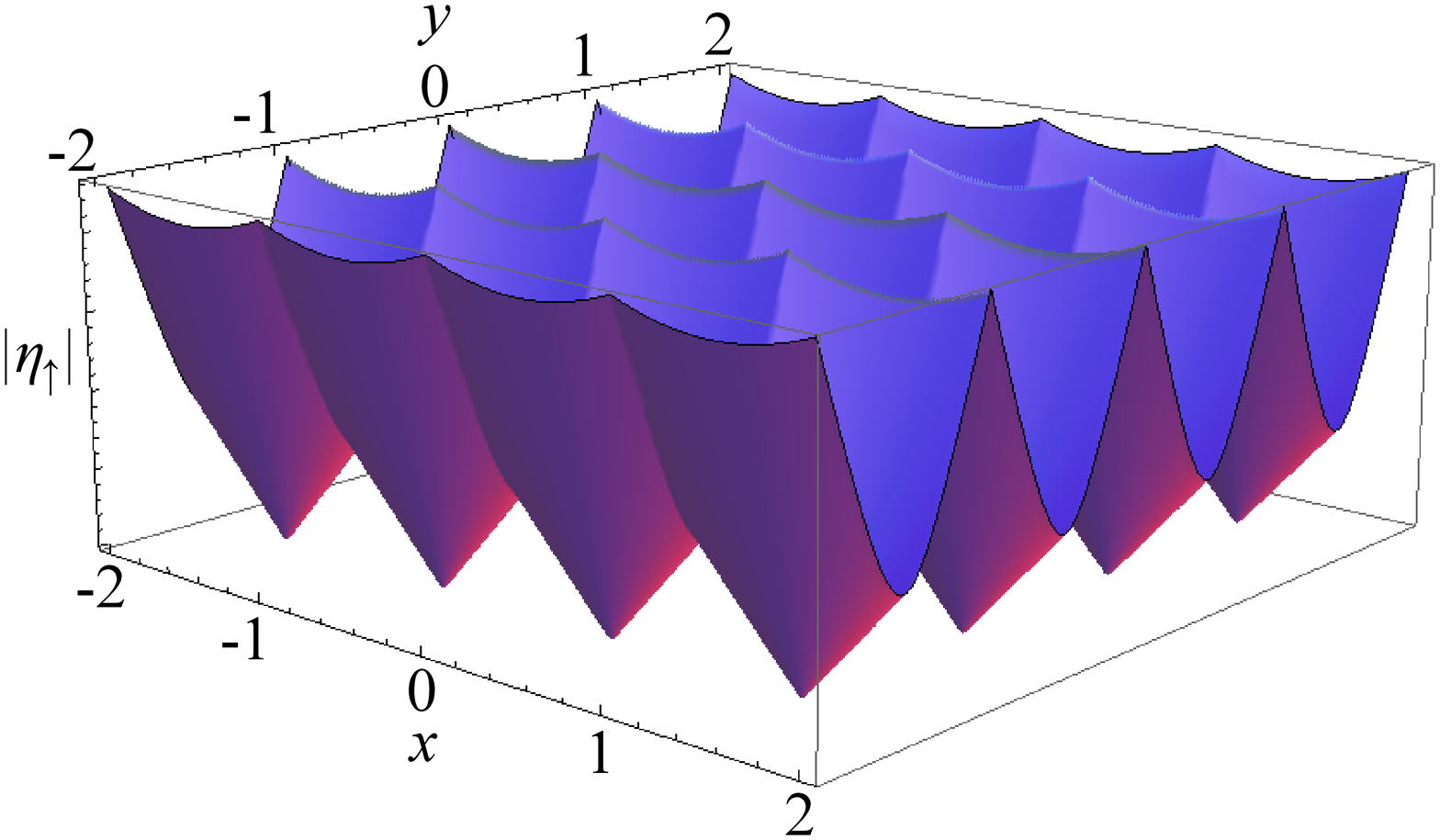}}
\hspace{0.01in}
\subfigure[{}]{\includegraphics[height=1.8in]{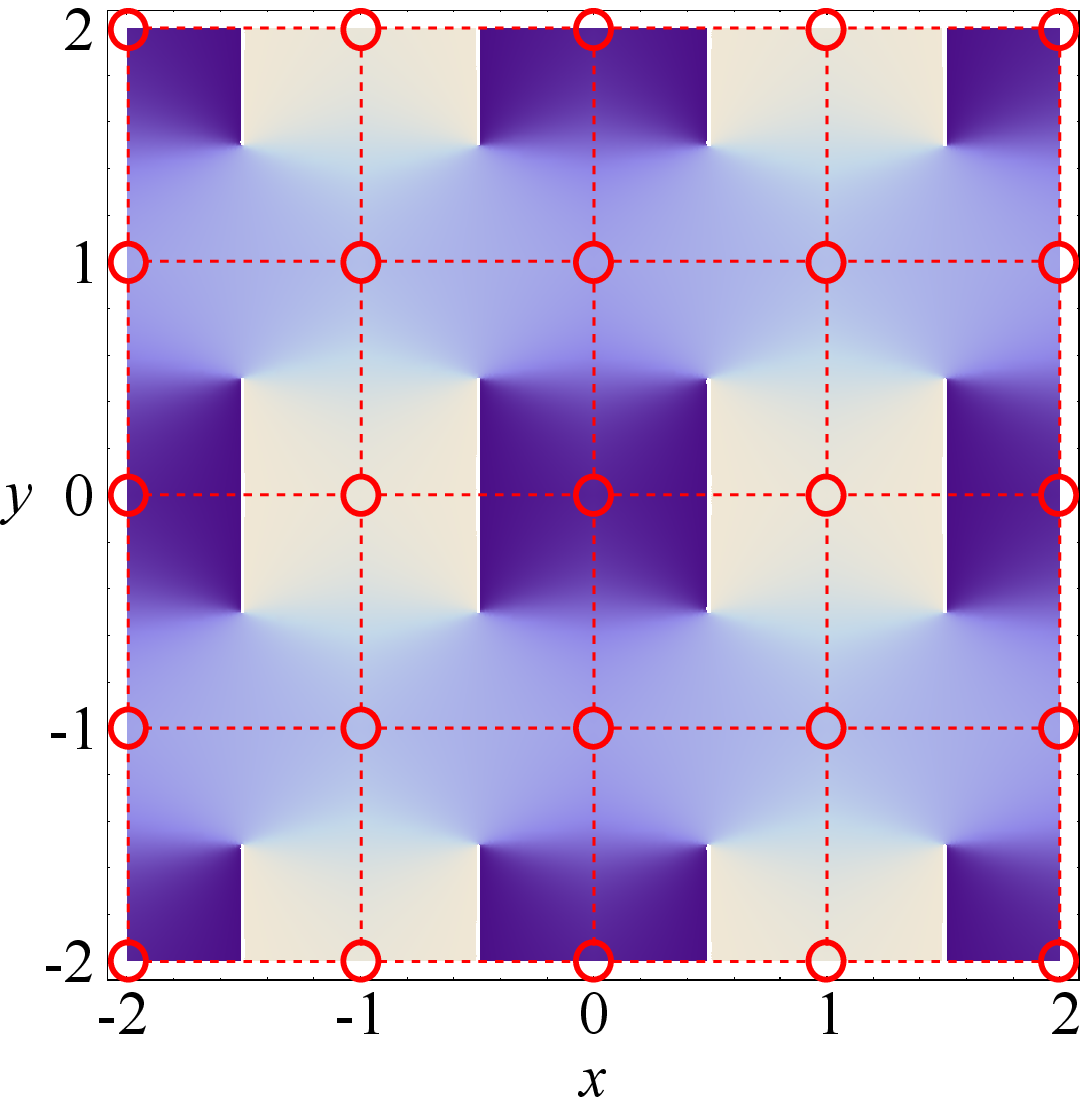}}
\caption{\small\label{boson_VL}The $S^z=1$ component $\eta_\uparrow$ of the triplet superconducting order parameter obtained by numerical mean-field minimization in the model (\ref{TIQWmodel}) with $\Delta=t$, $a=1.4$ and attractive interactions. (a) The magnitude $|\eta_\uparrow|$ in the best conventional condensate, expressed in arbitrary units. (b) The magnitude and (c) phase density plot of $\eta_\uparrow$ in the competing TR-invariant vortex lattice state with the same energy. Note that $\eta_\uparrow$ is defined only on discrete lattice sites $i=(x,y)$, where $x,y$ are integers. All other values (at real $x,y$) are obtained by linear interpolation. Similar condensed states of excitons are expected in the equivalent model with short-range repulsive interactions}
\end{figure}

The hallmark of triplet instabilities is condensation at large momenta that yields unusual forms of translation symmetry breaking. This is most appropriately studied on a lattice. As an illustration, consider a simple interacting Hamiltonian that captures the above physics:
\begin{eqnarray}\label{TIQWmodel}
H_{\textrm{QW}} &=& -t \sum_{\langle {\bf r}{\bf r}' \rangle} \left\lbrack 
   c_{{\bf r}}^{\dagger} \Bigl( e^{-i\tau^{z}\mathcal{A}_{{\bf r},{\bf r}'}}
   +\Delta_{{\bf r}{\bf r}'}^{\phantom{\dagger}}\tau^{x} \Bigr) c_{{\bf r}'}^{\phantom{\dagger}} + h.c. \right\rbrack \nonumber \\
&& - \mu \sum_{\bf r} c_{\bf r}^{\dagger}c_{\bf r}^{\phantom{\dagger}}
   + U \sum_{\bf r} (c_{\bf r}^\dagger c_{\bf r}^{\phantom{\dagger}})^2 \ .
\end{eqnarray}
This is not directly a model of a Kondo TIQW, but describes a tight-binding TIQW of electrons in two surface states, which are coupled to the static Rashba SU(2) gauge field (\ref{GF}). The notation is inherited from (\ref{LL2}). Preliminary numerical mean-field calculations of the order parameter have been carried out in the context of triplet inter-surface Cooper pairing, $U<0$. Order parameters with different spatial structures were treated as variational states aimed at minimizing the ground state energy of the corresponding Bogoliubov-de Gennes mean-field Hamiltonian. This minimization has revealed novel TR-invariant lattices of SU(2) spin-current vortices, which are the lattice version of the continuum limit vortex lattices discussed in Ref.\cite{Nikolic2014}. Vortex arrays appear to be the lowest energy configurations in some parameter regimes, despite the aggressive search for alternative orders. Fig.\ref{boson_VL} shows the structure of a typical SU(2) vortex lattice.

The above model is a simplistic rendition of the complicated hybridized 2D band-structure in Kondo TIQWs. However, it is designed to qualitatively capture the competition between interactions and Rashba spin-orbit coupling for the influence on strongly correlated phases in Kondo TIQWs. The prospect of having stable vortex lattices somewhere in the phase diagram is especially exciting because their quantum melting is expected to produce incompressible quantum liquids with likely non-Abelian fractional excitations \cite{Nikolic2012} when the number of particles per vortex is small. It has been argued through a quantum Lindeman criterion \cite{nikolic:144507, Nikolic2011a} that this first order transition preempts any second order transition out of the ordered phase, and therefore is a generic transition that can be driven by adjusting the gate voltage in a suitable heterostructure. What distinguishes the ensuing novel fractional states from the spin liquids mentioned before is their ``bosonic'' nature, an even rather than odd number of flux quanta attached to a particle. By identifying the structure of vortices in a parent vortex lattice, one can determine the type of quasiparticle fractional statistics in the vortex liquid.

\section{Conclusions}\label{secConcl}

In conclusion, we constructed microscopic slave boson models of the protected Kondo TI boundaries, and surveyed a variety of strong correlation phenomena that they exhibit. The slave boson model phenomenologically predicts the existence of several correlated and topologically enhanced phases, the analogues of which are found in the phase diagram of heavy fermion metals. In addition to the quantum critical point associated with a magnetic instability of the surface metal, strong Coulomb interactions and Kondo singlet fluctuations in the two-dimensional geometry can localize electrons in the $f$ orbitals and stabilize exotic phases such as algebraic spin liquids (charge-insulating, but spin-metallic TI crystal surface). Depending on the interface conditions, the metallic boundary of a Kondo TI can alternatively feature light electrons dominated by $d$ orbitals, whose dynamics exhibits obscure non-Fermi liquid transport properties. Kondo TI quantum wells can show even richer physics involving spin-triplet fluctuations, but their quasiparticle excitations are gapped. The condensation of spin triplets can produce unusual vortex lattice states. Quantum wells are tunable via the gate voltage in heterostructure devices, and a driven quantum melting of a vortex lattice can yield a novel fractional TI state in the quantum well.

\section{Acknowledgements}

The author is very grateful for numerous illuminating discussions with Collin Broholm and Wesley Fuhrman, whose experiment has motivated this study. This work was supported by the U.\ S.\ Department of Energy, Office of Basic Energy Sciences, Division of Materials Sciences and Engineering, under Award No.\ DE-FG02-08ER46544, and also in part by the National Science Foundation under Grant No. PHY-1066293 with hospitality of the Aspen Center for Physics.


\begin{thebibliography}{135}%
\makeatletter
\providecommand \@ifxundefined [1]{%
 \@ifx{#1\undefined}
}%
\providecommand \@ifnum [1]{%
 \ifnum #1\expandafter \@firstoftwo
 \else \expandafter \@secondoftwo
 \fi
}%
\providecommand \@ifx [1]{%
 \ifx #1\expandafter \@firstoftwo
 \else \expandafter \@secondoftwo
 \fi
}%
\providecommand \natexlab [1]{#1}%
\providecommand \enquote  [1]{``#1''}%
\providecommand \bibnamefont  [1]{#1}%
\providecommand \bibfnamefont [1]{#1}%
\providecommand \citenamefont [1]{#1}%
\providecommand \href@noop [0]{\@secondoftwo}%
\providecommand \href [0]{\begingroup \@sanitize@url \@href}%
\providecommand \@href[1]{\@@startlink{#1}\@@href}%
\providecommand \@@href[1]{\endgroup#1\@@endlink}%
\providecommand \@sanitize@url [0]{\catcode `\\12\catcode `\$12\catcode
  `\&12\catcode `\#12\catcode `\^12\catcode `\_12\catcode `\%12\relax}%
\providecommand \@@startlink[1]{}%
\providecommand \@@endlink[0]{}%
\providecommand \url  [0]{\begingroup\@sanitize@url \@url }%
\providecommand \@url [1]{\endgroup\@href {#1}{\urlprefix }}%
\providecommand \urlprefix  [0]{URL }%
\providecommand \Eprint [0]{\href }%
\providecommand \doibase [0]{http://dx.doi.org/}%
\providecommand \selectlanguage [0]{\@gobble}%
\providecommand \bibinfo  [0]{\@secondoftwo}%
\providecommand \bibfield  [0]{\@secondoftwo}%
\providecommand \translation [1]{[#1]}%
\providecommand \BibitemOpen [0]{}%
\providecommand \bibitemStop [0]{}%
\providecommand \bibitemNoStop [0]{.\EOS\space}%
\providecommand \EOS [0]{\spacefactor3000\relax}%
\providecommand \BibitemShut  [1]{\csname bibitem#1\endcsname}%
\let\auto@bib@innerbib\@empty
\bibitem [{\citenamefont {Hasan}\ and\ \citenamefont {Kane}(2010)}]{Hasan2010}%
  \BibitemOpen
  \bibfield  {author} {\bibinfo {author} {\bibfnamefont {M.~Z.}\ \bibnamefont
  {Hasan}}\ and\ \bibinfo {author} {\bibfnamefont {C.~L.}\ \bibnamefont
  {Kane}},\ }\href@noop {} {\bibfield  {journal} {\bibinfo  {journal} {Reviews
  of Modern Physics}\ }\textbf {\bibinfo {volume} {82}},\ \bibinfo {pages}
  {3045} (\bibinfo {year} {2010})}\BibitemShut {NoStop}%
\bibitem [{\citenamefont {Moore}(2010)}]{Moore2010}%
  \BibitemOpen
  \bibfield  {author} {\bibinfo {author} {\bibfnamefont {J.~E.}\ \bibnamefont
  {Moore}},\ }\href@noop {} {\bibfield  {journal} {\bibinfo  {journal}
  {Nature}\ }\textbf {\bibinfo {volume} {464}},\ \bibinfo {pages} {194}
  (\bibinfo {year} {2010})}\BibitemShut {NoStop}%
\bibitem [{\citenamefont {Qi}\ and\ \citenamefont {Zhang}(2011)}]{Qi2010a}%
  \BibitemOpen
  \bibfield  {author} {\bibinfo {author} {\bibfnamefont {X.-L.}\ \bibnamefont
  {Qi}}\ and\ \bibinfo {author} {\bibfnamefont {S.-C.}\ \bibnamefont {Zhang}},\
  }\href@noop {} {\bibfield  {journal} {\bibinfo  {journal} {Reviews of Modern
  Physics}\ }\textbf {\bibinfo {volume} {83}},\ \bibinfo {pages} {1057}
  (\bibinfo {year} {2011})}\BibitemShut {NoStop}%
\bibitem [{\citenamefont {Lin}\ \emph {et~al.}(2010{\natexlab{a}})\citenamefont
  {Lin}, \citenamefont {Wray}, \citenamefont {Xia}, \citenamefont {Xu},
  \citenamefont {Jia}, \citenamefont {Cava}, \citenamefont {Bansil},\ and\
  \citenamefont {Hasan}}]{Hsin2010}%
  \BibitemOpen
  \bibfield  {author} {\bibinfo {author} {\bibfnamefont {H.}~\bibnamefont
  {Lin}}, \bibinfo {author} {\bibfnamefont {L.~A.}\ \bibnamefont {Wray}},
  \bibinfo {author} {\bibfnamefont {Y.}~\bibnamefont {Xia}}, \bibinfo {author}
  {\bibfnamefont {S.}~\bibnamefont {Xu}}, \bibinfo {author} {\bibfnamefont
  {S.}~\bibnamefont {Jia}}, \bibinfo {author} {\bibfnamefont {R.~J.}\
  \bibnamefont {Cava}}, \bibinfo {author} {\bibfnamefont {A.}~\bibnamefont
  {Bansil}}, \ and\ \bibinfo {author} {\bibfnamefont {M.~Z.}\ \bibnamefont
  {Hasan}},\ }\href@noop {} {\bibfield  {journal} {\bibinfo  {journal} {Nature
  Materials}\ }\textbf {\bibinfo {volume} {9}},\ \bibinfo {pages} {546}
  (\bibinfo {year} {2010}{\natexlab{a}})}\BibitemShut {NoStop}%
\bibitem [{\citenamefont {Lin}\ \emph {et~al.}(2010{\natexlab{b}})\citenamefont
  {Lin}, \citenamefont {Wray}, \citenamefont {Xia}, \citenamefont {Xu},
  \citenamefont {Jia}, \citenamefont {Cava}, \citenamefont {Bansil},\ and\
  \citenamefont {Hasan}}]{Lin2010b}%
  \BibitemOpen
  \bibfield  {author} {\bibinfo {author} {\bibfnamefont {H.}~\bibnamefont
  {Lin}}, \bibinfo {author} {\bibfnamefont {L.~A.}\ \bibnamefont {Wray}},
  \bibinfo {author} {\bibfnamefont {Y.}~\bibnamefont {Xia}}, \bibinfo {author}
  {\bibfnamefont {S.-Y.}\ \bibnamefont {Xu}}, \bibinfo {author} {\bibfnamefont
  {S.}~\bibnamefont {Jia}}, \bibinfo {author} {\bibfnamefont {R.~J.}\
  \bibnamefont {Cava}}, \bibinfo {author} {\bibfnamefont {A.}~\bibnamefont
  {Bansil}}, \ and\ \bibinfo {author} {\bibfnamefont {M.~Z.}\ \bibnamefont
  {Hasan}},\ }\href@noop {} {\  (\bibinfo {year} {2010}{\natexlab{b}})},\
  \bibinfo {note} {arXiv:1004.0999v1}\BibitemShut {NoStop}%
\bibitem [{\citenamefont {Fu}\ \emph {et~al.}(2007)\citenamefont {Fu},
  \citenamefont {Kane},\ and\ \citenamefont {Mele}}]{Fu2007}%
  \BibitemOpen
  \bibfield  {author} {\bibinfo {author} {\bibfnamefont {L.}~\bibnamefont
  {Fu}}, \bibinfo {author} {\bibfnamefont {C.~L.}\ \bibnamefont {Kane}}, \ and\
  \bibinfo {author} {\bibfnamefont {E.~J.}\ \bibnamefont {Mele}},\ }\href@noop
  {} {\bibfield  {journal} {\bibinfo  {journal} {Physical Review Letters}\
  }\textbf {\bibinfo {volume} {98}},\ \bibinfo {pages} {106803} (\bibinfo
  {year} {2007})}\BibitemShut {NoStop}%
\bibitem [{\citenamefont {Moore}\ and\ \citenamefont
  {Balents}(2007)}]{Moore2007}%
  \BibitemOpen
  \bibfield  {author} {\bibinfo {author} {\bibfnamefont {J.~E.}\ \bibnamefont
  {Moore}}\ and\ \bibinfo {author} {\bibfnamefont {L.}~\bibnamefont
  {Balents}},\ }\href@noop {} {\bibfield  {journal} {\bibinfo  {journal}
  {Physical Review B}\ }\textbf {\bibinfo {volume} {75}},\ \bibinfo {pages}
  {121306(R)} (\bibinfo {year} {2007})}\BibitemShut {NoStop}%
\bibitem [{\citenamefont {Chen}\ \emph {et~al.}(2009)\citenamefont {Chen},
  \citenamefont {Analytis}, \citenamefont {Chu}, \citenamefont {Liu},
  \citenamefont {Mo}, \citenamefont {Qi}, \citenamefont {Zhang}, \citenamefont
  {Lu}, \citenamefont {Dai}, \citenamefont {Fang}, \citenamefont {Zhang},
  \citenamefont {Fisher}, \citenamefont {Hussain},\ and\ \citenamefont
  {Shen}}]{Chen2009b}%
  \BibitemOpen
  \bibfield  {author} {\bibinfo {author} {\bibfnamefont {Y.~L.}\ \bibnamefont
  {Chen}}, \bibinfo {author} {\bibfnamefont {J.~G.}\ \bibnamefont {Analytis}},
  \bibinfo {author} {\bibfnamefont {J.~H.}\ \bibnamefont {Chu}}, \bibinfo
  {author} {\bibfnamefont {K.}~\bibnamefont {Liu}}, \bibinfo {author}
  {\bibfnamefont {S.~K.}\ \bibnamefont {Mo}}, \bibinfo {author} {\bibfnamefont
  {X.~L.}\ \bibnamefont {Qi}}, \bibinfo {author} {\bibfnamefont {H.~J.}\
  \bibnamefont {Zhang}}, \bibinfo {author} {\bibfnamefont {H.}~\bibnamefont
  {Lu}}, \bibinfo {author} {\bibfnamefont {X.}~\bibnamefont {Dai}}, \bibinfo
  {author} {\bibfnamefont {Z.}~\bibnamefont {Fang}}, \bibinfo {author}
  {\bibfnamefont {S.~C.}\ \bibnamefont {Zhang}}, \bibinfo {author}
  {\bibfnamefont {I.~R.}\ \bibnamefont {Fisher}}, \bibinfo {author}
  {\bibfnamefont {Z.}~\bibnamefont {Hussain}}, \ and\ \bibinfo {author}
  {\bibfnamefont {Z.~X.}\ \bibnamefont {Shen}},\ }\href@noop {} {\bibfield
  {journal} {\bibinfo  {journal} {Science}\ }\textbf {\bibinfo {volume}
  {325}},\ \bibinfo {pages} {178} (\bibinfo {year} {2009})}\BibitemShut
  {NoStop}%
\bibitem [{\citenamefont {Abrahams}\ \emph {et~al.}(1979)\citenamefont
  {Abrahams}, \citenamefont {Anderson}, \citenamefont {Liccibpdello},\ and\
  \citenamefont {Ramakrishnan}}]{abrahams79}%
  \BibitemOpen
  \bibfield  {author} {\bibinfo {author} {\bibfnamefont {E.}~\bibnamefont
  {Abrahams}}, \bibinfo {author} {\bibfnamefont {P.~W.}\ \bibnamefont
  {Anderson}}, \bibinfo {author} {\bibfnamefont {D.~C.}\ \bibnamefont
  {Liccibpdello}}, \ and\ \bibinfo {author} {\bibfnamefont {T.~V.}\
  \bibnamefont {Ramakrishnan}},\ }\href@noop {} {\bibfield  {journal} {\bibinfo
   {journal} {Physical Review Letters}\ }\textbf {\bibinfo {volume} {42}},\
  \bibinfo {pages} {673} (\bibinfo {year} {1979})}\BibitemShut {NoStop}%
\bibitem [{\citenamefont {Menth}\ \emph {et~al.}(1969)\citenamefont {Menth},
  \citenamefont {Buehler},\ and\ \citenamefont {Geballe}}]{Menth1969}%
  \BibitemOpen
  \bibfield  {author} {\bibinfo {author} {\bibfnamefont {A.}~\bibnamefont
  {Menth}}, \bibinfo {author} {\bibfnamefont {E.}~\bibnamefont {Buehler}}, \
  and\ \bibinfo {author} {\bibfnamefont {T.~H.}\ \bibnamefont {Geballe}},\
  }\href@noop {} {\bibfield  {journal} {\bibinfo  {journal} {Physical Review
  Letters}\ }\textbf {\bibinfo {volume} {22}},\ \bibinfo {pages} {295}
  (\bibinfo {year} {1969})}\BibitemShut {NoStop}%
\bibitem [{\citenamefont {Nickerson}\ \emph {et~al.}(1971)\citenamefont
  {Nickerson}, \citenamefont {White}, \citenamefont {Lee}, \citenamefont
  {Bachmann}, \citenamefont {Geballe},\ and\ \citenamefont {{G. W.
  Hull}}}]{Nickerson1971}%
  \BibitemOpen
  \bibfield  {author} {\bibinfo {author} {\bibfnamefont {J.~C.}\ \bibnamefont
  {Nickerson}}, \bibinfo {author} {\bibfnamefont {R.~M.}\ \bibnamefont
  {White}}, \bibinfo {author} {\bibfnamefont {K.~N.}\ \bibnamefont {Lee}},
  \bibinfo {author} {\bibfnamefont {R.}~\bibnamefont {Bachmann}}, \bibinfo
  {author} {\bibfnamefont {T.~H.}\ \bibnamefont {Geballe}}, \ and\ \bibinfo
  {author} {\bibfnamefont {J.}~\bibnamefont {{G. W. Hull}}},\ }\href@noop {}
  {\bibfield  {journal} {\bibinfo  {journal} {Physical Review B}\ }\textbf
  {\bibinfo {volume} {3}},\ \bibinfo {pages} {2030} (\bibinfo {year}
  {1971})}\BibitemShut {NoStop}%
\bibitem [{\citenamefont {Hundley}\ \emph {et~al.}(1990)\citenamefont
  {Hundley}, \citenamefont {Canfield}, \citenamefont {Thompson}, \citenamefont
  {Fisk},\ and\ \citenamefont {Lawrence}}]{Hundley1990}%
  \BibitemOpen
  \bibfield  {author} {\bibinfo {author} {\bibfnamefont {M.~F.}\ \bibnamefont
  {Hundley}}, \bibinfo {author} {\bibfnamefont {P.~C.}\ \bibnamefont
  {Canfield}}, \bibinfo {author} {\bibfnamefont {J.~D.}\ \bibnamefont
  {Thompson}}, \bibinfo {author} {\bibfnamefont {Z.}~\bibnamefont {Fisk}}, \
  and\ \bibinfo {author} {\bibfnamefont {J.~M.}\ \bibnamefont {Lawrence}},\
  }\href@noop {} {\bibfield  {journal} {\bibinfo  {journal} {Physical Review
  B}\ }\textbf {\bibinfo {volume} {42}},\ \bibinfo {pages} {6842} (\bibinfo
  {year} {1990})}\BibitemShut {NoStop}%
\bibitem [{\citenamefont {Riseborough}(1992)}]{Riseborough1992}%
  \BibitemOpen
  \bibfield  {author} {\bibinfo {author} {\bibfnamefont {P.~S.}\ \bibnamefont
  {Riseborough}},\ }\href@noop {} {\bibfield  {journal} {\bibinfo  {journal}
  {Physical Review B}\ }\textbf {\bibinfo {volume} {45}},\ \bibinfo {pages}
  {13984} (\bibinfo {year} {1992})}\BibitemShut {NoStop}%
\bibitem [{\citenamefont {Alekseev}\ \emph {et~al.}(1993)\citenamefont
  {Alekseev}, \citenamefont {Mignot}, \citenamefont {Rossat-Mignod},
  \citenamefont {Lazukov},\ and\ \citenamefont {Sadikov}}]{Alekseev1993}%
  \BibitemOpen
  \bibfield  {author} {\bibinfo {author} {\bibfnamefont {P.~A.}\ \bibnamefont
  {Alekseev}}, \bibinfo {author} {\bibfnamefont {J.-M.}\ \bibnamefont
  {Mignot}}, \bibinfo {author} {\bibfnamefont {J.}~\bibnamefont
  {Rossat-Mignod}}, \bibinfo {author} {\bibfnamefont {V.~N.}\ \bibnamefont
  {Lazukov}}, \ and\ \bibinfo {author} {\bibfnamefont {I.~P.}\ \bibnamefont
  {Sadikov}},\ }\href@noop {} {\bibfield  {journal} {\bibinfo  {journal}
  {Physica B}\ }\textbf {\bibinfo {volume} {186--188}},\ \bibinfo {pages} {384}
  (\bibinfo {year} {1993})}\BibitemShut {NoStop}%
\bibitem [{\citenamefont {Nyhus}\ \emph {et~al.}(1995)\citenamefont {Nyhus},
  \citenamefont {Cooper}, \citenamefont {Fisk},\ and\ \citenamefont
  {Sarrao}}]{Nyhus1995}%
  \BibitemOpen
  \bibfield  {author} {\bibinfo {author} {\bibfnamefont {P.}~\bibnamefont
  {Nyhus}}, \bibinfo {author} {\bibfnamefont {S.~L.}\ \bibnamefont {Cooper}},
  \bibinfo {author} {\bibfnamefont {Z.}~\bibnamefont {Fisk}}, \ and\ \bibinfo
  {author} {\bibfnamefont {J.}~\bibnamefont {Sarrao}},\ }\href@noop {}
  {\bibfield  {journal} {\bibinfo  {journal} {Physical Review B}\ }\textbf
  {\bibinfo {volume} {52}},\ \bibinfo {pages} {R14308} (\bibinfo {year}
  {1995})}\BibitemShut {NoStop}%
\bibitem [{\citenamefont {Sera}\ \emph {et~al.}(1996)\citenamefont {Sera},
  \citenamefont {Kobayashi}, \citenamefont {Hiroi}, \citenamefont {Kobayashi},\
  and\ \citenamefont {Kunii}}]{Sera1996}%
  \BibitemOpen
  \bibfield  {author} {\bibinfo {author} {\bibfnamefont {M.}~\bibnamefont
  {Sera}}, \bibinfo {author} {\bibfnamefont {S.}~\bibnamefont {Kobayashi}},
  \bibinfo {author} {\bibfnamefont {M.}~\bibnamefont {Hiroi}}, \bibinfo
  {author} {\bibfnamefont {N.}~\bibnamefont {Kobayashi}}, \ and\ \bibinfo
  {author} {\bibfnamefont {S.}~\bibnamefont {Kunii}},\ }\href@noop {}
  {\bibfield  {journal} {\bibinfo  {journal} {Physical Review B}\ }\textbf
  {\bibinfo {volume} {54}},\ \bibinfo {pages} {R5207} (\bibinfo {year}
  {1996})}\BibitemShut {NoStop}%
\bibitem [{\citenamefont {Okamura}\ \emph {et~al.}(1998)\citenamefont
  {Okamura}, \citenamefont {Kimura}, \citenamefont {Shinozaki}, \citenamefont
  {Nanba}, \citenamefont {Iga}, \citenamefont {Shimizu},\ and\ \citenamefont
  {Takabatake}}]{Okamura1998}%
  \BibitemOpen
  \bibfield  {author} {\bibinfo {author} {\bibfnamefont {H.}~\bibnamefont
  {Okamura}}, \bibinfo {author} {\bibfnamefont {S.}~\bibnamefont {Kimura}},
  \bibinfo {author} {\bibfnamefont {H.}~\bibnamefont {Shinozaki}}, \bibinfo
  {author} {\bibfnamefont {T.}~\bibnamefont {Nanba}}, \bibinfo {author}
  {\bibfnamefont {F.}~\bibnamefont {Iga}}, \bibinfo {author} {\bibfnamefont
  {N.}~\bibnamefont {Shimizu}}, \ and\ \bibinfo {author} {\bibfnamefont
  {T.}~\bibnamefont {Takabatake}},\ }\href@noop {} {\bibfield  {journal}
  {\bibinfo  {journal} {Physical Review B}\ }\textbf {\bibinfo {volume} {58}},\
  \bibinfo {pages} {R7496} (\bibinfo {year} {1998})}\BibitemShut {NoStop}%
\bibitem [{\citenamefont {Bouvet}\ \emph {et~al.}(1998)\citenamefont {Bouvet},
  \citenamefont {Kasuya}, \citenamefont {Bonnet}, \citenamefont {Regnault},
  \citenamefont {Rossat-Mignod}, \citenamefont {Iga}, \citenamefont {Fak},\
  and\ \citenamefont {Severing}}]{Bouvet1998}%
  \BibitemOpen
  \bibfield  {author} {\bibinfo {author} {\bibfnamefont {A.}~\bibnamefont
  {Bouvet}}, \bibinfo {author} {\bibfnamefont {T.}~\bibnamefont {Kasuya}},
  \bibinfo {author} {\bibfnamefont {M.}~\bibnamefont {Bonnet}}, \bibinfo
  {author} {\bibfnamefont {L.~P.}\ \bibnamefont {Regnault}}, \bibinfo {author}
  {\bibfnamefont {J.}~\bibnamefont {Rossat-Mignod}}, \bibinfo {author}
  {\bibfnamefont {F.}~\bibnamefont {Iga}}, \bibinfo {author} {\bibfnamefont
  {B.}~\bibnamefont {Fak}}, \ and\ \bibinfo {author} {\bibfnamefont
  {A.}~\bibnamefont {Severing}},\ }\href@noop {} {\bibfield  {journal}
  {\bibinfo  {journal} {Journal of Physics: Condensed Matter}\ }\textbf
  {\bibinfo {volume} {10}},\ \bibinfo {pages} {5667} (\bibinfo {year}
  {1998})}\BibitemShut {NoStop}%
\bibitem [{\citenamefont {Gorshunov}\ \emph {et~al.}(1999)\citenamefont
  {Gorshunov}, \citenamefont {Sluchanko}, \citenamefont {Volkov}, \citenamefont
  {Dressel}, \citenamefont {Knebel}, \citenamefont {Loidl},\ and\ \citenamefont
  {Kunii}}]{Gorshunov1999}%
  \BibitemOpen
  \bibfield  {author} {\bibinfo {author} {\bibfnamefont {B.}~\bibnamefont
  {Gorshunov}}, \bibinfo {author} {\bibfnamefont {N.}~\bibnamefont
  {Sluchanko}}, \bibinfo {author} {\bibfnamefont {A.}~\bibnamefont {Volkov}},
  \bibinfo {author} {\bibfnamefont {M.}~\bibnamefont {Dressel}}, \bibinfo
  {author} {\bibfnamefont {G.}~\bibnamefont {Knebel}}, \bibinfo {author}
  {\bibfnamefont {A.}~\bibnamefont {Loidl}}, \ and\ \bibinfo {author}
  {\bibfnamefont {S.}~\bibnamefont {Kunii}},\ }\href@noop {} {\bibfield
  {journal} {\bibinfo  {journal} {Physical Review B}\ }\textbf {\bibinfo
  {volume} {59}},\ \bibinfo {pages} {1808} (\bibinfo {year}
  {1999})}\BibitemShut {NoStop}%
\bibitem [{\citenamefont {Riseborough}(2000)}]{Riseborough2000}%
  \BibitemOpen
  \bibfield  {author} {\bibinfo {author} {\bibfnamefont {P.~S.}\ \bibnamefont
  {Riseborough}},\ }\href@noop {} {\bibfield  {journal} {\bibinfo  {journal}
  {Annals of Physics}\ }\textbf {\bibinfo {volume} {9}},\ \bibinfo {pages}
  {813} (\bibinfo {year} {2000})}\BibitemShut {NoStop}%
\bibitem [{\citenamefont {Kim}\ \emph {et~al.}(2009)\citenamefont {Kim},
  \citenamefont {Ohsumi}, \citenamefont {Komesu}, \citenamefont {Sakai},
  \citenamefont {Morita}, \citenamefont {Takagi},\ and\ \citenamefont
  {Arima}}]{Kim2009}%
  \BibitemOpen
  \bibfield  {author} {\bibinfo {author} {\bibfnamefont {B.~J.}\ \bibnamefont
  {Kim}}, \bibinfo {author} {\bibfnamefont {H.}~\bibnamefont {Ohsumi}},
  \bibinfo {author} {\bibfnamefont {T.}~\bibnamefont {Komesu}}, \bibinfo
  {author} {\bibfnamefont {S.}~\bibnamefont {Sakai}}, \bibinfo {author}
  {\bibfnamefont {T.}~\bibnamefont {Morita}}, \bibinfo {author} {\bibfnamefont
  {H.}~\bibnamefont {Takagi}}, \ and\ \bibinfo {author} {\bibfnamefont
  {T.}~\bibnamefont {Arima}},\ }\href@noop {} {\bibfield  {journal} {\bibinfo
  {journal} {Science}\ }\textbf {\bibinfo {volume} {323}},\ \bibinfo {pages}
  {1329} (\bibinfo {year} {2009})}\BibitemShut {NoStop}%
\bibitem [{\citenamefont {Shitade}\ \emph {et~al.}(2009)\citenamefont
  {Shitade}, \citenamefont {Katsura}, \citenamefont {Kune\v{s}}, \citenamefont
  {Qi}, \citenamefont {Zhang},\ and\ \citenamefont {Nagaosa}}]{Shitade2009}%
  \BibitemOpen
  \bibfield  {author} {\bibinfo {author} {\bibfnamefont {A.}~\bibnamefont
  {Shitade}}, \bibinfo {author} {\bibfnamefont {H.}~\bibnamefont {Katsura}},
  \bibinfo {author} {\bibfnamefont {J.}~\bibnamefont {Kune\v{s}}}, \bibinfo
  {author} {\bibfnamefont {X.-L.}\ \bibnamefont {Qi}}, \bibinfo {author}
  {\bibfnamefont {S.-C.}\ \bibnamefont {Zhang}}, \ and\ \bibinfo {author}
  {\bibfnamefont {N.}~\bibnamefont {Nagaosa}},\ }\href@noop {} {\bibfield
  {journal} {\bibinfo  {journal} {Physical Review Letters}\ }\textbf {\bibinfo
  {volume} {256403}},\ \bibinfo {pages} {102} (\bibinfo {year}
  {2009})}\BibitemShut {NoStop}%
\bibitem [{\citenamefont {Pesin}\ and\ \citenamefont
  {Balents}(2010)}]{Pesin2010}%
  \BibitemOpen
  \bibfield  {author} {\bibinfo {author} {\bibfnamefont {D.}~\bibnamefont
  {Pesin}}\ and\ \bibinfo {author} {\bibfnamefont {L.}~\bibnamefont
  {Balents}},\ }\href@noop {} {\bibfield  {journal} {\bibinfo  {journal}
  {Nature Physics}\ }\textbf {\bibinfo {volume} {6}},\ \bibinfo {pages} {376}
  (\bibinfo {year} {2010})}\BibitemShut {NoStop}%
\bibitem [{\citenamefont {Wang}\ and\ \citenamefont
  {Senthil}(2011)}]{Wang2011b}%
  \BibitemOpen
  \bibfield  {author} {\bibinfo {author} {\bibfnamefont {F.}~\bibnamefont
  {Wang}}\ and\ \bibinfo {author} {\bibfnamefont {T.}~\bibnamefont {Senthil}},\
  }\href@noop {} {\bibfield  {journal} {\bibinfo  {journal} {Physical Review
  Letters}\ }\textbf {\bibinfo {volume} {106}},\ \bibinfo {pages} {136402}
  (\bibinfo {year} {2011})}\BibitemShut {NoStop}%
\bibitem [{\citenamefont {Kim}\ \emph {et~al.}(2012)\citenamefont {Kim},
  \citenamefont {Casa}, \citenamefont {Upton}, \citenamefont {Gog},
  \citenamefont {Kim}, \citenamefont {Mitchell}, \citenamefont {van
  Veenendaal}, \citenamefont {Daghofer}, \citenamefont {van~den Brink},
  \citenamefont {Khaliullin},\ and\ \citenamefont {Kim}}]{Kim2012c}%
  \BibitemOpen
  \bibfield  {author} {\bibinfo {author} {\bibfnamefont {J.}~\bibnamefont
  {Kim}}, \bibinfo {author} {\bibfnamefont {D.}~\bibnamefont {Casa}}, \bibinfo
  {author} {\bibfnamefont {M.~H.}\ \bibnamefont {Upton}}, \bibinfo {author}
  {\bibfnamefont {T.}~\bibnamefont {Gog}}, \bibinfo {author} {\bibfnamefont
  {Y.-J.}\ \bibnamefont {Kim}}, \bibinfo {author} {\bibfnamefont {J.~F.}\
  \bibnamefont {Mitchell}}, \bibinfo {author} {\bibfnamefont {M.}~\bibnamefont
  {van Veenendaal}}, \bibinfo {author} {\bibfnamefont {M.}~\bibnamefont
  {Daghofer}}, \bibinfo {author} {\bibfnamefont {J.}~\bibnamefont {van~den
  Brink}}, \bibinfo {author} {\bibfnamefont {G.}~\bibnamefont {Khaliullin}}, \
  and\ \bibinfo {author} {\bibfnamefont {B.~J.}\ \bibnamefont {Kim}},\
  }\href@noop {} {\bibfield  {journal} {\bibinfo  {journal} {Physical Review
  Letters}\ }\textbf {\bibinfo {volume} {108}},\ \bibinfo {pages} {177003}
  (\bibinfo {year} {2012})}\BibitemShut {NoStop}%
\bibitem [{\citenamefont {Gretarsson}\ \emph {et~al.}(2013)\citenamefont
  {Gretarsson}, \citenamefont {Clancy}, \citenamefont {Liu}, \citenamefont
  {Hill}, \citenamefont {Bozin}, \citenamefont {Singh}, \citenamefont {Manni},
  \citenamefont {Gegenwart}, \citenamefont {Kim}, \citenamefont {Said},
  \citenamefont {Casa}, \citenamefont {Gog}, \citenamefont {Upton},
  \citenamefont {Kim}, \citenamefont {Yu}, \citenamefont {Katukuri},
  \citenamefont {Hozoi}, \citenamefont {van~den Brink},\ and\ \citenamefont
  {Kim}}]{Gretarsson2013}%
  \BibitemOpen
  \bibfield  {author} {\bibinfo {author} {\bibfnamefont {H.}~\bibnamefont
  {Gretarsson}}, \bibinfo {author} {\bibfnamefont {J.~P.}\ \bibnamefont
  {Clancy}}, \bibinfo {author} {\bibfnamefont {X.}~\bibnamefont {Liu}},
  \bibinfo {author} {\bibfnamefont {J.~P.}\ \bibnamefont {Hill}}, \bibinfo
  {author} {\bibfnamefont {E.}~\bibnamefont {Bozin}}, \bibinfo {author}
  {\bibfnamefont {Y.}~\bibnamefont {Singh}}, \bibinfo {author} {\bibfnamefont
  {S.}~\bibnamefont {Manni}}, \bibinfo {author} {\bibfnamefont
  {P.}~\bibnamefont {Gegenwart}}, \bibinfo {author} {\bibfnamefont
  {J.}~\bibnamefont {Kim}}, \bibinfo {author} {\bibfnamefont {A.~H.}\
  \bibnamefont {Said}}, \bibinfo {author} {\bibfnamefont {D.}~\bibnamefont
  {Casa}}, \bibinfo {author} {\bibfnamefont {T.}~\bibnamefont {Gog}}, \bibinfo
  {author} {\bibfnamefont {M.~H.}\ \bibnamefont {Upton}}, \bibinfo {author}
  {\bibfnamefont {H.-S.}\ \bibnamefont {Kim}}, \bibinfo {author} {\bibfnamefont
  {J.}~\bibnamefont {Yu}}, \bibinfo {author} {\bibfnamefont {V.~M.}\
  \bibnamefont {Katukuri}}, \bibinfo {author} {\bibfnamefont {L.}~\bibnamefont
  {Hozoi}}, \bibinfo {author} {\bibfnamefont {J.}~\bibnamefont {van~den
  Brink}}, \ and\ \bibinfo {author} {\bibfnamefont {Y.-J.}\ \bibnamefont
  {Kim}},\ }\href@noop {} {\bibfield  {journal} {\bibinfo  {journal} {Physical
  Review Letters}\ }\textbf {\bibinfo {volume} {110}},\ \bibinfo {pages}
  {076402} (\bibinfo {year} {2013})}\BibitemShut {NoStop}%
\bibitem [{\citenamefont {Mazin}\ \emph {et~al.}(2013)\citenamefont {Mazin},
  \citenamefont {Manni}, \citenamefont {Foyevtsova}, \citenamefont {Jeschke},
  \citenamefont {Gegenwart},\ and\ \citenamefont {Valenti}}]{Mazin2013}%
  \BibitemOpen
  \bibfield  {author} {\bibinfo {author} {\bibfnamefont {I.~I.}\ \bibnamefont
  {Mazin}}, \bibinfo {author} {\bibfnamefont {S.}~\bibnamefont {Manni}},
  \bibinfo {author} {\bibfnamefont {K.}~\bibnamefont {Foyevtsova}}, \bibinfo
  {author} {\bibfnamefont {H.~O.}\ \bibnamefont {Jeschke}}, \bibinfo {author}
  {\bibfnamefont {P.}~\bibnamefont {Gegenwart}}, \ and\ \bibinfo {author}
  {\bibfnamefont {R.}~\bibnamefont {Valenti}},\ }\href@noop {} {\bibfield
  {journal} {\bibinfo  {journal} {Physical Review B}\ }\textbf {\bibinfo
  {volume} {88}},\ \bibinfo {pages} {035115} (\bibinfo {year}
  {2013})}\BibitemShut {NoStop}%
\bibitem [{\citenamefont {Dzero}\ \emph {et~al.}(2010)\citenamefont {Dzero},
  \citenamefont {Sun}, \citenamefont {Galitski},\ and\ \citenamefont
  {Coleman}}]{Dzero2010}%
  \BibitemOpen
  \bibfield  {author} {\bibinfo {author} {\bibfnamefont {M.}~\bibnamefont
  {Dzero}}, \bibinfo {author} {\bibfnamefont {K.}~\bibnamefont {Sun}}, \bibinfo
  {author} {\bibfnamefont {V.}~\bibnamefont {Galitski}}, \ and\ \bibinfo
  {author} {\bibfnamefont {P.}~\bibnamefont {Coleman}},\ }\href@noop {}
  {\bibfield  {journal} {\bibinfo  {journal} {Physical Review Letters}\
  }\textbf {\bibinfo {volume} {104}},\ \bibinfo {pages} {106408} (\bibinfo
  {year} {2010})}\BibitemShut {NoStop}%
\bibitem [{\citenamefont {Takimoto}(2011)}]{Takimoto2011}%
  \BibitemOpen
  \bibfield  {author} {\bibinfo {author} {\bibfnamefont {T.}~\bibnamefont
  {Takimoto}},\ }\href@noop {} {\bibfield  {journal} {\bibinfo  {journal}
  {Journal of the Physical Society of Japan}\ }\textbf {\bibinfo {volume}
  {80}},\ \bibinfo {pages} {123710} (\bibinfo {year} {2011})}\BibitemShut
  {NoStop}%
\bibitem [{\citenamefont {Dzero}\ \emph {et~al.}(2012)\citenamefont {Dzero},
  \citenamefont {Sun}, \citenamefont {Coleman},\ and\ \citenamefont
  {Galitski}}]{Dzero2012}%
  \BibitemOpen
  \bibfield  {author} {\bibinfo {author} {\bibfnamefont {M.}~\bibnamefont
  {Dzero}}, \bibinfo {author} {\bibfnamefont {K.}~\bibnamefont {Sun}}, \bibinfo
  {author} {\bibfnamefont {P.}~\bibnamefont {Coleman}}, \ and\ \bibinfo
  {author} {\bibfnamefont {V.}~\bibnamefont {Galitski}},\ }\href@noop {}
  {\bibfield  {journal} {\bibinfo  {journal} {Physical Review B}\ }\textbf
  {\bibinfo {volume} {85}},\ \bibinfo {pages} {045130} (\bibinfo {year}
  {2012})}\BibitemShut {NoStop}%
\bibitem [{\citenamefont {Alexandrov}\ \emph {et~al.}(2013)\citenamefont
  {Alexandrov}, \citenamefont {Dzero},\ and\ \citenamefont
  {Coleman}}]{Alex2013}%
  \BibitemOpen
  \bibfield  {author} {\bibinfo {author} {\bibfnamefont {V.}~\bibnamefont
  {Alexandrov}}, \bibinfo {author} {\bibfnamefont {M.}~\bibnamefont {Dzero}}, \
  and\ \bibinfo {author} {\bibfnamefont {P.}~\bibnamefont {Coleman}},\
  }\href@noop {} {\bibfield  {journal} {\bibinfo  {journal} {Physical Review
  Letters}\ }\textbf {\bibinfo {volume} {111}},\ \bibinfo {pages} {226403}
  (\bibinfo {year} {2013})}\BibitemShut {NoStop}%
\bibitem [{\citenamefont {Dzero}\ and\ \citenamefont
  {Galitski}(2013)}]{Dzero2013}%
  \BibitemOpen
  \bibfield  {author} {\bibinfo {author} {\bibfnamefont {M.}~\bibnamefont
  {Dzero}}\ and\ \bibinfo {author} {\bibfnamefont {V.}~\bibnamefont
  {Galitski}},\ }\href@noop {} {\  (\bibinfo {year} {2013})},\ \bibinfo {note}
  {arXiv:1304.7828}\BibitemShut {NoStop}%
\bibitem [{\citenamefont {Ye}\ \emph {et~al.}(2013)\citenamefont {Ye},
  \citenamefont {Allen},\ and\ \citenamefont {Sun}}]{Ye2013}%
  \BibitemOpen
  \bibfield  {author} {\bibinfo {author} {\bibfnamefont {M.}~\bibnamefont
  {Ye}}, \bibinfo {author} {\bibfnamefont {J.~W.}\ \bibnamefont {Allen}}, \
  and\ \bibinfo {author} {\bibfnamefont {K.}~\bibnamefont {Sun}},\ }\href@noop
  {} {\  (\bibinfo {year} {2013})},\ \bibinfo {note}
  {arXiv:1307.7191}\BibitemShut {NoStop}%
\bibitem [{\citenamefont {Hundley}\ \emph {et~al.}(1994)\citenamefont
  {Hundley}, \citenamefont {Thompson}, \citenamefont {Canfield},\ and\
  \citenamefont {Fisk}}]{Hundley1994}%
  \BibitemOpen
  \bibfield  {author} {\bibinfo {author} {\bibfnamefont {M.~F.}\ \bibnamefont
  {Hundley}}, \bibinfo {author} {\bibfnamefont {J.~D.}\ \bibnamefont
  {Thompson}}, \bibinfo {author} {\bibfnamefont {P.~C.}\ \bibnamefont
  {Canfield}}, \ and\ \bibinfo {author} {\bibfnamefont {Z.}~\bibnamefont
  {Fisk}},\ }\href@noop {} {\bibfield  {journal} {\bibinfo  {journal} {Physica
  B}\ }\textbf {\bibinfo {volume} {199-200}},\ \bibinfo {pages} {443} (\bibinfo
  {year} {1994})}\BibitemShut {NoStop}%
\bibitem [{\citenamefont {Kasuya}(1994)}]{Kasuya1994}%
  \BibitemOpen
  \bibfield  {author} {\bibinfo {author} {\bibfnamefont {T.}~\bibnamefont
  {Kasuya}},\ }\href@noop {} {\bibfield  {journal} {\bibinfo  {journal}
  {Europhysics Letters}\ }\textbf {\bibinfo {volume} {26}},\ \bibinfo {pages}
  {277} (\bibinfo {year} {1994})}\BibitemShut {NoStop}%
\bibitem [{\citenamefont {Fuhrman}\ \emph {et~al.}(2014)\citenamefont
  {Fuhrman}, \citenamefont {Leiner}, \citenamefont {Nikoli{\'c}}, \citenamefont
  {Granroth}, \citenamefont {Stone}, \citenamefont {Lumsden}, \citenamefont
  {DeBeer-Schmitt}, \citenamefont {Alekseev}, \citenamefont {Mignot},
  \citenamefont {Koohpayeh}, \citenamefont {Cottingham}, \citenamefont
  {Phelan}, \citenamefont {Schoop}, \citenamefont {McQueen},\ and\
  \citenamefont {Broholm}}]{Fuhrman2014}%
  \BibitemOpen
  \bibfield  {author} {\bibinfo {author} {\bibfnamefont {W.~T.}\ \bibnamefont
  {Fuhrman}}, \bibinfo {author} {\bibfnamefont {J.}~\bibnamefont {Leiner}},
  \bibinfo {author} {\bibfnamefont {P.}~\bibnamefont {Nikoli{\'c}}}, \bibinfo
  {author} {\bibfnamefont {G.~E.}\ \bibnamefont {Granroth}}, \bibinfo {author}
  {\bibfnamefont {M.~B.}\ \bibnamefont {Stone}}, \bibinfo {author}
  {\bibfnamefont {M.~D.}\ \bibnamefont {Lumsden}}, \bibinfo {author}
  {\bibfnamefont {L.}~\bibnamefont {DeBeer-Schmitt}}, \bibinfo {author}
  {\bibfnamefont {P.~A.}\ \bibnamefont {Alekseev}}, \bibinfo {author}
  {\bibfnamefont {J.-M.}\ \bibnamefont {Mignot}}, \bibinfo {author}
  {\bibfnamefont {S.~M.}\ \bibnamefont {Koohpayeh}}, \bibinfo {author}
  {\bibfnamefont {P.}~\bibnamefont {Cottingham}}, \bibinfo {author}
  {\bibfnamefont {W.~A.}\ \bibnamefont {Phelan}}, \bibinfo {author}
  {\bibfnamefont {L.}~\bibnamefont {Schoop}}, \bibinfo {author} {\bibfnamefont
  {T.~M.}\ \bibnamefont {McQueen}}, \ and\ \bibinfo {author} {\bibfnamefont
  {C.}~\bibnamefont {Broholm}},\ }\href@noop {} {\  (\bibinfo {year} {2014})},\
  \bibinfo {note} {arXiv:1407.2647}\BibitemShut {NoStop}%
\bibitem [{\citenamefont {Kasuya}\ \emph {et~al.}(1979)\citenamefont {Kasuya},
  \citenamefont {Takegahara},\ and\ \citenamefont {Fujita}}]{Kasuya1979}%
  \BibitemOpen
  \bibfield  {author} {\bibinfo {author} {\bibfnamefont {T.}~\bibnamefont
  {Kasuya}}, \bibinfo {author} {\bibfnamefont {K.}~\bibnamefont {Takegahara}},
  \ and\ \bibinfo {author} {\bibfnamefont {T.}~\bibnamefont {Fujita}},\
  }\href@noop {} {\bibfield  {journal} {\bibinfo  {journal} {Journal de
  Physique Colloques}\ }\textbf {\bibinfo {volume} {40}},\ \bibinfo {pages}
  {C5} (\bibinfo {year} {1979})}\BibitemShut {NoStop}%
\bibitem [{\citenamefont {Kebede}\ \emph {et~al.}(1996)\citenamefont {Kebede},
  \citenamefont {Aronson}, \citenamefont {Buford}, \citenamefont {Canfield},
  \citenamefont {Cho}, \citenamefont {Coles}, \citenamefont {Cooley},
  \citenamefont {Coulter}, \citenamefont {Fisk}, \citenamefont {Goettee},
  \citenamefont {Hults}, \citenamefont {Lacerda}, \citenamefont {McLendon},
  \citenamefont {Tiwari},\ and\ \citenamefont {Smith}}]{Kebede1996}%
  \BibitemOpen
  \bibfield  {author} {\bibinfo {author} {\bibfnamefont {A.}~\bibnamefont
  {Kebede}}, \bibinfo {author} {\bibfnamefont {M.~C.}\ \bibnamefont {Aronson}},
  \bibinfo {author} {\bibfnamefont {C.~M.}\ \bibnamefont {Buford}}, \bibinfo
  {author} {\bibfnamefont {P.~C.}\ \bibnamefont {Canfield}}, \bibinfo {author}
  {\bibfnamefont {J.~H.}\ \bibnamefont {Cho}}, \bibinfo {author} {\bibfnamefont
  {B.~R.}\ \bibnamefont {Coles}}, \bibinfo {author} {\bibfnamefont {J.~C.}\
  \bibnamefont {Cooley}}, \bibinfo {author} {\bibfnamefont {J.~Y.}\
  \bibnamefont {Coulter}}, \bibinfo {author} {\bibfnamefont {Z.}~\bibnamefont
  {Fisk}}, \bibinfo {author} {\bibfnamefont {J.~D.}\ \bibnamefont {Goettee}},
  \bibinfo {author} {\bibfnamefont {W.~L.}\ \bibnamefont {Hults}}, \bibinfo
  {author} {\bibfnamefont {A.}~\bibnamefont {Lacerda}}, \bibinfo {author}
  {\bibfnamefont {T.~D.}\ \bibnamefont {McLendon}}, \bibinfo {author}
  {\bibfnamefont {P.}~\bibnamefont {Tiwari}}, \ and\ \bibinfo {author}
  {\bibfnamefont {J.~L.}\ \bibnamefont {Smith}},\ }\href@noop {} {\bibfield
  {journal} {\bibinfo  {journal} {Physica B}\ }\textbf {\bibinfo {volume} {223
  \& 224}},\ \bibinfo {pages} {256} (\bibinfo {year} {1996})}\BibitemShut
  {NoStop}%
\bibitem [{\citenamefont {Flachbart}\ \emph {et~al.}(2001)\citenamefont
  {Flachbart}, \citenamefont {Gloos}, \citenamefont {Konovalova}, \citenamefont
  {Paderno}, \citenamefont {Reiffers}, \citenamefont {Samuely},\ and\
  \citenamefont {\v{S}vec}}]{Flachbart2001}%
  \BibitemOpen
  \bibfield  {author} {\bibinfo {author} {\bibfnamefont {K.}~\bibnamefont
  {Flachbart}}, \bibinfo {author} {\bibfnamefont {K.}~\bibnamefont {Gloos}},
  \bibinfo {author} {\bibfnamefont {E.}~\bibnamefont {Konovalova}}, \bibinfo
  {author} {\bibfnamefont {Y.}~\bibnamefont {Paderno}}, \bibinfo {author}
  {\bibfnamefont {M.}~\bibnamefont {Reiffers}}, \bibinfo {author}
  {\bibfnamefont {P.}~\bibnamefont {Samuely}}, \ and\ \bibinfo {author}
  {\bibfnamefont {P.}~\bibnamefont {\v{S}vec}},\ }\href@noop {} {\bibfield
  {journal} {\bibinfo  {journal} {Physical Review B}\ }\textbf {\bibinfo
  {volume} {64}},\ \bibinfo {pages} {085104} (\bibinfo {year}
  {2001})}\BibitemShut {NoStop}%
\bibitem [{\citenamefont {Barla}\ \emph {et~al.}(2005)\citenamefont {Barla},
  \citenamefont {Derr}, \citenamefont {Sanchez}, \citenamefont {Salce},
  \citenamefont {Lapertot}, \citenamefont {Doyle}, \citenamefont {R{\"u}ffer},
  \citenamefont {Lengsdorf}, \citenamefont {Abd-Elmeguid},\ and\ \citenamefont
  {Flouquet}}]{Barla2005}%
  \BibitemOpen
  \bibfield  {author} {\bibinfo {author} {\bibfnamefont {A.}~\bibnamefont
  {Barla}}, \bibinfo {author} {\bibfnamefont {J.}~\bibnamefont {Derr}},
  \bibinfo {author} {\bibfnamefont {J.~P.}\ \bibnamefont {Sanchez}}, \bibinfo
  {author} {\bibfnamefont {B.}~\bibnamefont {Salce}}, \bibinfo {author}
  {\bibfnamefont {G.}~\bibnamefont {Lapertot}}, \bibinfo {author}
  {\bibfnamefont {B.~P.}\ \bibnamefont {Doyle}}, \bibinfo {author}
  {\bibfnamefont {R.}~\bibnamefont {R{\"u}ffer}}, \bibinfo {author}
  {\bibfnamefont {R.}~\bibnamefont {Lengsdorf}}, \bibinfo {author}
  {\bibfnamefont {M.~M.}\ \bibnamefont {Abd-Elmeguid}}, \ and\ \bibinfo
  {author} {\bibfnamefont {J.}~\bibnamefont {Flouquet}},\ }\href@noop {}
  {\bibfield  {journal} {\bibinfo  {journal} {Physical Review Letters}\
  }\textbf {\bibinfo {volume} {94}},\ \bibinfo {pages} {166401} (\bibinfo
  {year} {2005})}\BibitemShut {NoStop}%
\bibitem [{\citenamefont {Derr}\ \emph {et~al.}(2008)\citenamefont {Derr},
  \citenamefont {Knebel}, \citenamefont {Braithwaite}, \citenamefont {Salce},
  \citenamefont {Flouquet}, \citenamefont {Flachbart}, \citenamefont
  {Gab{\'a}ni},\ and\ \citenamefont {Shitsevalova}}]{Derr2008}%
  \BibitemOpen
  \bibfield  {author} {\bibinfo {author} {\bibfnamefont {J.}~\bibnamefont
  {Derr}}, \bibinfo {author} {\bibfnamefont {G.}~\bibnamefont {Knebel}},
  \bibinfo {author} {\bibfnamefont {D.}~\bibnamefont {Braithwaite}}, \bibinfo
  {author} {\bibfnamefont {B.}~\bibnamefont {Salce}}, \bibinfo {author}
  {\bibfnamefont {J.}~\bibnamefont {Flouquet}}, \bibinfo {author}
  {\bibfnamefont {K.}~\bibnamefont {Flachbart}}, \bibinfo {author}
  {\bibfnamefont {S.}~\bibnamefont {Gab{\'a}ni}}, \ and\ \bibinfo {author}
  {\bibfnamefont {N.}~\bibnamefont {Shitsevalova}},\ }\href@noop {} {\bibfield
  {journal} {\bibinfo  {journal} {Physical Review B}\ }\textbf {\bibinfo
  {volume} {77}},\ \bibinfo {pages} {193107} (\bibinfo {year}
  {2008})}\BibitemShut {NoStop}%
\bibitem [{\citenamefont {Zhang}\ \emph {et~al.}(2013)\citenamefont {Zhang},
  \citenamefont {Butch}, \citenamefont {Syers}, \citenamefont {Ziemak},
  \citenamefont {Greene},\ and\ \citenamefont {Paglione}}]{Zhang2013}%
  \BibitemOpen
  \bibfield  {author} {\bibinfo {author} {\bibfnamefont {X.}~\bibnamefont
  {Zhang}}, \bibinfo {author} {\bibfnamefont {N.~P.}\ \bibnamefont {Butch}},
  \bibinfo {author} {\bibfnamefont {P.}~\bibnamefont {Syers}}, \bibinfo
  {author} {\bibfnamefont {S.}~\bibnamefont {Ziemak}}, \bibinfo {author}
  {\bibfnamefont {R.~L.}\ \bibnamefont {Greene}}, \ and\ \bibinfo {author}
  {\bibfnamefont {J.}~\bibnamefont {Paglione}},\ }\href@noop {} {\bibfield
  {journal} {\bibinfo  {journal} {Physical Review X}\ }\textbf {\bibinfo
  {volume} {3}},\ \bibinfo {pages} {011011} (\bibinfo {year}
  {2013})}\BibitemShut {NoStop}%
\bibitem [{\citenamefont {Wolgast}\ \emph {et~al.}(2013)\citenamefont
  {Wolgast}, \citenamefont {Kurdak}, \citenamefont {Sun}, \citenamefont
  {Allen}, \citenamefont {Kim},\ and\ \citenamefont {Fisk}}]{Wolgast2013}%
  \BibitemOpen
  \bibfield  {author} {\bibinfo {author} {\bibfnamefont {S.}~\bibnamefont
  {Wolgast}}, \bibinfo {author} {\bibfnamefont {C.}~\bibnamefont {Kurdak}},
  \bibinfo {author} {\bibfnamefont {K.}~\bibnamefont {Sun}}, \bibinfo {author}
  {\bibfnamefont {J.~W.}\ \bibnamefont {Allen}}, \bibinfo {author}
  {\bibfnamefont {D.-J.}\ \bibnamefont {Kim}}, \ and\ \bibinfo {author}
  {\bibfnamefont {Z.}~\bibnamefont {Fisk}},\ }\href@noop {} {\bibfield
  {journal} {\bibinfo  {journal} {Physical Review B}\ }\textbf {\bibinfo
  {volume} {88}},\ \bibinfo {pages} {180405(R)} (\bibinfo {year}
  {2013})}\BibitemShut {NoStop}%
\bibitem [{\citenamefont {Kim}\ \emph {et~al.}(2013)\citenamefont {Kim},
  \citenamefont {Thomas}, \citenamefont {Grant}, \citenamefont {Botimer},
  \citenamefont {Fisk},\ and\ \citenamefont {Xia}}]{Kim2013}%
  \BibitemOpen
  \bibfield  {author} {\bibinfo {author} {\bibfnamefont {D.~J.}\ \bibnamefont
  {Kim}}, \bibinfo {author} {\bibfnamefont {S.}~\bibnamefont {Thomas}},
  \bibinfo {author} {\bibfnamefont {T.}~\bibnamefont {Grant}}, \bibinfo
  {author} {\bibfnamefont {J.}~\bibnamefont {Botimer}}, \bibinfo {author}
  {\bibfnamefont {Z.}~\bibnamefont {Fisk}}, \ and\ \bibinfo {author}
  {\bibfnamefont {J.}~\bibnamefont {Xia}},\ }\href@noop {} {\bibfield
  {journal} {\bibinfo  {journal} {Scientific Reports}\ }\textbf {\bibinfo
  {volume} {3}},\ \bibinfo {pages} {3150} (\bibinfo {year} {2013})}\BibitemShut
  {NoStop}%
\bibitem [{\citenamefont {Kim}\ \emph {et~al.}(2014)\citenamefont {Kim},
  \citenamefont {Xia},\ and\ \citenamefont {Fisk}}]{Kim2013a}%
  \BibitemOpen
  \bibfield  {author} {\bibinfo {author} {\bibfnamefont {D.~J.}\ \bibnamefont
  {Kim}}, \bibinfo {author} {\bibfnamefont {J.}~\bibnamefont {Xia}}, \ and\
  \bibinfo {author} {\bibfnamefont {Z.}~\bibnamefont {Fisk}},\ }\href@noop {}
  {\bibfield  {journal} {\bibinfo  {journal} {Nature Materials}\ }\textbf
  {\bibinfo {volume} {13}},\ \bibinfo {pages} {466} (\bibinfo {year} {2014})},\
  \bibinfo {note} {arXiv:1307.0448}\BibitemShut {NoStop}%
\bibitem [{\citenamefont {Li}\ \emph {et~al.}(2013)\citenamefont {Li},
  \citenamefont {Xiang}, \citenamefont {Yu}, \citenamefont {Asaba},
  \citenamefont {Lawson}, \citenamefont {Cai}, \citenamefont {Tinsman},
  \citenamefont {Berkley}, \citenamefont {Wolgast}, \citenamefont {Eo},
  \citenamefont {Kim}, \citenamefont {Kurdak}, \citenamefont {Allen},
  \citenamefont {Sun}, \citenamefont {Chen}, \citenamefont {Wang},
  \citenamefont {Fisk},\ and\ \citenamefont {Li}}]{Xiang2013}%
  \BibitemOpen
  \bibfield  {author} {\bibinfo {author} {\bibfnamefont {G.}~\bibnamefont
  {Li}}, \bibinfo {author} {\bibfnamefont {Z.}~\bibnamefont {Xiang}}, \bibinfo
  {author} {\bibfnamefont {F.}~\bibnamefont {Yu}}, \bibinfo {author}
  {\bibfnamefont {T.}~\bibnamefont {Asaba}}, \bibinfo {author} {\bibfnamefont
  {B.}~\bibnamefont {Lawson}}, \bibinfo {author} {\bibfnamefont
  {P.}~\bibnamefont {Cai}}, \bibinfo {author} {\bibfnamefont {C.}~\bibnamefont
  {Tinsman}}, \bibinfo {author} {\bibfnamefont {A.}~\bibnamefont {Berkley}},
  \bibinfo {author} {\bibfnamefont {S.}~\bibnamefont {Wolgast}}, \bibinfo
  {author} {\bibfnamefont {Y.~S.}\ \bibnamefont {Eo}}, \bibinfo {author}
  {\bibfnamefont {D.-J.}\ \bibnamefont {Kim}}, \bibinfo {author} {\bibfnamefont
  {C.}~\bibnamefont {Kurdak}}, \bibinfo {author} {\bibfnamefont {J.~W.}\
  \bibnamefont {Allen}}, \bibinfo {author} {\bibfnamefont {K.}~\bibnamefont
  {Sun}}, \bibinfo {author} {\bibfnamefont {X.~H.}\ \bibnamefont {Chen}},
  \bibinfo {author} {\bibfnamefont {Y.~Y.}\ \bibnamefont {Wang}}, \bibinfo
  {author} {\bibfnamefont {Z.}~\bibnamefont {Fisk}}, \ and\ \bibinfo {author}
  {\bibfnamefont {L.}~\bibnamefont {Li}},\ }\href@noop {} {\  (\bibinfo {year}
  {2013})},\ \bibinfo {note} {arXiv:1306.5221}\BibitemShut {NoStop}%
\bibitem [{\citenamefont {R{\"o}{\ss}ler}\ \emph {et~al.}(2013)\citenamefont
  {R{\"o}{\ss}ler}, \citenamefont {Jang}, \citenamefont {Kim}, \citenamefont
  {Tjeng}, \citenamefont {Fisk}, \citenamefont {Steglich},\ and\ \citenamefont
  {Wirth}}]{Rossler2013}%
  \BibitemOpen
  \bibfield  {author} {\bibinfo {author} {\bibfnamefont {S.}~\bibnamefont
  {R{\"o}{\ss}ler}}, \bibinfo {author} {\bibfnamefont {T.-H.}\ \bibnamefont
  {Jang}}, \bibinfo {author} {\bibfnamefont {D.~J.}\ \bibnamefont {Kim}},
  \bibinfo {author} {\bibfnamefont {L.~H.}\ \bibnamefont {Tjeng}}, \bibinfo
  {author} {\bibfnamefont {Z.}~\bibnamefont {Fisk}}, \bibinfo {author}
  {\bibfnamefont {F.}~\bibnamefont {Steglich}}, \ and\ \bibinfo {author}
  {\bibfnamefont {S.}~\bibnamefont {Wirth}},\ }\href@noop {} {\  (\bibinfo
  {year} {2013})}\BibitemShut {NoStop}%
\bibitem [{\citenamefont {Denlinger}\ \emph {et~al.}(2013)\citenamefont
  {Denlinger}, \citenamefont {Allen}, \citenamefont {Kang}, \citenamefont
  {Sun}, \citenamefont {Kim}, \citenamefont {Shim}, \citenamefont {Min},
  \citenamefont {Kim},\ and\ \citenamefont {Fisk}}]{Denlinger2013}%
  \BibitemOpen
  \bibfield  {author} {\bibinfo {author} {\bibfnamefont {J.~D.}\ \bibnamefont
  {Denlinger}}, \bibinfo {author} {\bibfnamefont {J.~W.}\ \bibnamefont
  {Allen}}, \bibinfo {author} {\bibfnamefont {J.-S.}\ \bibnamefont {Kang}},
  \bibinfo {author} {\bibfnamefont {K.}~\bibnamefont {Sun}}, \bibinfo {author}
  {\bibfnamefont {J.-W.}\ \bibnamefont {Kim}}, \bibinfo {author} {\bibfnamefont
  {J.~H.}\ \bibnamefont {Shim}}, \bibinfo {author} {\bibfnamefont {B.~I.}\
  \bibnamefont {Min}}, \bibinfo {author} {\bibfnamefont {D.-J.}\ \bibnamefont
  {Kim}}, \ and\ \bibinfo {author} {\bibfnamefont {Z.}~\bibnamefont {Fisk}},\
  }\href@noop {} {\  (\bibinfo {year} {2013})},\ \bibinfo {note}
  {arXiv:1312.6637}\BibitemShut {NoStop}%
\bibitem [{\citenamefont {Thomas}\ \emph {et~al.}(2013)\citenamefont {Thomas},
  \citenamefont {Kim}, \citenamefont {Chung}, \citenamefont {Grant},
  \citenamefont {Fisk},\ and\ \citenamefont {Xia}}]{Thomas2013}%
  \BibitemOpen
  \bibfield  {author} {\bibinfo {author} {\bibfnamefont {S.}~\bibnamefont
  {Thomas}}, \bibinfo {author} {\bibfnamefont {D.~J.}\ \bibnamefont {Kim}},
  \bibinfo {author} {\bibfnamefont {S.~B.}\ \bibnamefont {Chung}}, \bibinfo
  {author} {\bibfnamefont {T.}~\bibnamefont {Grant}}, \bibinfo {author}
  {\bibfnamefont {Z.}~\bibnamefont {Fisk}}, \ and\ \bibinfo {author}
  {\bibfnamefont {J.}~\bibnamefont {Xia}},\ }\href@noop {} {\  (\bibinfo {year}
  {2013})},\ \bibinfo {note} {arXiv:1307.4133}\BibitemShut {NoStop}%
\bibitem [{\citenamefont {Neupane}\ \emph {et~al.}(2013)\citenamefont
  {Neupane}, \citenamefont {Alidoust}, \citenamefont {Xu}, \citenamefont
  {Kondo}, \citenamefont {Ishida}, \citenamefont {Kim}, \citenamefont {Liu},
  \citenamefont {Belopolski}, \citenamefont {Jo}, \citenamefont {Chang},
  \citenamefont {Jeng}, \citenamefont {Durakiewicz}, \citenamefont {Balicas},
  \citenamefont {Lin}, \citenamefont {Bansil}, \citenamefont {Shin},
  \citenamefont {Fisk},\ and\ \citenamefont {Hasan}}]{Neupane2013}%
  \BibitemOpen
  \bibfield  {author} {\bibinfo {author} {\bibfnamefont {M.}~\bibnamefont
  {Neupane}}, \bibinfo {author} {\bibfnamefont {N.}~\bibnamefont {Alidoust}},
  \bibinfo {author} {\bibfnamefont {S.}~\bibnamefont {Xu}}, \bibinfo {author}
  {\bibfnamefont {T.}~\bibnamefont {Kondo}}, \bibinfo {author} {\bibfnamefont
  {Y.}~\bibnamefont {Ishida}}, \bibinfo {author} {\bibfnamefont {D.-J.}\
  \bibnamefont {Kim}}, \bibinfo {author} {\bibfnamefont {C.}~\bibnamefont
  {Liu}}, \bibinfo {author} {\bibfnamefont {I.}~\bibnamefont {Belopolski}},
  \bibinfo {author} {\bibfnamefont {Y.}~\bibnamefont {Jo}}, \bibinfo {author}
  {\bibfnamefont {T.-R.}\ \bibnamefont {Chang}}, \bibinfo {author}
  {\bibfnamefont {H.-T.}\ \bibnamefont {Jeng}}, \bibinfo {author}
  {\bibfnamefont {T.}~\bibnamefont {Durakiewicz}}, \bibinfo {author}
  {\bibfnamefont {L.}~\bibnamefont {Balicas}}, \bibinfo {author} {\bibfnamefont
  {H.}~\bibnamefont {Lin}}, \bibinfo {author} {\bibfnamefont {A.}~\bibnamefont
  {Bansil}}, \bibinfo {author} {\bibfnamefont {S.}~\bibnamefont {Shin}},
  \bibinfo {author} {\bibfnamefont {Z.}~\bibnamefont {Fisk}}, \ and\ \bibinfo
  {author} {\bibfnamefont {M.~Z.}\ \bibnamefont {Hasan}},\ }\href@noop {}
  {\bibfield  {journal} {\bibinfo  {journal} {Nature Communications}\ ,\
  \bibinfo {pages} {2991}} (\bibinfo {year} {2013})}\BibitemShut {NoStop}%
\bibitem [{\citenamefont {Jiang}\ \emph {et~al.}(2013)\citenamefont {Jiang},
  \citenamefont {Li}, \citenamefont {Zhang}, \citenamefont {Sun}, \citenamefont
  {Chen}, \citenamefont {Ye}, \citenamefont {Xu}, \citenamefont {Ge},
  \citenamefont {Tan}, \citenamefont {Niu}, \citenamefont {Xia}, \citenamefont
  {Xie}, \citenamefont {Li}, \citenamefont {Chen}, \citenamefont {Wen},\ and\
  \citenamefont {Feng}}]{Jiang2013}%
  \BibitemOpen
  \bibfield  {author} {\bibinfo {author} {\bibfnamefont {J.}~\bibnamefont
  {Jiang}}, \bibinfo {author} {\bibfnamefont {S.}~\bibnamefont {Li}}, \bibinfo
  {author} {\bibfnamefont {T.}~\bibnamefont {Zhang}}, \bibinfo {author}
  {\bibfnamefont {Z.}~\bibnamefont {Sun}}, \bibinfo {author} {\bibfnamefont
  {F.}~\bibnamefont {Chen}}, \bibinfo {author} {\bibfnamefont {Z.~R.}\
  \bibnamefont {Ye}}, \bibinfo {author} {\bibfnamefont {M.}~\bibnamefont {Xu}},
  \bibinfo {author} {\bibfnamefont {Q.~Q.}\ \bibnamefont {Ge}}, \bibinfo
  {author} {\bibfnamefont {S.~Y.}\ \bibnamefont {Tan}}, \bibinfo {author}
  {\bibfnamefont {X.~H.}\ \bibnamefont {Niu}}, \bibinfo {author} {\bibfnamefont
  {M.}~\bibnamefont {Xia}}, \bibinfo {author} {\bibfnamefont {B.~P.}\
  \bibnamefont {Xie}}, \bibinfo {author} {\bibfnamefont {Y.~F.}\ \bibnamefont
  {Li}}, \bibinfo {author} {\bibfnamefont {X.~H.}\ \bibnamefont {Chen}},
  \bibinfo {author} {\bibfnamefont {H.~H.}\ \bibnamefont {Wen}}, \ and\
  \bibinfo {author} {\bibfnamefont {D.~L.}\ \bibnamefont {Feng}},\ }\href@noop
  {} {\bibfield  {journal} {\bibinfo  {journal} {Nature Communications}\
  }\textbf {\bibinfo {volume} {4}},\ \bibinfo {pages} {3010} (\bibinfo {year}
  {2013})}\BibitemShut {NoStop}%
\bibitem [{\citenamefont {Phelan}\ \emph {et~al.}(2014)\citenamefont {Phelan},
  \citenamefont {Koohpayeh}, \citenamefont {Cottingham}, \citenamefont
  {Freeland}, \citenamefont {Leiner}, \citenamefont {Broholm},\ and\
  \citenamefont {McQueen}}]{Phelan2014}%
  \BibitemOpen
  \bibfield  {author} {\bibinfo {author} {\bibfnamefont {W.~A.}\ \bibnamefont
  {Phelan}}, \bibinfo {author} {\bibfnamefont {S.~M.}\ \bibnamefont
  {Koohpayeh}}, \bibinfo {author} {\bibfnamefont {P.}~\bibnamefont
  {Cottingham}}, \bibinfo {author} {\bibfnamefont {J.~W.}\ \bibnamefont
  {Freeland}}, \bibinfo {author} {\bibfnamefont {J.~C.}\ \bibnamefont
  {Leiner}}, \bibinfo {author} {\bibfnamefont {C.~L.}\ \bibnamefont {Broholm}},
  \ and\ \bibinfo {author} {\bibfnamefont {T.~M.}\ \bibnamefont {McQueen}},\
  }\href@noop {} {\  (\bibinfo {year} {2014})},\ \bibinfo {note}
  {arXiv:1403.1462}\BibitemShut {NoStop}%
\bibitem [{\citenamefont {Werner}\ and\ \citenamefont
  {Assaad}(2013)}]{Werner2013}%
  \BibitemOpen
  \bibfield  {author} {\bibinfo {author} {\bibfnamefont {J.}~\bibnamefont
  {Werner}}\ and\ \bibinfo {author} {\bibfnamefont {F.~F.}\ \bibnamefont
  {Assaad}},\ }\href@noop {} {\bibfield  {journal} {\bibinfo  {journal}
  {Physical Review B}\ }\textbf {\bibinfo {volume} {88}},\ \bibinfo {pages}
  {035113} (\bibinfo {year} {2013})}\BibitemShut {NoStop}%
\bibitem [{\citenamefont {Kang}\ \emph {et~al.}(2013)\citenamefont {Kang},
  \citenamefont {Kim}, \citenamefont {Kim}, \citenamefont {Kang}, \citenamefont
  {Denlinger},\ and\ \citenamefont {Min}}]{Kang2013}%
  \BibitemOpen
  \bibfield  {author} {\bibinfo {author} {\bibfnamefont {C.-J.}\ \bibnamefont
  {Kang}}, \bibinfo {author} {\bibfnamefont {J.}~\bibnamefont {Kim}}, \bibinfo
  {author} {\bibfnamefont {K.}~\bibnamefont {Kim}}, \bibinfo {author}
  {\bibfnamefont {J.-S.}\ \bibnamefont {Kang}}, \bibinfo {author}
  {\bibfnamefont {J.~D.}\ \bibnamefont {Denlinger}}, \ and\ \bibinfo {author}
  {\bibfnamefont {B.~I.}\ \bibnamefont {Min}},\ }\href@noop {} {\  (\bibinfo
  {year} {2013})}\BibitemShut {NoStop}%
\bibitem [{\citenamefont {L{\"u}}\ and\ \citenamefont {Sheehy}(2013)}]{Lu2013}%
  \BibitemOpen
  \bibfield  {author} {\bibinfo {author} {\bibfnamefont {Q.-Q.}\ \bibnamefont
  {L{\"u}}}\ and\ \bibinfo {author} {\bibfnamefont {D.~E.}\ \bibnamefont
  {Sheehy}},\ }\href@noop {} {\bibfield  {journal} {\bibinfo  {journal}
  {Physical Review A}\ }\textbf {\bibinfo {volume} {88}},\ \bibinfo {pages}
  {043645} (\bibinfo {year} {2013})}\BibitemShut {NoStop}%
\bibitem [{\citenamefont {Si}\ \emph {et~al.}(2013)\citenamefont {Si},
  \citenamefont {Pixley}, \citenamefont {Nica}, \citenamefont {Yamamoto},
  \citenamefont {Goswami}, \citenamefont {Yu},\ and\ \citenamefont
  {Kirchner}}]{Si2013}%
  \BibitemOpen
  \bibfield  {author} {\bibinfo {author} {\bibfnamefont {Q.}~\bibnamefont
  {Si}}, \bibinfo {author} {\bibfnamefont {J.~H.}\ \bibnamefont {Pixley}},
  \bibinfo {author} {\bibfnamefont {E.}~\bibnamefont {Nica}}, \bibinfo {author}
  {\bibfnamefont {S.~J.}\ \bibnamefont {Yamamoto}}, \bibinfo {author}
  {\bibfnamefont {P.}~\bibnamefont {Goswami}}, \bibinfo {author} {\bibfnamefont
  {R.}~\bibnamefont {Yu}}, \ and\ \bibinfo {author} {\bibfnamefont
  {S.}~\bibnamefont {Kirchner}},\ }\href@noop {} {\  (\bibinfo {year}
  {2013})},\ \bibinfo {note} {arXiv:1312.0764}\BibitemShut {NoStop}%
\bibitem [{\citenamefont {Werner}\ and\ \citenamefont
  {Assaad}(2014)}]{Werner2014}%
  \BibitemOpen
  \bibfield  {author} {\bibinfo {author} {\bibfnamefont {J.}~\bibnamefont
  {Werner}}\ and\ \bibinfo {author} {\bibfnamefont {F.~F.}\ \bibnamefont
  {Assaad}},\ }\href@noop {} {\bibfield  {journal} {\bibinfo  {journal}
  {Physical Review B}\ }\textbf {\bibinfo {volume} {89}},\ \bibinfo {pages}
  {245119} (\bibinfo {year} {2014})}\BibitemShut {NoStop}%
\bibitem [{\citenamefont {Baruselli}\ and\ \citenamefont
  {Vojta}(2014)}]{Baruselli2014}%
  \BibitemOpen
  \bibfield  {author} {\bibinfo {author} {\bibfnamefont {P.~P.}\ \bibnamefont
  {Baruselli}}\ and\ \bibinfo {author} {\bibfnamefont {M.}~\bibnamefont
  {Vojta}},\ }\href@noop {} {\bibfield  {journal} {\bibinfo  {journal}
  {Physical Review B}\ }\textbf {\bibinfo {volume} {89}},\ \bibinfo {pages}
  {205105} (\bibinfo {year} {2014})}\BibitemShut {NoStop}%
\bibitem [{\citenamefont {Alexandrov}\ and\ \citenamefont
  {Coleman}(2014)}]{Alex2014}%
  \BibitemOpen
  \bibfield  {author} {\bibinfo {author} {\bibfnamefont {V.}~\bibnamefont
  {Alexandrov}}\ and\ \bibinfo {author} {\bibfnamefont {P.}~\bibnamefont
  {Coleman}},\ }\href@noop {} {\  (\bibinfo {year} {2014})},\ \bibinfo {note}
  {arXiv:1403.6819}\BibitemShut {NoStop}%
\bibitem [{\citenamefont {Legner}\ \emph {et~al.}(2014)\citenamefont {Legner},
  \citenamefont {R{\"u}egg},\ and\ \citenamefont {Sigrist}}]{Legner2014}%
  \BibitemOpen
  \bibfield  {author} {\bibinfo {author} {\bibfnamefont {M.}~\bibnamefont
  {Legner}}, \bibinfo {author} {\bibfnamefont {A.}~\bibnamefont {R{\"u}egg}}, \
  and\ \bibinfo {author} {\bibfnamefont {M.}~\bibnamefont {Sigrist}},\
  }\href@noop {} {\bibfield  {journal} {\bibinfo  {journal} {Physical Review
  B}\ }\textbf {\bibinfo {volume} {89}},\ \bibinfo {pages} {085110} (\bibinfo
  {year} {2014})}\BibitemShut {NoStop}%
\bibitem [{\citenamefont {Farberovich}\ \emph {et~al.}(1983)\citenamefont
  {Farberovich}, \citenamefont {Kurganskii}, \citenamefont {Sidorin},
  \citenamefont {Karin}, \citenamefont {Bobrikov}, \citenamefont {Nizhnikova},
  \citenamefont {Shelikh}, \citenamefont {Korsukova},\ and\ \citenamefont
  {Gurin}}]{Farberovich1983}%
  \BibitemOpen
  \bibfield  {author} {\bibinfo {author} {\bibfnamefont {O.~V.}\ \bibnamefont
  {Farberovich}}, \bibinfo {author} {\bibfnamefont {S.~I.}\ \bibnamefont
  {Kurganskii}}, \bibinfo {author} {\bibfnamefont {K.~K.}\ \bibnamefont
  {Sidorin}}, \bibinfo {author} {\bibfnamefont {M.~G.}\ \bibnamefont {Karin}},
  \bibinfo {author} {\bibfnamefont {V.~N.}\ \bibnamefont {Bobrikov}}, \bibinfo
  {author} {\bibfnamefont {G.~P.}\ \bibnamefont {Nizhnikova}}, \bibinfo
  {author} {\bibnamefont {Shelikh}}, \bibinfo {author} {\bibfnamefont {M.~M.}\
  \bibnamefont {Korsukova}}, \ and\ \bibinfo {author} {\bibfnamefont {V.~N.}\
  \bibnamefont {Gurin}},\ }\href@noop {} {\bibfield  {journal} {\bibinfo
  {journal} {Sov. Phys. Solid. State}\ }\textbf {\bibinfo {volume} {25}},\
  \bibinfo {pages} {404} (\bibinfo {year} {1983})}\BibitemShut {NoStop}%
\bibitem [{\citenamefont {Roy}\ \emph {et~al.}(2014)\citenamefont {Roy},
  \citenamefont {Sau}, \citenamefont {Dzero},\ and\ \citenamefont
  {Galitski}}]{Roy2014}%
  \BibitemOpen
  \bibfield  {author} {\bibinfo {author} {\bibfnamefont {B.}~\bibnamefont
  {Roy}}, \bibinfo {author} {\bibfnamefont {J.~D.}\ \bibnamefont {Sau}},
  \bibinfo {author} {\bibfnamefont {M.}~\bibnamefont {Dzero}}, \ and\ \bibinfo
  {author} {\bibfnamefont {V.}~\bibnamefont {Galitski}},\ }\href@noop {} {\
  (\bibinfo {year} {2014})},\ \bibinfo {note} {arXiv:1405.5526}\BibitemShut
  {NoStop}%
\bibitem [{\citenamefont {Efimkin}\ and\ \citenamefont
  {Galitski}(2014)}]{Efimkin2014}%
  \BibitemOpen
  \bibfield  {author} {\bibinfo {author} {\bibfnamefont {D.~K.}\ \bibnamefont
  {Efimkin}}\ and\ \bibinfo {author} {\bibfnamefont {V.}~\bibnamefont
  {Galitski}},\ }\href@noop {} {\  (\bibinfo {year} {2014})}\BibitemShut
  {NoStop}%
\bibitem [{\citenamefont {Steglich}\ \emph {et~al.}(1979)\citenamefont
  {Steglich}, \citenamefont {Aarts}, \citenamefont {Bredl}, \citenamefont
  {Lieke}, \citenamefont {Meschede}, \citenamefont {Franz},\ and\ \citenamefont
  {Sch{\"a}fer}}]{Steglich1979}%
  \BibitemOpen
  \bibfield  {author} {\bibinfo {author} {\bibfnamefont {F.}~\bibnamefont
  {Steglich}}, \bibinfo {author} {\bibfnamefont {J.}~\bibnamefont {Aarts}},
  \bibinfo {author} {\bibfnamefont {C.~D.}\ \bibnamefont {Bredl}}, \bibinfo
  {author} {\bibfnamefont {W.}~\bibnamefont {Lieke}}, \bibinfo {author}
  {\bibfnamefont {D.}~\bibnamefont {Meschede}}, \bibinfo {author}
  {\bibfnamefont {W.}~\bibnamefont {Franz}}, \ and\ \bibinfo {author}
  {\bibfnamefont {H.}~\bibnamefont {Sch{\"a}fer}},\ }\href@noop {} {\bibfield
  {journal} {\bibinfo  {journal} {Physical Review Letters}\ }\textbf {\bibinfo
  {volume} {43}},\ \bibinfo {pages} {1892} (\bibinfo {year}
  {1979})}\BibitemShut {NoStop}%
\bibitem [{\citenamefont {Taillefer}\ and\ \citenamefont
  {Lonzarich}(1988)}]{Taillefer1988}%
  \BibitemOpen
  \bibfield  {author} {\bibinfo {author} {\bibfnamefont {L.}~\bibnamefont
  {Taillefer}}\ and\ \bibinfo {author} {\bibfnamefont {G.~G.}\ \bibnamefont
  {Lonzarich}},\ }\href@noop {} {\bibfield  {journal} {\bibinfo  {journal}
  {Physical Review Letters}\ }\textbf {\bibinfo {volume} {60}},\ \bibinfo
  {pages} {1570} (\bibinfo {year} {1988})}\BibitemShut {NoStop}%
\bibitem [{\citenamefont {Maple}\ \emph {et~al.}(1994)\citenamefont {Maple},
  \citenamefont {Seaman}, \citenamefont {Gajewski}, \citenamefont
  {Dalichaouch}, \citenamefont {Barbetta}, \citenamefont {de~Andrade},
  \citenamefont {Mook}, \citenamefont {Lukefahr}, \citenamefont {Bernal},\ and\
  \citenamefont {MacLaughlin}}]{Maple1994}%
  \BibitemOpen
  \bibfield  {author} {\bibinfo {author} {\bibfnamefont {M.~B.}\ \bibnamefont
  {Maple}}, \bibinfo {author} {\bibfnamefont {C.~L.}\ \bibnamefont {Seaman}},
  \bibinfo {author} {\bibfnamefont {D.~A.}\ \bibnamefont {Gajewski}}, \bibinfo
  {author} {\bibfnamefont {Y.}~\bibnamefont {Dalichaouch}}, \bibinfo {author}
  {\bibfnamefont {V.~B.}\ \bibnamefont {Barbetta}}, \bibinfo {author}
  {\bibfnamefont {M.~C.}\ \bibnamefont {de~Andrade}}, \bibinfo {author}
  {\bibfnamefont {H.~A.}\ \bibnamefont {Mook}}, \bibinfo {author}
  {\bibfnamefont {H.~G.}\ \bibnamefont {Lukefahr}}, \bibinfo {author}
  {\bibfnamefont {O.}~\bibnamefont {Bernal}}, \ and\ \bibinfo {author}
  {\bibfnamefont {D.~E.}\ \bibnamefont {MacLaughlin}},\ }\href@noop {}
  {\bibfield  {journal} {\bibinfo  {journal} {Journal of Low Temperature
  Physics}\ }\textbf {\bibinfo {volume} {95}},\ \bibinfo {pages} {225}
  (\bibinfo {year} {1994})}\BibitemShut {NoStop}%
\bibitem [{\citenamefont {v.~L{\"o}hneysen}\ \emph {et~al.}(1994)\citenamefont
  {v.~L{\"o}hneysen}, \citenamefont {Pietrus}, \citenamefont {Portisch},
  \citenamefont {Schlager}, \citenamefont {Schr{\"o}der}, \citenamefont
  {Sieck},\ and\ \citenamefont {Trappmann}}]{Lohneysen1994}%
  \BibitemOpen
  \bibfield  {author} {\bibinfo {author} {\bibfnamefont {H.}~\bibnamefont
  {v.~L{\"o}hneysen}}, \bibinfo {author} {\bibfnamefont {T.}~\bibnamefont
  {Pietrus}}, \bibinfo {author} {\bibfnamefont {G.}~\bibnamefont {Portisch}},
  \bibinfo {author} {\bibfnamefont {H.~G.}\ \bibnamefont {Schlager}}, \bibinfo
  {author} {\bibfnamefont {A.}~\bibnamefont {Schr{\"o}der}}, \bibinfo {author}
  {\bibfnamefont {M.}~\bibnamefont {Sieck}}, \ and\ \bibinfo {author}
  {\bibfnamefont {T.}~\bibnamefont {Trappmann}},\ }\href@noop {} {\bibfield
  {journal} {\bibinfo  {journal} {Physical Review Letters}\ }\textbf {\bibinfo
  {volume} {72}},\ \bibinfo {pages} {3262} (\bibinfo {year}
  {1994})}\BibitemShut {NoStop}%
\bibitem [{\citenamefont {Mathur}\ \emph {et~al.}(1998)\citenamefont {Mathur},
  \citenamefont {Grosche}, \citenamefont {Julian}, \citenamefont {Walker},
  \citenamefont {Freye}, \citenamefont {Haselwimmer},\ and\ \citenamefont
  {Lonzarich}}]{Mathur1998}%
  \BibitemOpen
  \bibfield  {author} {\bibinfo {author} {\bibfnamefont {N.~D.}\ \bibnamefont
  {Mathur}}, \bibinfo {author} {\bibfnamefont {F.~M.}\ \bibnamefont {Grosche}},
  \bibinfo {author} {\bibfnamefont {S.~R.}\ \bibnamefont {Julian}}, \bibinfo
  {author} {\bibfnamefont {I.~R.}\ \bibnamefont {Walker}}, \bibinfo {author}
  {\bibfnamefont {D.~M.}\ \bibnamefont {Freye}}, \bibinfo {author}
  {\bibfnamefont {R.~K.~W.}\ \bibnamefont {Haselwimmer}}, \ and\ \bibinfo
  {author} {\bibfnamefont {G.~G.}\ \bibnamefont {Lonzarich}},\ }\href@noop {}
  {\bibfield  {journal} {\bibinfo  {journal} {Nature}\ }\textbf {\bibinfo
  {volume} {394}},\ \bibinfo {pages} {39} (\bibinfo {year} {1998})}\BibitemShut
  {NoStop}%
\bibitem [{\citenamefont {Schroder}\ \emph {et~al.}(1998)\citenamefont
  {Schroder}, \citenamefont {Aeppli}, \citenamefont {Bucher}, \citenamefont
  {Ramazashvili},\ and\ \citenamefont {Coleman}}]{Schroder1998}%
  \BibitemOpen
  \bibfield  {author} {\bibinfo {author} {\bibfnamefont {A.}~\bibnamefont
  {Schroder}}, \bibinfo {author} {\bibfnamefont {G.}~\bibnamefont {Aeppli}},
  \bibinfo {author} {\bibfnamefont {E.}~\bibnamefont {Bucher}}, \bibinfo
  {author} {\bibfnamefont {R.}~\bibnamefont {Ramazashvili}}, \ and\ \bibinfo
  {author} {\bibfnamefont {P.}~\bibnamefont {Coleman}},\ }\href@noop {}
  {\bibfield  {journal} {\bibinfo  {journal} {Phys. Rev. Lett.}\ }\textbf
  {\bibinfo {volume} {80}},\ \bibinfo {pages} {5623} (\bibinfo {year}
  {1998})}\BibitemShut {NoStop}%
\bibitem [{\citenamefont {Stockert}\ \emph {et~al.}(1998)\citenamefont
  {Stockert}, \citenamefont {v.~L{\"o}hneysen}, \citenamefont {Rosch},
  \citenamefont {Pyka},\ and\ \citenamefont {Loewenhaupt}}]{Stockert1998}%
  \BibitemOpen
  \bibfield  {author} {\bibinfo {author} {\bibfnamefont {O.}~\bibnamefont
  {Stockert}}, \bibinfo {author} {\bibfnamefont {H.}~\bibnamefont
  {v.~L{\"o}hneysen}}, \bibinfo {author} {\bibfnamefont {A.}~\bibnamefont
  {Rosch}}, \bibinfo {author} {\bibfnamefont {N.}~\bibnamefont {Pyka}}, \ and\
  \bibinfo {author} {\bibfnamefont {M.}~\bibnamefont {Loewenhaupt}},\
  }\href@noop {} {\bibfield  {journal} {\bibinfo  {journal} {Physical Review
  Letters}\ }\textbf {\bibinfo {volume} {80}},\ \bibinfo {pages} {5627}
  (\bibinfo {year} {1998})}\BibitemShut {NoStop}%
\bibitem [{\citenamefont {Stewart}(2001)}]{Stewart2001}%
  \BibitemOpen
  \bibfield  {author} {\bibinfo {author} {\bibfnamefont {G.~R.}\ \bibnamefont
  {Stewart}},\ }\href@noop {} {\bibfield  {journal} {\bibinfo  {journal}
  {Reviews of Modern Physics}\ }\textbf {\bibinfo {volume} {73}},\ \bibinfo
  {pages} {797} (\bibinfo {year} {2001})}\BibitemShut {NoStop}%
\bibitem [{\citenamefont {McCollam}\ \emph {et~al.}(2005)\citenamefont
  {McCollam}, \citenamefont {Daou}, \citenamefont {Julian}, \citenamefont
  {Bergemann}, \citenamefont {Flouquet},\ and\ \citenamefont
  {Aoki}}]{McCollam2005}%
  \BibitemOpen
  \bibfield  {author} {\bibinfo {author} {\bibfnamefont {A.}~\bibnamefont
  {McCollam}}, \bibinfo {author} {\bibfnamefont {R.}~\bibnamefont {Daou}},
  \bibinfo {author} {\bibfnamefont {S.}~\bibnamefont {Julian}}, \bibinfo
  {author} {\bibfnamefont {C.}~\bibnamefont {Bergemann}}, \bibinfo {author}
  {\bibfnamefont {J.}~\bibnamefont {Flouquet}}, \ and\ \bibinfo {author}
  {\bibfnamefont {D.}~\bibnamefont {Aoki}},\ }\href@noop {} {\bibfield
  {journal} {\bibinfo  {journal} {Physica B: Condensed Matter}\ }\textbf
  {\bibinfo {volume} {359}},\ \bibinfo {pages} {1} (\bibinfo {year}
  {2005})}\BibitemShut {NoStop}%
\bibitem [{\citenamefont {Kadowaki}\ \emph {et~al.}(2006)\citenamefont
  {Kadowaki}, \citenamefont {Tabata}, \citenamefont {Sato}, \citenamefont
  {Aso}, \citenamefont {Raymond},\ and\ \citenamefont
  {Kawarazaki}}]{Kadowaki2006}%
  \BibitemOpen
  \bibfield  {author} {\bibinfo {author} {\bibfnamefont {H.}~\bibnamefont
  {Kadowaki}}, \bibinfo {author} {\bibfnamefont {Y.}~\bibnamefont {Tabata}},
  \bibinfo {author} {\bibfnamefont {M.}~\bibnamefont {Sato}}, \bibinfo {author}
  {\bibfnamefont {N.}~\bibnamefont {Aso}}, \bibinfo {author} {\bibfnamefont
  {S.}~\bibnamefont {Raymond}}, \ and\ \bibinfo {author} {\bibfnamefont
  {S.}~\bibnamefont {Kawarazaki}},\ }\href@noop {} {\bibfield  {journal}
  {\bibinfo  {journal} {Physical Review Letters}\ }\textbf {\bibinfo {volume}
  {96}},\ \bibinfo {pages} {016401} (\bibinfo {year} {2006})}\BibitemShut
  {NoStop}%
\bibitem [{\citenamefont {Fr{\"o}hlich}\ and\ \citenamefont
  {Studer}(1992)}]{Frohlich1992}%
  \BibitemOpen
  \bibfield  {author} {\bibinfo {author} {\bibfnamefont {J.}~\bibnamefont
  {Fr{\"o}hlich}}\ and\ \bibinfo {author} {\bibfnamefont {U.~M.}\ \bibnamefont
  {Studer}},\ }\href@noop {} {\bibfield  {journal} {\bibinfo  {journal}
  {Communications in Mathematical Physics}\ }\textbf {\bibinfo {volume}
  {148}},\ \bibinfo {pages} {553} (\bibinfo {year} {1992})}\BibitemShut
  {NoStop}%
\bibitem [{\citenamefont {Nikolic}\ \emph {et~al.}(2013)\citenamefont
  {Nikolic}, \citenamefont {Duric},\ and\ \citenamefont
  {Tesanovic}}]{Nikolic2011a}%
  \BibitemOpen
  \bibfield  {author} {\bibinfo {author} {\bibfnamefont {P.}~\bibnamefont
  {Nikolic}}, \bibinfo {author} {\bibfnamefont {T.}~\bibnamefont {Duric}}, \
  and\ \bibinfo {author} {\bibfnamefont {Z.}~\bibnamefont {Tesanovic}},\
  }\href@noop {} {\bibfield  {journal} {\bibinfo  {journal} {Physical Review
  Letters}\ }\textbf {\bibinfo {volume} {110}},\ \bibinfo {pages} {176804}
  (\bibinfo {year} {2013})}\BibitemShut {NoStop}%
\bibitem [{\citenamefont {Nikolic}\ and\ \citenamefont
  {Tesanovic}(2013{\natexlab{a}})}]{Nikolic2012a}%
  \BibitemOpen
  \bibfield  {author} {\bibinfo {author} {\bibfnamefont {P.}~\bibnamefont
  {Nikolic}}\ and\ \bibinfo {author} {\bibfnamefont {Z.}~\bibnamefont
  {Tesanovic}},\ }\href@noop {} {\bibfield  {journal} {\bibinfo  {journal}
  {Physical Review B}\ }\textbf {\bibinfo {volume} {87}},\ \bibinfo {pages}
  {104514} (\bibinfo {year} {2013}{\natexlab{a}})}\BibitemShut {NoStop}%
\bibitem [{\citenamefont {Nikolic}\ and\ \citenamefont
  {Tesanovic}(2013{\natexlab{b}})}]{Nikolic2012b}%
  \BibitemOpen
  \bibfield  {author} {\bibinfo {author} {\bibfnamefont {P.}~\bibnamefont
  {Nikolic}}\ and\ \bibinfo {author} {\bibfnamefont {Z.}~\bibnamefont
  {Tesanovic}},\ }\href@noop {} {\bibfield  {journal} {\bibinfo  {journal}
  {Physical Review B}\ }\textbf {\bibinfo {volume} {87}},\ \bibinfo {pages}
  {134511} (\bibinfo {year} {2013}{\natexlab{b}})}\BibitemShut {NoStop}%
\bibitem [{\citenamefont {Nikolic}(2013)}]{Nikolic2012}%
  \BibitemOpen
  \bibfield  {author} {\bibinfo {author} {\bibfnamefont {P.}~\bibnamefont
  {Nikolic}},\ }\href@noop {} {\bibfield  {journal} {\bibinfo  {journal}
  {Physical Review B}\ }\textbf {\bibinfo {volume} {87}},\ \bibinfo {pages}
  {245120} (\bibinfo {year} {2013})}\BibitemShut {NoStop}%
\bibitem [{\citenamefont {Kasumov}\ \emph {et~al.}(1996)\citenamefont
  {Kasumov}, \citenamefont {Kononenko}, \citenamefont {Matveev}, \citenamefont
  {Borsenko}, \citenamefont {Tulin}, \citenamefont {Vdovin},\ and\
  \citenamefont {Khodos}}]{Kasumov1996}%
  \BibitemOpen
  \bibfield  {author} {\bibinfo {author} {\bibfnamefont {A.~Y.}\ \bibnamefont
  {Kasumov}}, \bibinfo {author} {\bibfnamefont {O.~V.}\ \bibnamefont
  {Kononenko}}, \bibinfo {author} {\bibfnamefont {V.~N.}\ \bibnamefont
  {Matveev}}, \bibinfo {author} {\bibfnamefont {T.~B.}\ \bibnamefont
  {Borsenko}}, \bibinfo {author} {\bibfnamefont {V.~A.}\ \bibnamefont {Tulin}},
  \bibinfo {author} {\bibfnamefont {E.~E.}\ \bibnamefont {Vdovin}}, \ and\
  \bibinfo {author} {\bibfnamefont {I.~I.}\ \bibnamefont {Khodos}},\
  }\href@noop {} {\bibfield  {journal} {\bibinfo  {journal} {Physical Review
  Letters}\ }\textbf {\bibinfo {volume} {77}},\ \bibinfo {pages} {3029}
  (\bibinfo {year} {1996})}\BibitemShut {NoStop}%
\bibitem [{\citenamefont {Koren}\ \emph {et~al.}(2011)\citenamefont {Koren},
  \citenamefont {Kirzhner}, \citenamefont {Lahoud}, \citenamefont {Chashka},\
  and\ \citenamefont {Kanigel}}]{Koren2011}%
  \BibitemOpen
  \bibfield  {author} {\bibinfo {author} {\bibfnamefont {G.}~\bibnamefont
  {Koren}}, \bibinfo {author} {\bibfnamefont {T.}~\bibnamefont {Kirzhner}},
  \bibinfo {author} {\bibfnamefont {E.}~\bibnamefont {Lahoud}}, \bibinfo
  {author} {\bibfnamefont {K.~B.}\ \bibnamefont {Chashka}}, \ and\ \bibinfo
  {author} {\bibfnamefont {A.}~\bibnamefont {Kanigel}},\ }\href@noop {}
  {\bibfield  {journal} {\bibinfo  {journal} {Physical Review B}\ }\textbf
  {\bibinfo {volume} {84}},\ \bibinfo {pages} {224521} (\bibinfo {year}
  {2011})}\BibitemShut {NoStop}%
\bibitem [{\citenamefont {Qu}\ \emph {et~al.}(2012)\citenamefont {Qu},
  \citenamefont {Yang}, \citenamefont {Shen}, \citenamefont {Ding},
  \citenamefont {Chen}, \citenamefont {Ji}, \citenamefont {Liu}, \citenamefont
  {Fan}, \citenamefont {Jing}, \citenamefont {Yang},\ and\ \citenamefont
  {Lu}}]{Qu2011}%
  \BibitemOpen
  \bibfield  {author} {\bibinfo {author} {\bibfnamefont {F.}~\bibnamefont
  {Qu}}, \bibinfo {author} {\bibfnamefont {F.}~\bibnamefont {Yang}}, \bibinfo
  {author} {\bibfnamefont {J.}~\bibnamefont {Shen}}, \bibinfo {author}
  {\bibfnamefont {Y.}~\bibnamefont {Ding}}, \bibinfo {author} {\bibfnamefont
  {J.}~\bibnamefont {Chen}}, \bibinfo {author} {\bibfnamefont {Z.}~\bibnamefont
  {Ji}}, \bibinfo {author} {\bibfnamefont {G.}~\bibnamefont {Liu}}, \bibinfo
  {author} {\bibfnamefont {J.}~\bibnamefont {Fan}}, \bibinfo {author}
  {\bibfnamefont {X.}~\bibnamefont {Jing}}, \bibinfo {author} {\bibfnamefont
  {C.}~\bibnamefont {Yang}}, \ and\ \bibinfo {author} {\bibfnamefont
  {L.}~\bibnamefont {Lu}},\ }\href@noop {} {\bibfield  {journal} {\bibinfo
  {journal} {Scientific Reports}\ }\textbf {\bibinfo {volume} {2}},\ \bibinfo
  {pages} {339} (\bibinfo {year} {2012})}\BibitemShut {NoStop}%
\bibitem [{\citenamefont {Sacepe}\ \emph {et~al.}(2011)\citenamefont {Sacepe},
  \citenamefont {Oostinga}, \citenamefont {Li}, \citenamefont {Ubaldini},
  \citenamefont {Couto}, \citenamefont {Giannini},\ and\ \citenamefont
  {Morpurgo}}]{Sacepe2011}%
  \BibitemOpen
  \bibfield  {author} {\bibinfo {author} {\bibfnamefont {B.}~\bibnamefont
  {Sacepe}}, \bibinfo {author} {\bibfnamefont {J.~B.}\ \bibnamefont
  {Oostinga}}, \bibinfo {author} {\bibfnamefont {J.}~\bibnamefont {Li}},
  \bibinfo {author} {\bibfnamefont {A.}~\bibnamefont {Ubaldini}}, \bibinfo
  {author} {\bibfnamefont {N.~J.~G.}\ \bibnamefont {Couto}}, \bibinfo {author}
  {\bibfnamefont {E.}~\bibnamefont {Giannini}}, \ and\ \bibinfo {author}
  {\bibfnamefont {A.~F.}\ \bibnamefont {Morpurgo}},\ }\href@noop {} {\bibfield
  {journal} {\bibinfo  {journal} {Nature Communications}\ }\textbf {\bibinfo
  {volume} {2}},\ \bibinfo {pages} {575} (\bibinfo {year} {2011})}\BibitemShut
  {NoStop}%
\bibitem [{\citenamefont {Yang}\ \emph {et~al.}(2011)\citenamefont {Yang},
  \citenamefont {Ding}, \citenamefont {Qu}, \citenamefont {Shen}, \citenamefont
  {Chen}, \citenamefont {Wei}, \citenamefont {Ji}, \citenamefont {Liu},
  \citenamefont {Fan}, \citenamefont {Yang}, \citenamefont {Xiang},\ and\
  \citenamefont {Lu}}]{Yang2011}%
  \BibitemOpen
  \bibfield  {author} {\bibinfo {author} {\bibfnamefont {F.}~\bibnamefont
  {Yang}}, \bibinfo {author} {\bibfnamefont {Y.}~\bibnamefont {Ding}}, \bibinfo
  {author} {\bibfnamefont {F.}~\bibnamefont {Qu}}, \bibinfo {author}
  {\bibfnamefont {J.}~\bibnamefont {Shen}}, \bibinfo {author} {\bibfnamefont
  {J.}~\bibnamefont {Chen}}, \bibinfo {author} {\bibfnamefont {Z.}~\bibnamefont
  {Wei}}, \bibinfo {author} {\bibfnamefont {Z.}~\bibnamefont {Ji}}, \bibinfo
  {author} {\bibfnamefont {G.}~\bibnamefont {Liu}}, \bibinfo {author}
  {\bibfnamefont {J.}~\bibnamefont {Fan}}, \bibinfo {author} {\bibfnamefont
  {C.}~\bibnamefont {Yang}}, \bibinfo {author} {\bibfnamefont {T.}~\bibnamefont
  {Xiang}}, \ and\ \bibinfo {author} {\bibfnamefont {L.}~\bibnamefont {Lu}},\
  }\href@noop {} {\bibfield  {journal} {\bibinfo  {journal} {Physical Review
  B}\ }\textbf {\bibinfo {volume} {85}},\ \bibinfo {pages} {104508} (\bibinfo
  {year} {2011})}\BibitemShut {NoStop}%
\bibitem [{\citenamefont {Zhang}\ \emph {et~al.}(2011)\citenamefont {Zhang},
  \citenamefont {Wang}, \citenamefont {DaSilva}, \citenamefont {Lee},
  \citenamefont {Gutierrez}, \citenamefont {Chan}, \citenamefont {Jain},\ and\
  \citenamefont {Samarth}}]{Zhang2011}%
  \BibitemOpen
  \bibfield  {author} {\bibinfo {author} {\bibfnamefont {D.}~\bibnamefont
  {Zhang}}, \bibinfo {author} {\bibfnamefont {J.}~\bibnamefont {Wang}},
  \bibinfo {author} {\bibfnamefont {A.~M.}\ \bibnamefont {DaSilva}}, \bibinfo
  {author} {\bibfnamefont {J.~S.}\ \bibnamefont {Lee}}, \bibinfo {author}
  {\bibfnamefont {H.~R.}\ \bibnamefont {Gutierrez}}, \bibinfo {author}
  {\bibfnamefont {M.~H.~W.}\ \bibnamefont {Chan}}, \bibinfo {author}
  {\bibfnamefont {J.}~\bibnamefont {Jain}}, \ and\ \bibinfo {author}
  {\bibfnamefont {N.}~\bibnamefont {Samarth}},\ }\href@noop {} {\bibfield
  {journal} {\bibinfo  {journal} {Physical Review B}\ }\textbf {\bibinfo
  {volume} {84}},\ \bibinfo {pages} {165120} (\bibinfo {year}
  {2011})}\BibitemShut {NoStop}%
\bibitem [{\citenamefont {Veldhorst}\ \emph {et~al.}(2012)\citenamefont
  {Veldhorst}, \citenamefont {Snelder}, \citenamefont {Hoek}, \citenamefont
  {Gang}, \citenamefont {Guduru}, \citenamefont {Wang}, \citenamefont
  {Zeitler}, \citenamefont {van~der Wiel}, \citenamefont {Golubov},
  \citenamefont {Hilgenkamp},\ and\ \citenamefont {Brinkman}}]{Veldhorst2012}%
  \BibitemOpen
  \bibfield  {author} {\bibinfo {author} {\bibfnamefont {M.}~\bibnamefont
  {Veldhorst}}, \bibinfo {author} {\bibfnamefont {M.}~\bibnamefont {Snelder}},
  \bibinfo {author} {\bibfnamefont {M.}~\bibnamefont {Hoek}}, \bibinfo {author}
  {\bibfnamefont {T.}~\bibnamefont {Gang}}, \bibinfo {author} {\bibfnamefont
  {V.~K.}\ \bibnamefont {Guduru}}, \bibinfo {author} {\bibfnamefont {X.~L.}\
  \bibnamefont {Wang}}, \bibinfo {author} {\bibfnamefont {U.}~\bibnamefont
  {Zeitler}}, \bibinfo {author} {\bibfnamefont {W.~G.}\ \bibnamefont {van~der
  Wiel}}, \bibinfo {author} {\bibfnamefont {A.~A.}\ \bibnamefont {Golubov}},
  \bibinfo {author} {\bibfnamefont {H.}~\bibnamefont {Hilgenkamp}}, \ and\
  \bibinfo {author} {\bibfnamefont {A.}~\bibnamefont {Brinkman}},\ }\href@noop
  {} {\bibfield  {journal} {\bibinfo  {journal} {Nature Materials}\ }\textbf
  {\bibinfo {volume} {11}},\ \bibinfo {pages} {417} (\bibinfo {year}
  {2012})}\BibitemShut {NoStop}%
\bibitem [{\citenamefont {Wang}\ \emph
  {et~al.}(2012{\natexlab{a}})\citenamefont {Wang}, \citenamefont {Chang},
  \citenamefont {Li}, \citenamefont {He}, \citenamefont {Zhang}, \citenamefont
  {Singh}, \citenamefont {Ma}, \citenamefont {Samarth}, \citenamefont {Xie},
  \citenamefont {Xue},\ and\ \citenamefont {Chan}}]{Wang2012}%
  \BibitemOpen
  \bibfield  {author} {\bibinfo {author} {\bibfnamefont {J.}~\bibnamefont
  {Wang}}, \bibinfo {author} {\bibfnamefont {C.-Z.}\ \bibnamefont {Chang}},
  \bibinfo {author} {\bibfnamefont {H.}~\bibnamefont {Li}}, \bibinfo {author}
  {\bibfnamefont {K.}~\bibnamefont {He}}, \bibinfo {author} {\bibfnamefont
  {D.}~\bibnamefont {Zhang}}, \bibinfo {author} {\bibfnamefont
  {M.}~\bibnamefont {Singh}}, \bibinfo {author} {\bibfnamefont {X.-C.}\
  \bibnamefont {Ma}}, \bibinfo {author} {\bibfnamefont {N.}~\bibnamefont
  {Samarth}}, \bibinfo {author} {\bibfnamefont {M.}~\bibnamefont {Xie}},
  \bibinfo {author} {\bibfnamefont {Q.-K.}\ \bibnamefont {Xue}}, \ and\
  \bibinfo {author} {\bibfnamefont {M.~H.~W.}\ \bibnamefont {Chan}},\
  }\href@noop {} {\bibfield  {journal} {\bibinfo  {journal} {Physical Review
  B}\ }\textbf {\bibinfo {volume} {85}},\ \bibinfo {pages} {045415} (\bibinfo
  {year} {2012}{\natexlab{a}})}\BibitemShut {NoStop}%
\bibitem [{\citenamefont {Wang}\ \emph
  {et~al.}(2012{\natexlab{b}})\citenamefont {Wang}, \citenamefont {Liu},
  \citenamefont {Xu}, \citenamefont {Yang}, \citenamefont {Miao}, \citenamefont
  {Yao}, \citenamefont {Gao}, \citenamefont {{Chenyi Shen2}}, \citenamefont
  {Ma}, \citenamefont {Chen}, \citenamefont {Xu}, \citenamefont {Liu},
  \citenamefont {Zhang}, \citenamefont {Qian}, \citenamefont {Jia},\ and\
  \citenamefont {Xue}}]{Wang2012a}%
  \BibitemOpen
  \bibfield  {author} {\bibinfo {author} {\bibfnamefont {M.-X.}\ \bibnamefont
  {Wang}}, \bibinfo {author} {\bibfnamefont {C.}~\bibnamefont {Liu}}, \bibinfo
  {author} {\bibfnamefont {J.-P.}\ \bibnamefont {Xu}}, \bibinfo {author}
  {\bibfnamefont {F.}~\bibnamefont {Yang}}, \bibinfo {author} {\bibfnamefont
  {L.}~\bibnamefont {Miao}}, \bibinfo {author} {\bibfnamefont {M.-Y.}\
  \bibnamefont {Yao}}, \bibinfo {author} {\bibfnamefont {C.~L.}\ \bibnamefont
  {Gao}}, \bibinfo {author} {\bibnamefont {{Chenyi Shen2}}}, \bibinfo {author}
  {\bibfnamefont {X.}~\bibnamefont {Ma}}, \bibinfo {author} {\bibfnamefont
  {X.}~\bibnamefont {Chen}}, \bibinfo {author} {\bibfnamefont {Z.-A.}\
  \bibnamefont {Xu}}, \bibinfo {author} {\bibfnamefont {Y.}~\bibnamefont
  {Liu}}, \bibinfo {author} {\bibfnamefont {S.-C.}\ \bibnamefont {Zhang}},
  \bibinfo {author} {\bibfnamefont {D.}~\bibnamefont {Qian}}, \bibinfo {author}
  {\bibfnamefont {J.-F.}\ \bibnamefont {Jia}}, \ and\ \bibinfo {author}
  {\bibfnamefont {Q.-K.}\ \bibnamefont {Xue}},\ }\href@noop {} {\bibfield
  {journal} {\bibinfo  {journal} {Science}\ }\textbf {\bibinfo {volume}
  {336}},\ \bibinfo {pages} {52} (\bibinfo {year}
  {2012}{\natexlab{b}})}\BibitemShut {NoStop}%
\bibitem [{\citenamefont {Williams}\ \emph {et~al.}(2012)\citenamefont
  {Williams}, \citenamefont {Bestwick}, \citenamefont {Gallagher},
  \citenamefont {Hong}, \citenamefont {Cui}, \citenamefont {Bleich},
  \citenamefont {Analytis}, \citenamefont {Fisher},\ and\ \citenamefont
  {Goldhaber-Gordon}}]{Williams2012}%
  \BibitemOpen
  \bibfield  {author} {\bibinfo {author} {\bibfnamefont {J.~R.}\ \bibnamefont
  {Williams}}, \bibinfo {author} {\bibfnamefont {A.~J.}\ \bibnamefont
  {Bestwick}}, \bibinfo {author} {\bibfnamefont {P.}~\bibnamefont {Gallagher}},
  \bibinfo {author} {\bibfnamefont {S.~S.}\ \bibnamefont {Hong}}, \bibinfo
  {author} {\bibfnamefont {Y.}~\bibnamefont {Cui}}, \bibinfo {author}
  {\bibfnamefont {A.~S.}\ \bibnamefont {Bleich}}, \bibinfo {author}
  {\bibfnamefont {J.~G.}\ \bibnamefont {Analytis}}, \bibinfo {author}
  {\bibfnamefont {I.~R.}\ \bibnamefont {Fisher}}, \ and\ \bibinfo {author}
  {\bibfnamefont {D.}~\bibnamefont {Goldhaber-Gordon}},\ }\href@noop {}
  {\bibfield  {journal} {\bibinfo  {journal} {Physical Review Letters}\
  }\textbf {\bibinfo {volume} {109}},\ \bibinfo {pages} {056803} (\bibinfo
  {year} {2012})}\BibitemShut {NoStop}%
\bibitem [{\citenamefont {Yang}\ \emph {et~al.}(2012)\citenamefont {Yang},
  \citenamefont {Qu}, \citenamefont {Shen}, \citenamefont {Ding}, \citenamefont
  {Chen}, \citenamefont {Ji}, \citenamefont {Liu}, \citenamefont {Fan},
  \citenamefont {Yang}, \citenamefont {Fu},\ and\ \citenamefont
  {Lu}}]{Yang2012}%
  \BibitemOpen
  \bibfield  {author} {\bibinfo {author} {\bibfnamefont {F.}~\bibnamefont
  {Yang}}, \bibinfo {author} {\bibfnamefont {F.}~\bibnamefont {Qu}}, \bibinfo
  {author} {\bibfnamefont {J.}~\bibnamefont {Shen}}, \bibinfo {author}
  {\bibfnamefont {Y.}~\bibnamefont {Ding}}, \bibinfo {author} {\bibfnamefont
  {J.}~\bibnamefont {Chen}}, \bibinfo {author} {\bibfnamefont {Z.}~\bibnamefont
  {Ji}}, \bibinfo {author} {\bibfnamefont {G.}~\bibnamefont {Liu}}, \bibinfo
  {author} {\bibfnamefont {J.}~\bibnamefont {Fan}}, \bibinfo {author}
  {\bibfnamefont {C.}~\bibnamefont {Yang}}, \bibinfo {author} {\bibfnamefont
  {L.}~\bibnamefont {Fu}}, \ and\ \bibinfo {author} {\bibfnamefont
  {L.}~\bibnamefont {Lu}},\ }\href@noop {} {\bibfield  {journal} {\bibinfo
  {journal} {Physical Review B}\ }\textbf {\bibinfo {volume} {86}},\ \bibinfo
  {pages} {134504} (\bibinfo {year} {2012})}\BibitemShut {NoStop}%
\bibitem [{\citenamefont {Fu}\ and\ \citenamefont {Kane}(2008)}]{Fu2008}%
  \BibitemOpen
  \bibfield  {author} {\bibinfo {author} {\bibfnamefont {L.}~\bibnamefont
  {Fu}}\ and\ \bibinfo {author} {\bibfnamefont {C.~L.}\ \bibnamefont {Kane}},\
  }\href@noop {} {\bibfield  {journal} {\bibinfo  {journal} {Physical Review
  Letters}\ }\textbf {\bibinfo {volume} {100}},\ \bibinfo {pages} {096407}
  (\bibinfo {year} {2008})}\BibitemShut {NoStop}%
\bibitem [{\citenamefont {Kitaev}(2003)}]{Kitaev2003}%
  \BibitemOpen
  \bibfield  {author} {\bibinfo {author} {\bibfnamefont {A.}~\bibnamefont
  {Kitaev}},\ }\href@noop {} {\bibfield  {journal} {\bibinfo  {journal} {Annals
  of Physics}\ }\textbf {\bibinfo {volume} {303}},\ \bibinfo {pages} {2}
  (\bibinfo {year} {2003})}\BibitemShut {NoStop}%
\bibitem [{\citenamefont {Nayak}\ \emph {et~al.}(2008)\citenamefont {Nayak},
  \citenamefont {Simon}, \citenamefont {Stern}, \citenamefont {Freedman},\ and\
  \citenamefont {Sarma}}]{Nayak2008}%
  \BibitemOpen
  \bibfield  {author} {\bibinfo {author} {\bibfnamefont {C.}~\bibnamefont
  {Nayak}}, \bibinfo {author} {\bibfnamefont {S.~H.}\ \bibnamefont {Simon}},
  \bibinfo {author} {\bibfnamefont {A.}~\bibnamefont {Stern}}, \bibinfo
  {author} {\bibfnamefont {M.}~\bibnamefont {Freedman}}, \ and\ \bibinfo
  {author} {\bibfnamefont {S.~D.}\ \bibnamefont {Sarma}},\ }\href@noop {}
  {\bibfield  {journal} {\bibinfo  {journal} {Reviews of Modern Physics}\
  }\textbf {\bibinfo {volume} {80}},\ \bibinfo {pages} {1083} (\bibinfo {year}
  {2008})}\BibitemShut {NoStop}%
\bibitem [{\citenamefont {Doniach}(1977)}]{Doniach1977}%
  \BibitemOpen
  \bibfield  {author} {\bibinfo {author} {\bibfnamefont {S.}~\bibnamefont
  {Doniach}},\ }\href@noop {} {\bibfield  {journal} {\bibinfo  {journal}
  {Physica B+C}\ }\textbf {\bibinfo {volume} {91}},\ \bibinfo {pages} {231}
  (\bibinfo {year} {1977})}\BibitemShut {NoStop}%
\bibitem [{\citenamefont {Nozi{\`e}res}(1985)}]{Nozieres1985a}%
  \BibitemOpen
  \bibfield  {author} {\bibinfo {author} {\bibfnamefont {P.}~\bibnamefont
  {Nozi{\`e}res}},\ }\href@noop {} {\bibfield  {journal} {\bibinfo  {journal}
  {Annals of Physics (Paris)}\ }\textbf {\bibinfo {volume} {10}},\ \bibinfo
  {pages} {19} (\bibinfo {year} {1985})}\BibitemShut {NoStop}%
\bibitem [{\citenamefont {Varma}\ and\ \citenamefont
  {Yafet}(1976)}]{Varma1976a}%
  \BibitemOpen
  \bibfield  {author} {\bibinfo {author} {\bibfnamefont {C.~M.}\ \bibnamefont
  {Varma}}\ and\ \bibinfo {author} {\bibfnamefont {Y.}~\bibnamefont {Yafet}},\
  }\href@noop {} {\bibfield  {journal} {\bibinfo  {journal} {Physical Review
  B}\ }\textbf {\bibinfo {volume} {13}},\ \bibinfo {pages} {2950} (\bibinfo
  {year} {1976})}\BibitemShut {NoStop}%
\bibitem [{\citenamefont {Varma}(1985)}]{Varma1985}%
  \BibitemOpen
  \bibfield  {author} {\bibinfo {author} {\bibfnamefont {C.~M.}\ \bibnamefont
  {Varma}},\ }\href@noop {} {\bibfield  {journal} {\bibinfo  {journal}
  {Physical Review Letters}\ }\textbf {\bibinfo {volume} {55}},\ \bibinfo
  {pages} {2723} (\bibinfo {year} {1985})}\BibitemShut {NoStop}%
\bibitem [{\citenamefont {Varma}\ \emph {et~al.}(1986)\citenamefont {Varma},
  \citenamefont {Weber},\ and\ \citenamefont {Randall}}]{Varma1986}%
  \BibitemOpen
  \bibfield  {author} {\bibinfo {author} {\bibfnamefont {C.~M.}\ \bibnamefont
  {Varma}}, \bibinfo {author} {\bibfnamefont {W.}~\bibnamefont {Weber}}, \ and\
  \bibinfo {author} {\bibfnamefont {L.~J.}\ \bibnamefont {Randall}},\
  }\href@noop {} {\bibfield  {journal} {\bibinfo  {journal} {Physical Review
  B}\ }\textbf {\bibinfo {volume} {33}},\ \bibinfo {pages} {1015} (\bibinfo
  {year} {1986})}\BibitemShut {NoStop}%
\bibitem [{\citenamefont {Auerbach}\ and\ \citenamefont
  {Levin}(1986)}]{Auerbach1986}%
  \BibitemOpen
  \bibfield  {author} {\bibinfo {author} {\bibfnamefont {A.}~\bibnamefont
  {Auerbach}}\ and\ \bibinfo {author} {\bibfnamefont {K.}~\bibnamefont
  {Levin}},\ }\href@noop {} {\bibfield  {journal} {\bibinfo  {journal}
  {Physical Review Letters}\ }\textbf {\bibinfo {volume} {57}},\ \bibinfo
  {pages} {877} (\bibinfo {year} {1986})}\BibitemShut {NoStop}%
\bibitem [{\citenamefont {Millis}\ and\ \citenamefont
  {Lee}(1987)}]{Millis1987a}%
  \BibitemOpen
  \bibfield  {author} {\bibinfo {author} {\bibfnamefont {A.~J.}\ \bibnamefont
  {Millis}}\ and\ \bibinfo {author} {\bibfnamefont {P.~A.}\ \bibnamefont
  {Lee}},\ }\href@noop {} {\bibfield  {journal} {\bibinfo  {journal} {Physical
  Review B}\ }\textbf {\bibinfo {volume} {35}},\ \bibinfo {pages} {3394}
  (\bibinfo {year} {1987})}\BibitemShut {NoStop}%
\bibitem [{\citenamefont {Hertz}(1976)}]{Hertz1976}%
  \BibitemOpen
  \bibfield  {author} {\bibinfo {author} {\bibfnamefont {J.~A.}\ \bibnamefont
  {Hertz}},\ }\href@noop {} {\bibfield  {journal} {\bibinfo  {journal}
  {Physical Review B}\ }\textbf {\bibinfo {volume} {14}},\ \bibinfo {pages}
  {1165} (\bibinfo {year} {1976})}\BibitemShut {NoStop}%
\bibitem [{\citenamefont {Moriya}(1985)}]{Moriya1985}%
  \BibitemOpen
  \bibfield  {author} {\bibinfo {author} {\bibfnamefont {T.}~\bibnamefont
  {Moriya}},\ }\href@noop {} {\emph {\bibinfo {title} {{Spin Fluctuations in
  Itinerant Electron Magnetism}}}},\ {Springer Series in Solid-State Sciences}\
  (\bibinfo  {publisher} {Springer},\ \bibinfo {year} {1985})\BibitemShut
  {NoStop}%
\bibitem [{\citenamefont {Millis}(1993)}]{Millis1993}%
  \BibitemOpen
  \bibfield  {author} {\bibinfo {author} {\bibfnamefont {A.~J.}\ \bibnamefont
  {Millis}},\ }\href@noop {} {\bibfield  {journal} {\bibinfo  {journal}
  {Physical Review B}\ }\textbf {\bibinfo {volume} {48}},\ \bibinfo {pages}
  {7183} (\bibinfo {year} {1993})}\BibitemShut {NoStop}%
\bibitem [{\citenamefont {Maple}\ \emph {et~al.}(1995)\citenamefont {Maple},
  \citenamefont {de~Andrade}, \citenamefont {Herrmann}, \citenamefont
  {Dalichaouch}, \citenamefont {Gajewski}, \citenamefont {Seaman},
  \citenamefont {Chau}, \citenamefont {Movshovich}, \citenamefont {Aronson},\
  and\ \citenamefont {Osborn}}]{Maple1995}%
  \BibitemOpen
  \bibfield  {author} {\bibinfo {author} {\bibfnamefont {M.~B.}\ \bibnamefont
  {Maple}}, \bibinfo {author} {\bibfnamefont {M.~C.}\ \bibnamefont
  {de~Andrade}}, \bibinfo {author} {\bibfnamefont {J.}~\bibnamefont
  {Herrmann}}, \bibinfo {author} {\bibfnamefont {Y.}~\bibnamefont
  {Dalichaouch}}, \bibinfo {author} {\bibfnamefont {D.~A.}\ \bibnamefont
  {Gajewski}}, \bibinfo {author} {\bibfnamefont {C.~L.}\ \bibnamefont
  {Seaman}}, \bibinfo {author} {\bibfnamefont {R.}~\bibnamefont {Chau}},
  \bibinfo {author} {\bibfnamefont {R.}~\bibnamefont {Movshovich}}, \bibinfo
  {author} {\bibfnamefont {M.~C.}\ \bibnamefont {Aronson}}, \ and\ \bibinfo
  {author} {\bibfnamefont {R.}~\bibnamefont {Osborn}},\ }\href@noop {}
  {\bibfield  {journal} {\bibinfo  {journal} {Journal of Low Temperature
  Physics}\ }\textbf {\bibinfo {volume} {99}},\ \bibinfo {pages} {223}
  (\bibinfo {year} {1995})}\BibitemShut {NoStop}%
\bibitem [{\citenamefont {Aronson}\ \emph {et~al.}(1995)\citenamefont
  {Aronson}, \citenamefont {Osborn}, \citenamefont {Robinson}, \citenamefont
  {Lynn}, \citenamefont {Chau}, \citenamefont {Seaman},\ and\ \citenamefont
  {Maple}}]{Aronson1995}%
  \BibitemOpen
  \bibfield  {author} {\bibinfo {author} {\bibfnamefont {M.~C.}\ \bibnamefont
  {Aronson}}, \bibinfo {author} {\bibfnamefont {R.}~\bibnamefont {Osborn}},
  \bibinfo {author} {\bibfnamefont {R.~A.}\ \bibnamefont {Robinson}}, \bibinfo
  {author} {\bibfnamefont {J.~W.}\ \bibnamefont {Lynn}}, \bibinfo {author}
  {\bibfnamefont {R.}~\bibnamefont {Chau}}, \bibinfo {author} {\bibfnamefont
  {C.~L.}\ \bibnamefont {Seaman}}, \ and\ \bibinfo {author} {\bibfnamefont
  {M.~B.}\ \bibnamefont {Maple}},\ }\href@noop {} {\bibfield  {journal}
  {\bibinfo  {journal} {Physical Review Letters}\ }\textbf {\bibinfo {volume}
  {75}},\ \bibinfo {pages} {725} (\bibinfo {year} {1995})}\BibitemShut
  {NoStop}%
\bibitem [{\citenamefont {Steglich}\ \emph {et~al.}(1997)\citenamefont
  {Steglich}, \citenamefont {Gegenwart}, \citenamefont {Helfrich},
  \citenamefont {Langhammer}, \citenamefont {Hellmann}, \citenamefont
  {Donnevert}, \citenamefont {Geibel}, \citenamefont {Lang}, \citenamefont
  {Sparn}, \citenamefont {Assmus}, \citenamefont {Stewart},\ and\ \citenamefont
  {Ochiai}}]{Steglich1997}%
  \BibitemOpen
  \bibfield  {author} {\bibinfo {author} {\bibfnamefont {F.}~\bibnamefont
  {Steglich}}, \bibinfo {author} {\bibfnamefont {P.}~\bibnamefont {Gegenwart}},
  \bibinfo {author} {\bibfnamefont {R.}~\bibnamefont {Helfrich}}, \bibinfo
  {author} {\bibfnamefont {C.}~\bibnamefont {Langhammer}}, \bibinfo {author}
  {\bibfnamefont {P.}~\bibnamefont {Hellmann}}, \bibinfo {author}
  {\bibfnamefont {L.}~\bibnamefont {Donnevert}}, \bibinfo {author}
  {\bibfnamefont {C.}~\bibnamefont {Geibel}}, \bibinfo {author} {\bibfnamefont
  {M.}~\bibnamefont {Lang}}, \bibinfo {author} {\bibfnamefont {G.}~\bibnamefont
  {Sparn}}, \bibinfo {author} {\bibfnamefont {W.}~\bibnamefont {Assmus}},
  \bibinfo {author} {\bibfnamefont {G.}~\bibnamefont {Stewart}}, \ and\
  \bibinfo {author} {\bibfnamefont {A.}~\bibnamefont {Ochiai}},\ }\href@noop {}
  {\bibfield  {journal} {\bibinfo  {journal} {Zeitschrift f{\"u}r Physik B}\
  }\textbf {\bibinfo {volume} {103}},\ \bibinfo {pages} {235} (\bibinfo {year}
  {1997})}\BibitemShut {NoStop}%
\bibitem [{\citenamefont {Aoki}\ \emph {et~al.}(1997)\citenamefont {Aoki},
  \citenamefont {Urakawa}, \citenamefont {Sugawara}, \citenamefont {Sato},
  \citenamefont {Fukuhara},\ and\ \citenamefont {Maezawa}}]{Aoki1997}%
  \BibitemOpen
  \bibfield  {author} {\bibinfo {author} {\bibfnamefont {Y.}~\bibnamefont
  {Aoki}}, \bibinfo {author} {\bibfnamefont {J.}~\bibnamefont {Urakawa}},
  \bibinfo {author} {\bibfnamefont {H.}~\bibnamefont {Sugawara}}, \bibinfo
  {author} {\bibfnamefont {H.}~\bibnamefont {Sato}}, \bibinfo {author}
  {\bibfnamefont {T.}~\bibnamefont {Fukuhara}}, \ and\ \bibinfo {author}
  {\bibfnamefont {K.}~\bibnamefont {Maezawa}},\ }\href@noop {} {\bibfield
  {journal} {\bibinfo  {journal} {Journal of the Physical Society of Japan}\
  }\textbf {\bibinfo {volume} {66}},\ \bibinfo {pages} {2993} (\bibinfo {year}
  {1997})}\BibitemShut {NoStop}%
\bibitem [{\citenamefont {Schroder}\ \emph {et~al.}(2000)\citenamefont
  {Schroder}, \citenamefont {Aeppli}, \citenamefont {Coldea}, \citenamefont
  {Adams}, \citenamefont {Stockert}, \citenamefont {von Lohneysen},
  \citenamefont {Bucher}, \citenamefont {Ramazashvili},\ and\ \citenamefont
  {Coleman}}]{Schroder2000}%
  \BibitemOpen
  \bibfield  {author} {\bibinfo {author} {\bibfnamefont {A.}~\bibnamefont
  {Schroder}}, \bibinfo {author} {\bibfnamefont {G.}~\bibnamefont {Aeppli}},
  \bibinfo {author} {\bibfnamefont {R.}~\bibnamefont {Coldea}}, \bibinfo
  {author} {\bibfnamefont {M.}~\bibnamefont {Adams}}, \bibinfo {author}
  {\bibfnamefont {O.}~\bibnamefont {Stockert}}, \bibinfo {author}
  {\bibfnamefont {H.}~\bibnamefont {von Lohneysen}}, \bibinfo {author}
  {\bibfnamefont {E.}~\bibnamefont {Bucher}}, \bibinfo {author} {\bibfnamefont
  {R.}~\bibnamefont {Ramazashvili}}, \ and\ \bibinfo {author} {\bibfnamefont
  {P.}~\bibnamefont {Coleman}},\ }\href@noop {} {\bibfield  {journal} {\bibinfo
   {journal} {Nature}\ }\textbf {\bibinfo {volume} {407}},\ \bibinfo {pages}
  {351} (\bibinfo {year} {2000})}\BibitemShut {NoStop}%
\bibitem [{\citenamefont {Grosche}\ \emph {et~al.}(2000)\citenamefont
  {Grosche}, \citenamefont {Agarwal}, \citenamefont {Julian}, \citenamefont
  {Wilson}, \citenamefont {Haselwimmer}, \citenamefont {Lister}, \citenamefont
  {Mathur}, \citenamefont {Carter}, \citenamefont {Saxena},\ and\ \citenamefont
  {Lonzarich}}]{Grosche2000}%
  \BibitemOpen
  \bibfield  {author} {\bibinfo {author} {\bibfnamefont {F.~M.}\ \bibnamefont
  {Grosche}}, \bibinfo {author} {\bibfnamefont {P.}~\bibnamefont {Agarwal}},
  \bibinfo {author} {\bibfnamefont {S.~R.}\ \bibnamefont {Julian}}, \bibinfo
  {author} {\bibfnamefont {N.~J.}\ \bibnamefont {Wilson}}, \bibinfo {author}
  {\bibfnamefont {R.~K.~W.}\ \bibnamefont {Haselwimmer}}, \bibinfo {author}
  {\bibfnamefont {S.~J.~S.}\ \bibnamefont {Lister}}, \bibinfo {author}
  {\bibfnamefont {N.~D.}\ \bibnamefont {Mathur}}, \bibinfo {author}
  {\bibfnamefont {F.~V.}\ \bibnamefont {Carter}}, \bibinfo {author}
  {\bibfnamefont {S.~S.}\ \bibnamefont {Saxena}}, \ and\ \bibinfo {author}
  {\bibfnamefont {G.~G.}\ \bibnamefont {Lonzarich}},\ }\href@noop {} {\bibfield
   {journal} {\bibinfo  {journal} {Journal of Physics: Condensed Matter}\
  }\textbf {\bibinfo {volume} {12}},\ \bibinfo {pages} {L553} (\bibinfo {year}
  {2000})}\BibitemShut {NoStop}%
\bibitem [{\citenamefont {Trovarelli}\ \emph {et~al.}(2000)\citenamefont
  {Trovarelli}, \citenamefont {Geibel}, \citenamefont {Mederle}, \citenamefont
  {Langhammer}, \citenamefont {Grosche}, \citenamefont {Gegenwart},
  \citenamefont {Lang}, \citenamefont {Sparn},\ and\ \citenamefont
  {Steglich}}]{Trovarelli2000}%
  \BibitemOpen
  \bibfield  {author} {\bibinfo {author} {\bibfnamefont {O.}~\bibnamefont
  {Trovarelli}}, \bibinfo {author} {\bibfnamefont {C.}~\bibnamefont {Geibel}},
  \bibinfo {author} {\bibfnamefont {S.}~\bibnamefont {Mederle}}, \bibinfo
  {author} {\bibfnamefont {C.}~\bibnamefont {Langhammer}}, \bibinfo {author}
  {\bibfnamefont {F.~M.}\ \bibnamefont {Grosche}}, \bibinfo {author}
  {\bibfnamefont {P.}~\bibnamefont {Gegenwart}}, \bibinfo {author}
  {\bibfnamefont {M.}~\bibnamefont {Lang}}, \bibinfo {author} {\bibfnamefont
  {G.}~\bibnamefont {Sparn}}, \ and\ \bibinfo {author} {\bibfnamefont
  {F.}~\bibnamefont {Steglich}},\ }\href@noop {} {\bibfield  {journal}
  {\bibinfo  {journal} {Physical Review Letters}\ }\textbf {\bibinfo {volume}
  {85}},\ \bibinfo {pages} {626} (\bibinfo {year} {2000})}\BibitemShut
  {NoStop}%
\bibitem [{\citenamefont {Fisher}\ \emph {et~al.}(2002)\citenamefont {Fisher},
  \citenamefont {Bouquet}, \citenamefont {Phillips}, \citenamefont {Hundley},
  \citenamefont {Pagliuso}, \citenamefont {Sarrao}, \citenamefont {Fisk},\ and\
  \citenamefont {Thompson}}]{Fisher2002x}%
  \BibitemOpen
  \bibfield  {author} {\bibinfo {author} {\bibfnamefont {R.~A.}\ \bibnamefont
  {Fisher}}, \bibinfo {author} {\bibfnamefont {F.}~\bibnamefont {Bouquet}},
  \bibinfo {author} {\bibfnamefont {N.~E.}\ \bibnamefont {Phillips}}, \bibinfo
  {author} {\bibfnamefont {M.~F.}\ \bibnamefont {Hundley}}, \bibinfo {author}
  {\bibfnamefont {P.~G.}\ \bibnamefont {Pagliuso}}, \bibinfo {author}
  {\bibfnamefont {J.~L.}\ \bibnamefont {Sarrao}}, \bibinfo {author}
  {\bibfnamefont {Z.}~\bibnamefont {Fisk}}, \ and\ \bibinfo {author}
  {\bibfnamefont {J.~D.}\ \bibnamefont {Thompson}},\ }\href@noop {} {\bibfield
  {journal} {\bibinfo  {journal} {Physical Review B}\ }\textbf {\bibinfo
  {volume} {65}},\ \bibinfo {pages} {224509} (\bibinfo {year}
  {2002})}\BibitemShut {NoStop}%
\bibitem [{\citenamefont {Custers}\ \emph {et~al.}(2003)\citenamefont
  {Custers}, \citenamefont {Gegenwart}, \citenamefont {Wilhelm}, \citenamefont
  {Neumaier}, \citenamefont {Tokiwa}, \citenamefont {Trovarelli}, \citenamefont
  {Geibel}, \citenamefont {Steglich}, \citenamefont {Pepin},\ and\
  \citenamefont {Coleman}}]{Custers2003}%
  \BibitemOpen
  \bibfield  {author} {\bibinfo {author} {\bibfnamefont {J.}~\bibnamefont
  {Custers}}, \bibinfo {author} {\bibfnamefont {P.}~\bibnamefont {Gegenwart}},
  \bibinfo {author} {\bibfnamefont {H.}~\bibnamefont {Wilhelm}}, \bibinfo
  {author} {\bibfnamefont {K.}~\bibnamefont {Neumaier}}, \bibinfo {author}
  {\bibfnamefont {Y.}~\bibnamefont {Tokiwa}}, \bibinfo {author} {\bibfnamefont
  {O.}~\bibnamefont {Trovarelli}}, \bibinfo {author} {\bibfnamefont
  {C.}~\bibnamefont {Geibel}}, \bibinfo {author} {\bibfnamefont
  {F.}~\bibnamefont {Steglich}}, \bibinfo {author} {\bibfnamefont
  {C.}~\bibnamefont {Pepin}}, \ and\ \bibinfo {author} {\bibfnamefont
  {P.}~\bibnamefont {Coleman}},\ }\href@noop {} {\bibfield  {journal} {\bibinfo
   {journal} {Nature}\ }\textbf {\bibinfo {volume} {424}},\ \bibinfo {pages}
  {524} (\bibinfo {year} {2003})}\BibitemShut {NoStop}%
\bibitem [{\citenamefont {Si}\ \emph {et~al.}(2001)\citenamefont {Si},
  \citenamefont {Rabello}, \citenamefont {Ingersent},\ and\ \citenamefont
  {Smith}}]{Si2001}%
  \BibitemOpen
  \bibfield  {author} {\bibinfo {author} {\bibfnamefont {Q.}~\bibnamefont
  {Si}}, \bibinfo {author} {\bibfnamefont {S.}~\bibnamefont {Rabello}},
  \bibinfo {author} {\bibfnamefont {K.}~\bibnamefont {Ingersent}}, \ and\
  \bibinfo {author} {\bibfnamefont {J.~L.}\ \bibnamefont {Smith}},\ }\href@noop
  {} {\bibfield  {journal} {\bibinfo  {journal} {Nature}\ }\textbf {\bibinfo
  {volume} {413}},\ \bibinfo {pages} {804} (\bibinfo {year}
  {2001})}\BibitemShut {NoStop}%
\bibitem [{\citenamefont {Si}(2006)}]{Si2006}%
  \BibitemOpen
  \bibfield  {author} {\bibinfo {author} {\bibfnamefont {Q.}~\bibnamefont
  {Si}},\ }\href@noop {} {\bibfield  {journal} {\bibinfo  {journal} {Physica
  B}\ }\textbf {\bibinfo {volume} {378-380}},\ \bibinfo {pages} {23} (\bibinfo
  {year} {2006})},\ \bibinfo {note} {kondo lattices; quantum
  criticality}\BibitemShut {NoStop}%
\bibitem [{\citenamefont {Coleman}\ and\ \citenamefont
  {Pepin}(2003)}]{Coleman2003y}%
  \BibitemOpen
  \bibfield  {author} {\bibinfo {author} {\bibfnamefont {P.}~\bibnamefont
  {Coleman}}\ and\ \bibinfo {author} {\bibfnamefont {C.}~\bibnamefont
  {Pepin}},\ }\href@noop {} {\bibfield  {journal} {\bibinfo  {journal} {Acta
  Physica Polonica B}\ }\textbf {\bibinfo {volume} {34}},\ \bibinfo {pages}
  {691} (\bibinfo {year} {2003})}\BibitemShut {NoStop}%
\bibitem [{\citenamefont {P{\'e}pin}(2005)}]{Pepin2005}%
  \BibitemOpen
  \bibfield  {author} {\bibinfo {author} {\bibfnamefont {C.}~\bibnamefont
  {P{\'e}pin}},\ }\href@noop {} {\bibfield  {journal} {\bibinfo  {journal}
  {Physical Review Letters}\ }\textbf {\bibinfo {volume} {94}},\ \bibinfo
  {pages} {066402} (\bibinfo {year} {2005})}\BibitemShut {NoStop}%
\bibitem [{\citenamefont {Rech}\ \emph {et~al.}(2006)\citenamefont {Rech},
  \citenamefont {Coleman}, \citenamefont {Zarand},\ and\ \citenamefont
  {Parcollet}}]{Rech2006}%
  \BibitemOpen
  \bibfield  {author} {\bibinfo {author} {\bibfnamefont {J.}~\bibnamefont
  {Rech}}, \bibinfo {author} {\bibfnamefont {P.}~\bibnamefont {Coleman}},
  \bibinfo {author} {\bibfnamefont {G.}~\bibnamefont {Zarand}}, \ and\ \bibinfo
  {author} {\bibfnamefont {O.}~\bibnamefont {Parcollet}},\ }\href@noop {}
  {\bibfield  {journal} {\bibinfo  {journal} {Physical Review Letters}\
  }\textbf {\bibinfo {volume} {96}},\ \bibinfo {pages} {016601} (\bibinfo
  {year} {2006})}\BibitemShut {NoStop}%
\bibitem [{\citenamefont {Senthil}\ \emph {et~al.}(2004)\citenamefont
  {Senthil}, \citenamefont {Vojta},\ and\ \citenamefont
  {Sachdev}}]{Senthil2004b}%
  \BibitemOpen
  \bibfield  {author} {\bibinfo {author} {\bibfnamefont {T.}~\bibnamefont
  {Senthil}}, \bibinfo {author} {\bibfnamefont {M.}~\bibnamefont {Vojta}}, \
  and\ \bibinfo {author} {\bibfnamefont {S.}~\bibnamefont {Sachdev}},\
  }\href@noop {} {\bibfield  {journal} {\bibinfo  {journal} {Physical Review
  B}\ }\textbf {\bibinfo {volume} {69}},\ \bibinfo {pages} {035111} (\bibinfo
  {year} {2004})}\BibitemShut {NoStop}%
\bibitem [{\citenamefont {Senthil}\ \emph {et~al.}(2005)\citenamefont
  {Senthil}, \citenamefont {Sachdev},\ and\ \citenamefont
  {Vojta}}]{Senthil2005c}%
  \BibitemOpen
  \bibfield  {author} {\bibinfo {author} {\bibfnamefont {T.}~\bibnamefont
  {Senthil}}, \bibinfo {author} {\bibfnamefont {S.}~\bibnamefont {Sachdev}}, \
  and\ \bibinfo {author} {\bibfnamefont {M.}~\bibnamefont {Vojta}},\
  }\href@noop {} {\bibfield  {journal} {\bibinfo  {journal} {Physica B:
  Condensed Matter}\ }\textbf {\bibinfo {volume} {359}},\ \bibinfo {pages} {9}
  (\bibinfo {year} {2005})}\BibitemShut {NoStop}%
\bibitem [{\citenamefont {Sachdev}\ \emph {et~al.}(2012)\citenamefont
  {Sachdev}, \citenamefont {Metlitski},\ and\ \citenamefont
  {Punk}}]{Metlitski2012}%
  \BibitemOpen
  \bibfield  {author} {\bibinfo {author} {\bibfnamefont {S.}~\bibnamefont
  {Sachdev}}, \bibinfo {author} {\bibfnamefont {M.~A.}\ \bibnamefont
  {Metlitski}}, \ and\ \bibinfo {author} {\bibfnamefont {M.}~\bibnamefont
  {Punk}},\ }\href@noop {} {\bibfield  {journal} {\bibinfo  {journal} {Journal
  of Physics: Condensed Matter}\ }\textbf {\bibinfo {volume} {24}},\ \bibinfo
  {pages} {294205} (\bibinfo {year} {2012})}\BibitemShut {NoStop}%
\bibitem [{\citenamefont {Metlitski}\ \emph {et~al.}(2014)\citenamefont
  {Metlitski}, \citenamefont {Mross}, \citenamefont {Sachdev},\ and\
  \citenamefont {Senthil}}]{Metlitski2014}%
  \BibitemOpen
  \bibfield  {author} {\bibinfo {author} {\bibfnamefont {M.~A.}\ \bibnamefont
  {Metlitski}}, \bibinfo {author} {\bibfnamefont {D.~F.}\ \bibnamefont
  {Mross}}, \bibinfo {author} {\bibfnamefont {S.}~\bibnamefont {Sachdev}}, \
  and\ \bibinfo {author} {\bibfnamefont {T.}~\bibnamefont {Senthil}},\
  }\href@noop {} {\  (\bibinfo {year} {2014})},\ \bibinfo {note}
  {arXiv:1403.3694}\BibitemShut {NoStop}%
\bibitem [{\citenamefont {Senthil}\ and\ \citenamefont
  {Fisher}(2000)}]{senthil00}%
  \BibitemOpen
  \bibfield  {author} {\bibinfo {author} {\bibfnamefont {T.}~\bibnamefont
  {Senthil}}\ and\ \bibinfo {author} {\bibfnamefont {M.~P.~A.}\ \bibnamefont
  {Fisher}},\ }\href@noop {} {\bibfield  {journal} {\bibinfo  {journal}
  {Physical Review B}\ }\textbf {\bibinfo {volume} {62}},\ \bibinfo {pages}
  {7850} (\bibinfo {year} {2000})}\BibitemShut {NoStop}%
\bibitem [{\citenamefont {Balents}\ \emph {et~al.}(2005)\citenamefont
  {Balents}, \citenamefont {Bartosch}, \citenamefont {Burkov}, \citenamefont
  {Sachdev},\ and\ \citenamefont {Sengupta}}]{balents05}%
  \BibitemOpen
  \bibfield  {author} {\bibinfo {author} {\bibfnamefont {L.}~\bibnamefont
  {Balents}}, \bibinfo {author} {\bibfnamefont {L.}~\bibnamefont {Bartosch}},
  \bibinfo {author} {\bibfnamefont {A.}~\bibnamefont {Burkov}}, \bibinfo
  {author} {\bibfnamefont {S.}~\bibnamefont {Sachdev}}, \ and\ \bibinfo
  {author} {\bibfnamefont {K.}~\bibnamefont {Sengupta}},\ }\href@noop {}
  {\bibfield  {journal} {\bibinfo  {journal} {Physical Review B}\ }\textbf
  {\bibinfo {volume} {71}},\ \bibinfo {pages} {144508} (\bibinfo {year}
  {2005})}\BibitemShut {NoStop}%
\bibitem [{\citenamefont {Metlitski}\ \emph {et~al.}(2013)\citenamefont
  {Metlitski}, \citenamefont {Kane},\ and\ \citenamefont
  {Fisher}}]{Metlitski2013}%
  \BibitemOpen
  \bibfield  {author} {\bibinfo {author} {\bibfnamefont {M.~A.}\ \bibnamefont
  {Metlitski}}, \bibinfo {author} {\bibfnamefont {C.~L.}\ \bibnamefont {Kane}},
  \ and\ \bibinfo {author} {\bibfnamefont {M.~P.~A.}\ \bibnamefont {Fisher}},\
  }\href@noop {} {\  (\bibinfo {year} {2013})}\BibitemShut {NoStop}%
\bibitem [{\citenamefont {Bonderson}\ \emph {et~al.}(2013)\citenamefont
  {Bonderson}, \citenamefont {Nayak},\ and\ \citenamefont
  {Qi}}]{Bonderson2013}%
  \BibitemOpen
  \bibfield  {author} {\bibinfo {author} {\bibfnamefont {P.}~\bibnamefont
  {Bonderson}}, \bibinfo {author} {\bibfnamefont {C.}~\bibnamefont {Nayak}}, \
  and\ \bibinfo {author} {\bibfnamefont {X.-L.}\ \bibnamefont {Qi}},\
  }\href@noop {} {\bibfield  {journal} {\bibinfo  {journal} {Journal of
  Statistical Mechanics: Theory and Experiment}\ ,\ \bibinfo {pages} {P09016}}
  (\bibinfo {year} {2013})}\BibitemShut {NoStop}%
\bibitem [{\citenamefont {Wang}\ \emph {et~al.}(2014)\citenamefont {Wang},
  \citenamefont {Potter},\ and\ \citenamefont {Senthil}}]{Senthil2014}%
  \BibitemOpen
  \bibfield  {author} {\bibinfo {author} {\bibfnamefont {C.}~\bibnamefont
  {Wang}}, \bibinfo {author} {\bibfnamefont {A.~C.}\ \bibnamefont {Potter}}, \
  and\ \bibinfo {author} {\bibfnamefont {T.}~\bibnamefont {Senthil}},\
  }\href@noop {} {\bibfield  {journal} {\bibinfo  {journal} {Science}\ }\textbf
  {\bibinfo {volume} {343}},\ \bibinfo {pages} {6171} (\bibinfo {year}
  {2014})}\BibitemShut {NoStop}%
\bibitem [{\citenamefont {Senthil}(2014)}]{Senthil2014a}%
  \BibitemOpen
  \bibfield  {author} {\bibinfo {author} {\bibfnamefont {T.}~\bibnamefont
  {Senthil}},\ }\href@noop {} {\  (\bibinfo {year} {2014})},\ \bibinfo {note}
  {arXiv:1405.4015}\BibitemShut {NoStop}%
\bibitem [{\citenamefont {Halperin}\ \emph {et~al.}(1993)\citenamefont
  {Halperin}, \citenamefont {Lee},\ and\ \citenamefont {Read}}]{Halperin1993}%
  \BibitemOpen
  \bibfield  {author} {\bibinfo {author} {\bibfnamefont {B.~I.}\ \bibnamefont
  {Halperin}}, \bibinfo {author} {\bibfnamefont {P.~A.}\ \bibnamefont {Lee}}, \
  and\ \bibinfo {author} {\bibfnamefont {N.}~\bibnamefont {Read}},\ }\href@noop
  {} {\bibfield  {journal} {\bibinfo  {journal} {Physical Review B}\ }\textbf
  {\bibinfo {volume} {47}},\ \bibinfo {pages} {7312} (\bibinfo {year}
  {1993})}\BibitemShut {NoStop}%
\bibitem [{\citenamefont {Altshuler}\ \emph {et~al.}(1994)\citenamefont
  {Altshuler}, \citenamefont {Ioffe},\ and\ \citenamefont
  {Millis}}]{Altshuler1994}%
  \BibitemOpen
  \bibfield  {author} {\bibinfo {author} {\bibfnamefont {B.~L.}\ \bibnamefont
  {Altshuler}}, \bibinfo {author} {\bibfnamefont {L.~B.}\ \bibnamefont
  {Ioffe}}, \ and\ \bibinfo {author} {\bibfnamefont {A.~J.}\ \bibnamefont
  {Millis}},\ }\href@noop {} {\bibfield  {journal} {\bibinfo  {journal}
  {Physical Review B}\ }\textbf {\bibinfo {volume} {50}},\ \bibinfo {pages}
  {14048} (\bibinfo {year} {1994})}\BibitemShut {NoStop}%
\bibitem [{\citenamefont {Kim}\ \emph {et~al.}(1994)\citenamefont {Kim},
  \citenamefont {Furusaki}, \citenamefont {Wen},\ and\ \citenamefont
  {Lee}}]{Kim1994a}%
  \BibitemOpen
  \bibfield  {author} {\bibinfo {author} {\bibfnamefont {Y.~B.}\ \bibnamefont
  {Kim}}, \bibinfo {author} {\bibfnamefont {A.}~\bibnamefont {Furusaki}},
  \bibinfo {author} {\bibfnamefont {X.-G.}\ \bibnamefont {Wen}}, \ and\
  \bibinfo {author} {\bibfnamefont {P.~A.}\ \bibnamefont {Lee}},\ }\href@noop
  {} {\bibfield  {journal} {\bibinfo  {journal} {Physical Review B}\ }\textbf
  {\bibinfo {volume} {50}},\ \bibinfo {pages} {17917} (\bibinfo {year}
  {1994})}\BibitemShut {NoStop}%
\bibitem [{\citenamefont {Kim}\ \emph {et~al.}(1995{\natexlab{a}})\citenamefont
  {Kim}, \citenamefont {Lee}, \citenamefont {Wen},\ and\ \citenamefont
  {Stamp}}]{Kim1995}%
  \BibitemOpen
  \bibfield  {author} {\bibinfo {author} {\bibfnamefont {Y.~B.}\ \bibnamefont
  {Kim}}, \bibinfo {author} {\bibfnamefont {P.~A.}\ \bibnamefont {Lee}},
  \bibinfo {author} {\bibfnamefont {X.~G.}\ \bibnamefont {Wen}}, \ and\
  \bibinfo {author} {\bibfnamefont {P.~C.~E.}\ \bibnamefont {Stamp}},\
  }\href@noop {} {\bibfield  {journal} {\bibinfo  {journal} {Physical Review
  B}\ }\textbf {\bibinfo {volume} {51}},\ \bibinfo {pages} {10779} (\bibinfo
  {year} {1995}{\natexlab{a}})}\BibitemShut {NoStop}%
\bibitem [{\citenamefont {Kim}\ \emph {et~al.}(1995{\natexlab{b}})\citenamefont
  {Kim}, \citenamefont {Lee},\ and\ \citenamefont {Wen}}]{Kim1995a}%
  \BibitemOpen
  \bibfield  {author} {\bibinfo {author} {\bibfnamefont {Y.~B.}\ \bibnamefont
  {Kim}}, \bibinfo {author} {\bibfnamefont {P.~A.}\ \bibnamefont {Lee}}, \ and\
  \bibinfo {author} {\bibfnamefont {X.~G.}\ \bibnamefont {Wen}},\ }\href@noop
  {} {\bibfield  {journal} {\bibinfo  {journal} {Physical Review B}\ }\textbf
  {\bibinfo {volume} {52}},\ \bibinfo {pages} {17275} (\bibinfo {year}
  {1995}{\natexlab{b}})}\BibitemShut {NoStop}%
\bibitem [{\citenamefont {Kim}\ and\ \citenamefont {Lee}(1996)}]{Kim1996}%
  \BibitemOpen
  \bibfield  {author} {\bibinfo {author} {\bibfnamefont {Y.~B.}\ \bibnamefont
  {Kim}}\ and\ \bibinfo {author} {\bibfnamefont {P.~A.}\ \bibnamefont {Lee}},\
  }\href@noop {} {\bibfield  {journal} {\bibinfo  {journal} {Physical Review
  B}\ }\textbf {\bibinfo {volume} {54}},\ \bibinfo {pages} {2715} (\bibinfo
  {year} {1996})}\BibitemShut {NoStop}%
\bibitem [{\citenamefont {Nikolic}(2011)}]{Nikolic2010}%
  \BibitemOpen
  \bibfield  {author} {\bibinfo {author} {\bibfnamefont {P.}~\bibnamefont
  {Nikolic}},\ }\href@noop {} {\bibfield  {journal} {\bibinfo  {journal}
  {Physical Review B}\ }\textbf {\bibinfo {volume} {83}},\ \bibinfo {pages}
  {064523} (\bibinfo {year} {2011})}\BibitemShut {NoStop}%
\bibitem [{\citenamefont {Nikoli{\'c}}(2014)}]{Nikolic2014}%
  \BibitemOpen
  \bibfield  {author} {\bibinfo {author} {\bibfnamefont {P.}~\bibnamefont
  {Nikoli{\'c}}},\ }\href@noop {} {\  (\bibinfo {year} {2014})},\ \bibinfo
  {note} {arXiv:1406.1198}\BibitemShut {NoStop}%
\bibitem [{\citenamefont {Nikolic}(2009)}]{nikolic:144507}%
  \BibitemOpen
  \bibfield  {author} {\bibinfo {author} {\bibfnamefont {P.}~\bibnamefont
  {Nikolic}},\ }\href@noop {} {\bibfield  {journal} {\bibinfo  {journal}
  {Physical Review B}\ }\textbf {\bibinfo {volume} {79}},\ \bibinfo {pages}
  {144507} (\bibinfo {year} {2009})}\BibitemShut {NoStop}%
\end{thebibliography}

%

\end{document}